\documentclass[fleqn,usenatbib]{mnras}

\usepackage[T1]{fontenc}

\DeclareRobustCommand{\VAN}[3]{#2}
\let\VANthebibliography\thebibliography
\def\thebibliography{\DeclareRobustCommand{\VAN}[3]{##3}\VANthebibliography}

\usepackage{graphicx}	
\usepackage{amsmath}	
\usepackage{amssymb}	
\usepackage{subfig}

\usepackage{newtxtext,newtxmath}
\usepackage{tablefootnote}

\title[Evolution of Eu and Ba]{Evolution of neutron capture elements in dwarf galaxies}

\author[M. Molero et al.]{
Marta Molero,$^{1}$\thanks{E-mail: marta.molero@phd.units.it}
Donatella Romano,$^{2}$
Moritz Reichert,$^{5}$
Francesca Matteucci,$^{1,3,4}$
Almudena Arcones,$^{5,6}$
\newauthor
Gabriele Cescutti,$^{3}$
Paolo Simonetti,$^{1,3}$
Camilla Juul Hansen,$^{7,8}$
Gustavo A. Lanfranchi$^{9}$
\\
$^{1}$Dipartimento di Fisica, Sezione di Astronomia, Università degli studi di Trieste, Via G.B. Tiepolo 11, I-34143 Trieste, Italy\\
$^{2}$INAF, Osservatorio di Astrofisica e Scienza dello Spazio, Via Gobetti 93/3, I-40129 Bologna, Italy\\
$^{3}$INAF, Osservatorio Astronomico di Trieste, Via Tiepolo 11, I-34131 Trieste, Italy\\
$^{4}$INFN, Sezione di Trieste, Via Valerio 2, I.34127 Trieste, Italy\\
$^{5}$Institut f\"ur Kernphysik, Technische Universit\"at Darmstadt, Schlossgartenstr. 2, Darmstadt 64289, Germany\\
$^{6}$GSI Helmholtzzentrum f\"ur Schwerionenforschung GmbH, Planckstr. 1, Darmstadt 64291, Germany\\
$^{7}$Max-Planck-institut f\"ur Astronomie, K\"onigstuhl 17, D-69117 Heidelberg, Germany\\
$^{8}$Copenhagen University, Dark Cosmology Centre, The Niels Bohr Institute, Vibenshuset, Lyngbyvej 2, DK-2100 Copenhagen, Denmark\\
$^{9}$N\'ucleo de Astrof\'\i sica, Universidade Cidade de S\~ao Paulo , R. Galv\~ao Bueno 868, Liberdade, 01506-000, S\~ao Paulo, SP, Brazil 
}

\date{Accepted XXX. Received YYY; in original form ZZZ}

\pubyear{2015}

\begin{document}
\label{firstpage}
\pagerange{\pageref{firstpage}--\pageref{lastpage}}
\maketitle

\begin{abstract}
We study the evolution of Eu and Ba abundances in local group dwarf spheroidal and ultra faint dwarf galaxies by means of detailed chemical evolution models and compare our results with new sets of homogeneous abundances. The adopted models include gas infall and outflow and have been previously tested. We investigate several production scenarios for r-process elements: merging neutron stars and magneto-rotational driven supernovae. Production of Ba through the main s-process acting in low- and intermediate- mass stars is considered as well. We also test different sets of nucleosynthesis yields. For merging neutron stars we adopt either a constant and short delay time for merging or a delay time distribution function. Our simulations show that: i) if r-process elements are produced only by a quick source, it is possible to reproduce the [Eu/Fe] vs [Fe/H], but those models fail in reproducing the [Ba/Fe] vs [Fe/H]. ii) If r-process elements are produced only with longer delays the opposite happens. iii) If both a quick source and a delayed one are adopted, such as magneto-rotational driven supernovae and merging neutron stars with a delay time distribution, the [Eu/Fe] abundance pattern is successfully reproduced, but models still fail in reproducing the [Ba/Fe]. iv) On the other hand, the characteristic abundances of Reticulum II can be reproduced only if both the Eu and the r-process fraction of Ba are produced on short and constant time delays during a single merging event. We discuss also other possible interpretations, including an inhomogeneous mixing of gas which might characterize this galaxy.

\end{abstract}




\section{Introduction}

The majority of heavy elements beyond the iron peak originate via neutron capture. Neutron capture processes can be \textit{slow} (s-process) or \textit{rapid} (r-process) with respect to the $\beta$-decay in nuclei. These two processes require different astrophysical conditions, in terms of neutron density and temperature, and therefore they occur in different sites. 

It has now been confirmed that the main s-process component takes place in the asymptotic giant branch (AGB) phase of low and intermediate mass stars (LIMS) (\citealp{1999busso}), via the reaction $\mathrm{^{13}C(\alpha,n)^{16}O}$. The s-process can also occur in massive stars as a "weak" s-process (\citealp{1989langer}; \citealp{1990Prantzos}), in this case the neutrons are produced through the reaction $\mathrm{^{22}Ne(\alpha,n)^{25}Mg}$. The efficiency of the weak s-process depends on metallicity, and for metallicity lower than $\sim10^{-4}$ the s-process becomes negligible (\citealp{1991Raiteri}; \citealp{2003Limongi}). However, recent studies have shown that rotation can increase the efficiency of the s-process in massive stars, in particular at low metallicities, where stars are expected to be more compact and to rotate faster (\citealp{2011Chiappini}; \citealp{2016Frischknnecht}; \citealp{2018Limongi}; \citealp{2019Rizzuti}). 

On the other hand, the major astrophysical r-process site is still under debate, with possible candidate sites being supernovae or neutron star mergers (see e.g.: \citealp{2011thielemann}; \citealp{2018frebel}; \citealp{2019cowan}). For many years the occurrence of r-process has been associated with core-collapse supernovae (CC-SNe; \citealp{1994woosley, 2012horowitz}). However, the prompt explosion mechanism which was believed to eject extremely neutron-rich matter, has been completely ruled out by earlier hydrodynamical nucleosynthesis calculations. Simulations showed not only difficulties in reproducing the high entropy needed to reproduce the solar r-process abundances (\citealp{Wanajo_2006}; \citealp{2007arcones}), but also that the neutrino winds which follow the supernovae explosion are only slightly neutron-rich or even proton-rich (\citealp{2006frolich}; \citealp{2010fischer}; \citealp{2013arcones}), providing insufficient conditions for the production of heavy r-process elements.

Among massive stars, CC-SNe induced by strong magnetic fields and/or fast rotation of the stellar core (magneto-rotational driven supernovae, MR-SNe) seem to also provide a source for the r-process (\citealp{Winteler2012}; \citealp{Nishimura2015}; \citealp{Nishimura2017}; \citealp{2018mosta}; \citealp{2018halevi}; \citealp{2021reichert}). However, the required rotation rates and magnetic energies restrict the mechanism to a minority of progenitor stars: only 1\% of all stars with initial mass larger than 10 $\mathrm{M_\odot}$ may have the necessary conditions to host strong enough magnetic fields, according to \cite{WoosleyHeger2006}. Nevertheless, the rarity of progenitors with this required initial conditions can provide an explanation for the observed scatter in the abundances of r-process elements for low metallicity stars.

Another possibility is represented by a rare class of CC-SNe, known as "collapsars", which are the supernova-triggering collapse of rapidly rotating massive stars. According to \cite{2019siegel}, collapsar accretion disks yield sufficient r-process elements to explain the observed abundances in the Universe. However, further investigations are needed to confirm this scenario, which still remains very poorly constrained observationally.

Merging neutron stars (MNS) have been supposed to be powerful sources of r-process matter and this has been proved thanks to the observation of the kilonova AT2017gfo, following the gravitational wave event GW170817 (\citealp{2017abbott}; \citealp{Watson2019}). However, although both the r-process yields and the estimated rate of this phenomena seem to point towards MNS as the main r-process astronomical source, galactic chemical evolution models have problems to reproduce the r-process abundance pattern in the Galaxy if MNS are considered the only producers. \cite{Matteucci2014} introduced MNS in a chemical evolution model, showing that MNS can be the only r-process producers in the Galaxy, if a short and constant delay time for merging is adopted. If a more realistic delay time distribution (DTD) for merging is assumed (\citealp{Simonetti2019}; \citealp{2019cot}; \citealp{Molero2021}), then an additional r-process source must be included, especially at low metallicities. This is also in agreement with the large star-to-star heavy element abundance spread seen in observations at low metallicity (e.g., \citealp{2007francois, 2012hansen}).

Among Milky Way satellites it is possible to distinguish between classical and ultra-faint dwarf spheroidal galaxies. Classical dwarf spheroidal galaxies (dSphs) are among the least luminous and most dark matter dominated galaxies observed. They can be classified as early type galaxies, since they are characterized by low present time gas mass and iron-poor stars (\citealp{2009koch}; \citealp{2009tolstoy}). Ultra-faint dwarf spheroidals (UFDs) have very similar physical properties to dSph galaxies, but are characterized by even smaller average surface brightness and effective radii (see \cite{2019simon} for a recent review). Both dSphs and UFDs are characterized by old ($>$ 10 Gyr) or intermediate-aged stellar population, with only some of them hosting younger stars, which is an indication of a recent star formation activity.

With the goal of better understanding both the r- and s- process production sites at low metallicity, we study the chemical evolution of Eu, taken as a typical r-process element, and Ba abundances in 6 dSphs and 2 UFDs for which homogeneous abundances have been published by \cite{Reichert2020}. We compute chemical evolution models in which we include LIMS as main s-process producers and test different scenarios for the production of Eu and r-process fraction of Ba: MNS and MRD-SNe. 

The paper is organized as follows: in Section \ref{observational data} we describe the observational data; Sections \ref{chemical evolution model} and \ref{sec:Nucleosynthesis Prescriptions} describe respectively the chemical evolution model and the adopted stellar yields; in Section \ref{results} we present our results for Sculptor, Fornax and Reticulum II. Finally, in Section \ref{conclusions} we summarize our conclusions.

\section{OBSERVATIONAL DATA}
\label{observational data}

We have modelled the chemical evolution of 6 dSph and 2 UFD galaxies, which are: B\"ootes I (Boo I), Carina (Car), Fornax (For), Leo I (Leo), Reticulum II (Reticulum II), Sculptor (Scl), Sagittarius (Sgr) and Ursa Minor I (Umi I). In the main text of this paper we will focus on the results for Sculptor, Fornax and Reticulum II. Our results for the other galaxies are provided as Supplementary Material (online only). We chose Sculptor and Fornax since their results are representative of those obtained for all the other dSphs, and Reticulum II because of its peculiar heavy elements abundances.

We have chosen abundances data of \cite{Reichert2020} for all galaxies, while for the metallicity distribution functions (MDF) we adopted collections of data from the SAGA database (\citealp{2008suda}) for all galaxies, except for Sculptor, for which the observational MDF is taken from \cite{Romano2013}. All abundances are scaled to the solar photosphere abundances of \cite{2009asplund}, which is the one adopted in our chemical evolution code. For comparison, we adopt the star formation histories (SFH) as derived by color-magnitude diagrams (CMD) fitting analysis of several authors (\citealp{Hernandez2000}; \citealp{Dolphin2002}; \citealp{deBoer2012}; \citealp{fornaxdeBoer2012}; \citealp{Brown2014}; \citealp{deBoer2015}). In particular, we assumed the same number and duration of the SF episodes of the CMDs.

\section{THE CHEMICAL EVOLUTION MODEL}
\label{chemical evolution model}

We use an updated version of the model presented by \cite{Lanfranchi&Matteucci04} to describe the chemical evolution of both UFDs and dSphs. Galaxies form by infall of primordial gas in a pre-existing diffuse dark matter (DM) halo. The model is a one zone with instantaneous and complete mixing of gas. The stellar lifetimes are taken into account, thus relaxing the instantaneous recycling approximation (IRA). The model is able to follow the evolution of 31 elements, from H to Eu, during $14$ Gyr. 

The evolution with time of the gas mass in the form of the element $i$, $\mathrm{M_{gas,i}}(t)$, within the ISM is:

\begin{equation}
\begin{split}
  \dot{M}_{\mathrm{gas},i}(t) = & -\psi(t)X_i(t) + (\dot{M}_{\mathrm{gas},i})_{\mathrm{inf}}-(\dot{M}_{\mathrm{gas},i})_{\mathrm{out}} + \dot R_{i}(t),
\end{split}
\end{equation}
\\
where $\mathrm{X_i(t)=M_{gas,i}(t)/M_{gas}}(t)$ is the abundance by mass of the element $i$ at the time $t$ and $\mathrm{M_{gas}(t)}$ is the total gas mass of the galaxy.

The terms on the right-hand side of the equation are:

\begin{itemize}
    \item The first term is the rate at which chemical elements are subtracted by the ISM to be included in stars. $\mathrm{\psi(t)}$ is the star formation rate (SFR), which has the following form (Schmidt-Kennicutt law with $k=1$, \citealp{schmidt, kennicutt}):
    
    \begin{equation}
        \psi(t)=\nu {M_{\mathrm{gas}}}^k,
    \end{equation}
    \\
    where $\nu$ is the star formation efficiency which is expressed in Gyr$^{-1}$ and represents the inverse of the time needed to convert all the gas into stars.
    \item The second term is the rate at which chemical elements are accreted through infall of gas. It is given by the following relation:
    
    \begin{equation}
        (\dot{M}_{\mathrm{gas},i})_{\mathrm{inf}}=aX_{i,\mathrm{inf}}e^{-t/\tau_{\mathrm{inf}}},
    \end{equation}
    \\
    where $a$ is a normalization constant constrained to reproduce the present time total infall mass; $\mathrm{X_{i,inf}}$ describes the chemical abundance of the element $i$ of the infalling gas (here assumed to be primordial); and $\tau_{\mathrm{inf}}$ is the infall time-scale, defined as the time at which half of the total mass of the galaxy has been assembled.
    \item The third term is the rate at which chemical elements are lost through galactic winds. It is assumed to be proportional to the SFR:
    
    \begin{equation}
        (\dot{M}_{\mathrm{gas},i})_{\mathrm{out}}=-\omega\psi(t),
    \end{equation}
    \\
    where $\omega$ is a free dimensionless parameter called the mass loading factor. In our model we assumed $\omega$ to be equal for all the chemical elements. Galactic winds develop when the thermal energy of the gas $\mathrm{E_{gas}^{th}(t)}$, heated by SN explosions, exceeds its binding energy $\mathrm{E_{gas}^b(t)}$ (see \citealp{1994matteucci, 1998bradamante, Vincenzo2014}):
    
    \begin{equation}
        E_{\mathrm{gas}}^{\mathrm{th}}(t) \geq E_{\mathrm{gas}}^\mathrm{b}(t).
    \end{equation}
    \\
    The thermal energy of the gas is produced by SN explosions (of all types) and by stellar winds, while the binding energy of the gas is computed as:
    
    \begin{equation}
        E_{\mathrm{gas}}^\mathrm{b}(t)=W_\mathrm{L}(t)+W_{\mathrm{LD}}(t),
    \end{equation}
    \\
    where $\mathrm{W_L(t)}$ is the potential well due to the luminous matter and $\mathrm{W_{LD}(t)}$ represents the potential well due to the interaction between dark and luminous matter. This last term mainly depends on the mass of the dark matter halo ($\mathrm{M_{DM}}$) as well as on the ratio $\mathrm{S=R_L/R_{DM}}$ between the galaxy effective radius ($\mathrm{R_L}$) and the radius of the dark matter core ($\mathrm{R_{DM}}$) (see \citealp{1992bertin}).
    
    \item The last term $\mathrm{R_i(t)}$ represents the fraction of matter which is returned by stars into the ISM through stellar winds, SN explosions and MNS, in the form of the element $i$. In other words, it represents the rate at which each chemical element is restored into the ISM by all stars dying at the time $t$. $\mathrm{R_i(t)}$ depends also on the initial mass function  (IMF). Here we adopt a \cite{salpeter} IMF for all galaxies:
    
    \begin{equation}
    \varphi(t)=0.17m^{-(1+1.35)},
    \end{equation}
    \\
    normalised to unity between $0.1$ and $100 \mathrm{M_\odot}$. 
\end{itemize}

\section{Nucleosynthesis Prescriptions}
\label{sec:Nucleosynthesis Prescriptions} 

For all the stars sufficiently massive to die in a Hubble time, the following stellar yields have been adopted:

\begin{itemize}
\item For low and intermediate mass stars (LIMS) we included the metallicity-dependent stellar yields of \cite{karakas2010};
\item For massive stars we assumed yields of \cite{Kobayashi2006};
\item For Type Ia SNe we included yields of \cite{Iwamoto1999}. We adopted the single-degenerate scenario for SNeIa, in which SNe arise from the explosion via C-deflagration of a C-O white dwarf in a close binary system as it has reached the Chandrasekhar mass due to accretion from its red giant companion.
\end{itemize}

The same stellar yields have been adopted in \cite{Vincenzo2015}. Also, a complete and detailed description of those yields can be found in \cite{Romano2010}.

\subsection{Eu and Ba yields}
\label{sec: Eu/Ba yields}

For both Eu and the r-process fraction of Ba we considered two different production sites: MNS and MRD-SNe. We assume r-process elements to be produced by (i) only MNS, (ii) only MRD-SNe, (iii) both MNS and MRD-SNe. 

In our simulations, MNS are systems of two $1.4 \mathrm{M_\odot}$ neutron stars with progenitors in the $9-50 \mathrm{M_\odot}$ mass range. In order to include the production of r-process elements from MNS in our chemical evolution code, we need to specify the following parameters (see \citealp{Matteucci2014}):

\begin{itemize}
    \item the mass of each elements which is produced per merging event, $\mathrm{Y_{Eu}^{MNS}}$ and $\mathrm{Y_{Ba}^{MNS}}$;
    \item the time delay between the formation of the double neutron star system and the merging event, $\tau;$
    \item the fraction of neutron stars in binaries that produce a MNS, $\mathrm{\alpha_{MNS}}$.
\end{itemize}

For what it concerns the yields of r-process elements from MNS, they have been obtained by assuming that there is a scaling relation between them and those of Sr. The adopted scaling factors are equal to $0.03$ for Eu and to $3.16$ for Ba, and have been found from the solar system r-process contribution, as determined by \cite{Simmerer2004}. For the yields of Sr, we adopted the value found by \cite{Watson2019} in the reanalysis of the spectra of the kilonova AT2017gfo which followed the neutron-star merger GW170817, equal to $(1-5)\times10^{-5} \mathrm{M_\odot}$ Those yields have also been multiplied by two different factors ($1\times10^1$, $1\times10^2$) in order to take into account the uncertainties that could affect them, because of their model assumptions as well as the scatter of Sr compared to Eu in old stars. The yields of Sr with the scaled yields of Eu and Ba are reported in Table \ref{tab: Yields of Sr, Eu, Ba}. 

For the time delay $\tau$, we adopt two different approaches. The first one consists in assuming a constant delay time between the formation of the neutron stars binary system and the merging event (as first done by \citealp{2004argast} and later by \citealp{Matteucci2014}), while the second one consists in adopting a distribution of delay times (DTD). In the first case we adopt a delay time $\tau=1 \mathrm{Myr}$. This is equivalent to assume that all neutron stars binary systems would merge on the same time-scale, which is short and constant. In the second case, we adopt the following DTD (see \citealp{Simonetti2019} for a more detailed discussion and \citealp{2021greggio} where a refinement of the derivation of DTDs for MNS can be found):

\begin{equation}
    f(\tau)\propto 
    \begin{cases}
    \ 0 \qquad \text{if} \qquad \tau < 10 \mathrm{Myr} \\
    \ p_1 \qquad \text{if} \qquad  10 < \tau < 40 \mathrm{Myr} \\
    \ p_2\tau^{0.25\beta -0.75}(M_{m}^{0.75(\beta + 2.33)}-M_{M}^{0.75(\beta + 2.33)}) \\ 
    \ \text{if} \qquad  40 \mathrm{Myr} < \tau < 13.7 \mathrm{Gyr}
    \end{cases}
\label{eq: DTD}
\end{equation}
\\
where $\beta=-0.9$ is the parameter which characterizes the shape of the initial separation function; $p_1=3.521$ and $p_2=0.065$ have been chosen in order to obtain a continuous and normalized function; $M_m$ and $M_M$ are the minimum and maximum total mass of the system, respectively. The first portion of the distribution ends with the formation of the first double neutron star system. 10 Myr is in fact the nuclear lifetime of a typical massive star. The second portion refers to systems which merge soon after the formation of the double neutron star system. This portion of the distribution is described by a flat plateau, up to the lifetime of the minimum mass progenitor of a neutron star. The third part of the distribution is the distribution of the gravitational delay times and pertains to those systems for which the time delay is dominated by gravitational radiation.

The parameter $\mathrm{\alpha_{MNS}}$ is the probability of the MNS event. For a DTD with $\beta=-0.9$ it is equal to $\mathrm{\alpha_{MNS}=5.42\times10^{-2}}$ for spiral galaxies (\citealp{Molero2021}). This value has been fixed in order to reproduce the observed present time MNS rate in the Milky Way as the one predicted by \cite{Kalogera2004}, equal to $\sim80_{-60}^{+200} \mathrm{Myr^{-1}}$. It is reasonable to presume that in dwarf galaxies the present time rate of MNS is lower than the one in the Milky Way because of the lower SFR. Moreover, we adopted a lower probability of MNS in order to take into account the less efficient r-process material enrichment which characterizes dwarf galaxies. We set $\mathrm{\alpha_{MNS}=2.15\times10^{-2}}$ and based our consideration on the work of \cite{2019bonetti}, according to which in low mass galaxies neutron stars binary systems tend to merge with a large off-set from the host galaxy, because of the kicks imparted by the two SN explosions. As stated by the authors, the immediate consequence of a merger location detached from the disc plane, is a dilution of the amount of r-process material retained by the galaxy.  

For the production of r-process elements from MR-SNe, we select a set of yields from different nucleosynthetic studies as reported in Table \ref{tab: Yields of Eu from MRD-SNe}. Another possibility could have been that of adopting as Ba yields those obtained by scaling the Eu yields of the studies reported in the Table by taking into account the solar system r-process contribution, as done for MNS. We checked that this choice would not have been significantly affect our results. Moreover, we run also models in which we assume that Eu and r-process Ba are produced by massive stars in the $(12-30) \mathrm{M_\odot}$ mass range with the yields of \cite{Cescutti2006} (their model 1). Details of those yields are reported in Table \ref{tab: Yields of Ba Cescutti+06}. Finally, for the Ba s-process component, which is the predominant one, we have adopted yields of \cite{Busso2001} for LIMS of $(1.5-3.0) \mathrm{M_\odot}$. Those yields have a strong dependence on the initial metallicity of the stars. For stars of $(1.0-1.5) \mathrm{M_\odot}$ we have adopted yields of \cite{Cescutti2015}, which are obtained simply by scaling with the yields of \cite{Busso2001} to stars of $1.5 \mathrm{M_\odot}$. 

MRD-SNe may be indeed important contributors to the enrichment of heavy elements. However, the required rotation rates and magnetic energies restrict the mechanism to a minority of progenitor stars (\citealp{Nishimura2017, 2018mosta, 2021reichert}). \cite{WoosleyHeger2006} speculated that approximately $1\%$ of all stars with initial mass $\geq10 \mathrm{M_\odot}$ have the necessary conditions to host strong enough magnetic fields. Therefore, here, we assume that only $1\%-2\%$ of all stars with initial mass in the $(10-80) \mathrm{M_\odot}$ range would explode as a MRD-SNe. Furthermore, it has been suggested that these events occur more frequently at low metallicities, because of the lower opacity that result in higher rotation rates and, as a consequence, stronger magnetic fields (see e.g., \citealp{2011Brott}, \citealp{2017Thielemann}). Therefore, we also test models in which the production of r-process elements from MRD-SNe is active only at metallicity $Z\leq10^{-3}$, as suggested also in \citealp{Winteler2012} and \cite{Cescutti2015}.

Details of the different models that we run are reported in Table \ref{tab: models_Scl}. 

\begin{table}
\centering
\hspace{-0.5 cm}
\caption{\label{tab: Yields of Sr, Eu, Ba}Yields of Sr, Eu and Ba from MNS adopted in this work. Yields of Sr are those measured by \protect\cite{Watson2019} while those of Eu and Ba have been obtained as described in the text.}
\begin{tabular}{ccc}
\hline
  $\mathrm{Y_{Sr}^{MNS}}$ $(\mathrm{M_{\odot}})$ & $\mathrm{Y_{Eu}^{MNS}}$ $(\mathrm{M_{\odot}})$ & $\mathrm{Y_{Ba}^{MNS}}$ $\mathrm{(M_{\odot}})$ \\
\hline
  $(1-5)\times10^{-5}$ & $3.0\times10^{-7}-1.5\times10^{-6}$ & $3.2\times10^{-6}-1.58\times10^{-5}$ \\
  $(1-5)\times10^{-4}$ & $3.0\times10^{-6}-1.5\times10^{-5}$ & $3.2\times10^{-5}-1.58\times10^{-4}$ \\
  $(1-5)\times10^{-3}$ & $3.0\times10^{-5}-1.5\times10^{-4}$ & $3.2\times10^{-4}-1.58\times10^{-3}$ \\
\hline
\end{tabular}%
\end{table}


\begin{table}
\centering
\hspace{-0.5 cm}
\caption{\label{tab: Yields of Eu from MRD-SNe}Yields of r-process elements from MRD-SNe adopted in this work.}
\begin{tabular}{llcc}
\hline
   $\mathrm{Y_{Eu}^{MNS}}$ $(\mathrm{M_{\odot}})$ & $\mathrm{Y_{Ba}^{MNS}}$ $\mathrm{(M_{\odot}})$ & Model & Reference \\
\hline
   $1.11\times10^{-5}$ & $2.10\times10^{-4}$ & -- & \cite{Winteler2012}\tablefootnote{The values are based on a recent recalculation that was presented in \cite{2020cote}.}\\
   $1.56\times10^{-6}$ & $2.72\times10^{-6}$ & B12$\beta$0.25 & \cite{Nishimura2015}\\
   $6.85\times10^{-6}$ & $2.58\times10^{-4}$ & L0.10 & \cite{Nishimura2017}\\
  $2.81\times10^{-6}$ & $1.23\times10^{-4}$ & L0.60 & "\\
  $4.69\times10^{-7}$ & $7.66\times10^{-6}$ & L0.75 & "\\
  $5.19\times10^{-6}$ & $2.07\times10^{-5}$ & 35OC-Rs & \cite{2021reichert}\\
\hline
\end{tabular}%
\end{table}

\begin{table}
\centering
\hspace{-0.5 cm}
\caption{\label{tab: Yields of Ba Cescutti+06}Yields of Eu and r-process Ba of \protect\cite{Cescutti2006} for massive stars in the $(12-30) \mathrm{M_\odot}$ mass range.}
\begin{tabular}{ccc}
\hline
  $\mathrm{M_{star}}$ $\mathrm{(M_{\odot}})$ & $\mathrm{Y_{Ba}}$ $\mathrm{(M_{\odot}})$ & $\mathrm{Y_{Eu}}$ $(\mathrm{M_{\odot}})$\\
\hline
  $12$ & $9.0\times10^{-7}$ & $4.5\times10^{-8}$\\
  $15$ & $3.0\times10^{-8}$ & $3.0\times10^{-9}$\\
  $30$ & $1.0\times10^{-9}$ & $5.0\times10^{-10}$\\
\hline
\end{tabular}%
\end{table}

\begin{table*}
\centering
\caption{\label{tab: models_Scl} Summary table of the nucleosynthesis prescriptions adopted by different models. In the $1^{st}$ column in is reported the name of the model, in the $2^{nd}$ column it is specified if there is production from MNS, in the $3^{rd}$ it is specified if we adopted a DTD for MNS, in the $4^{th}$ column it is reported the adopted yield of Eu from MNS, in the $5^{th}$ column the yield of Ba from MNS, in the $6^{th}$ it is specified if there is production from MRD-SNe, in the $7^{th}$ column it is reported the yield of both Eu and Ba from MRD-SNe, in the $8^{th}$ column the percentage of stars in the $(10-80) \mathrm{M_\odot}$ mass range which explode as MRD-SNe and in the last column the range of metallicities in which the MRD-SNe channel is active for the Eu and/or Ba production.}
\begin{tabular}{lcccccccc}
\hline
  Model & MNS & DTD & $\mathrm{Y_{Eu}^{MNS}}$ $\mathrm{(M_\odot)}$ & $\mathrm{Y_{Ba}^{MNS}}$ $\mathrm{(M_\odot)}$ & MRD & $\mathrm{Y_{Eu,Ba}^{MRD}}$ & $\%$ & Z \\
\hline
 C54 & yes & no & $3.00\times10^{-5}-1.50\times10^{-4}$ & $3.20\times10^{-4}-1.58\times10^{-3}$ & no & -- & -- & --\\
 C65 & yes & no & $3.00\times10^{-6}-1.50\times10^{-5}$ & $3.20\times10^{-5}-1.58\times10^{-4}$ & no & -- & -- & --\\
 C76 & yes & no & $3.00\times10^{-7}-1.50\times10^{-6}$ & $3.20\times10^{-6}-1.58\times10^{-5}$ & no & -- & -- & --\\
 \\
 D54 & yes & yes & $3.00\times10^{-5}-1.50\times10^{-4}$ & $3.20\times10^{-4}-1.58\times10^{-3}$ & no & -- & -- & --\\
 D65 & yes & yes & $3.00\times10^{-6}-1.50\times10^{-5}$ & $3.20\times10^{-5}-1.58\times10^{-4}$ & no & -- & -- & --\\
 D76 & yes & yes & $3.00\times10^{-7}-1.50\times10^{-6}$ & $3.20\times10^{-6}-1.58\times10^{-5}$ & no & -- & -- & --\\
 \\
 W12 & no & -- & -- & -- & yes & \cite{Winteler2012} & 1 & all\\
 N15 & no & -- & -- & -- & yes & \cite{Nishimura2015} & 1 & all\\
 N17a & no & -- & -- & -- & yes & \cite{Nishimura2017} L0.10 & 1 & all\\
 N17b & no & -- & -- & -- & yes & \cite{Nishimura2017} L0.60 & 1 & all\\
 N17c & no & -- & -- & -- & yes & \cite{Nishimura2017} L0.75 & 1-2 & all\\
 R21 & no & -- & -- & -- & yes & \cite{2021reichert} & 1 & all\\
 \\
 N15Z & no & -- & -- & -- & yes & \cite{Nishimura2015} & 1-2 & $<10^{-3}$\\
 N17cZ & no & -- & -- & -- & yes & \cite{Nishimura2017} L0.75 & 1-2 & $<10^{-3}$\\
 R21Z & no & -- & -- & -- & yes & \cite{2021reichert} & 1-2 & $<10^{-3}$\\
 \\
 CN54 & yes & no & $3.00\times10^{-5}-1.50\times10^{-4}$ & $3.20\times10^{-4}-1.58\times10^{-3}$ & yes & \cite{Nishimura2017} L0.75 & 1 & all\\
 CN65 & yes & no & $3.00\times10^{-6}-1.50\times10^{-5}$ & $3.20\times10^{-5}-1.58\times10^{-4}$ & yes & \cite{Nishimura2017} L0.75 & 1 & all\\
 CN76 & yes & no & $3.00\times10^{-7}-1.50\times10^{-6}$ & $3.20\times10^{-6}-1.58\times10^{-5}$ & yes & \cite{Nishimura2017} L0.75 & 1 & all\\
 CR54 & yes & no & $3.00\times10^{-5}-1.50\times10^{-4}$ & $3.20\times10^{-4}-1.58\times10^{-3}$ & yes & \cite{2021reichert} & 1 & all\\
 CR65 & yes & no & $3.00\times10^{-6}-1.50\times10^{-5}$ & $3.20\times10^{-5}-1.58\times10^{-4}$ & yes & \cite{2021reichert} & 1 & all\\
 CR76 & yes & no & $3.00\times10^{-7}-1.50\times10^{-6}$ & $3.20\times10^{-6}-1.58\times10^{-5}$ & yes & \cite{2021reichert} & 1 & all\\
 \\
 DN65 & yes & yes & $3.00\times10^{-6}-1.50\times10^{-5}$ & $3.20\times10^{-5}-1.58\times10^{-4}$ & yes & \cite{Nishimura2017} L0.75 & 1 & all\\
 DN65Z & yes & yes & $3.00\times10^{-6}-1.50\times10^{-5}$ & $3.20\times10^{-5}-1.58\times10^{-4}$ & yes & \cite{Nishimura2017} L0.75 & 1 & $<10^{-3}$\\
 DR54 & yes & yes & $3.00\times10^{-5}-1.50\times10^{-4}$ & $3.20\times10^{-4}-1.58\times10^{-3}$ & yes & \cite{2021reichert} & 1 & all\\
\hline
\end{tabular}%
\end{table*}

\begin{table*}
\centering
\caption{\label{tab: Input parameters}Input parameters of the chemical evolution model for specific dSphs and UFDs. In the $1^{st}$ column it is reported the name of the galaxy, in the $2^{nd}$ column the infall mass, in the $3^{rd}$ column the efficiency of star formation, in the $4^{th}$ column the infall time-scale, in the $5^{th}$ column the wind parameter and in the last three columns the number, time and duration of the bursts of star formation, respectively. The first three galaxies are discussed at length in the main text, while results for the others are presented as Supplementary Material in the online version of the journal.}
\begin{tabular}{lccccccc}
\hline
  Galaxy & $\mathrm{M_{infall}}$ $(\mathrm{M_{\odot}})$ & $\nu$ $(\mathrm{Gyr^{-1}})$ & $\mathrm{\tau_{infall}}$ $(\mathrm{Gyr})$ & $\omega$ & $\mathrm{n}$ & $\mathrm{t}$ $(\mathrm{Gyr})$ & $\mathrm{d}$ $(\mathrm{Gyr})$\\
\hline
 Fornax (For) & $5.0 \times 10^{8}$ & $0.1$ & $3$ & $1$ & $1$ & $0$ & $14$\\
 Sculptor (Scl) & $1.0 \times 10^{8}$ & $0.2$ & $0.5$ & $9$ & $1$ & $0$ & $7$\\
 Reticulum II (RetII) & $1.0 \times 10^{5}$ & $0.01$ & $0.05$ & $6$ & $1$ & $0$ & $1$\\
 Bootes I (BooI) & $1.1 \times 10^{7}$ & $0.005$ & $0.05$ & $12$ & $1$ & $0$ & $1$\\
 Carina (Car) & $5.0 \times 10^{8}$ & $0.15$ & $0.5$ & $5$ & $4$ & $1-3-8-10$ & $2-2-2-2$\\
 Sagittarius (Sgr) & $2.1 \times 10^{9}$ & $1$ & $0.5$ & $9$ & $2$ & $0-4.5$ & $4-2.5$\\
 Sextan (Sex) & $5.0 \times 10^{8}$ & $0.005$ & $0.5$ & $11$ & $1$ & $0$ & $8$\\
 Ursa Minor (UMi) & $5.0 \times 10^{8}$ & $0.05$ & $0.5$ & $11$ & $1$ & $0$ & $3$\\
\hline
\end{tabular}%
\end{table*}

\section{Results}
\label{results}

For each galaxy we set the input parameters of the chemical evolution model in order to reproduce the star formation, the observed MDF and the [Mg/Fe] vs [Fe/H] pattern. Except that for Fornax and Reticulum II, in general we follow previous literature results which provide an estimate of the parameters of the chemical evolution models able to reproduce the relevant data for each galaxy. The input parameters of chemical evolution models adopted in this work are reported in Table \ref{tab: Input parameters}.

After tuning our models with the observed data, we analysed the production of neutron capture elements by comparing the results of our models for the evolution of [Eu/Fe], [Ba/Fe] and [Ba/Eu] vs [Fe/H] with observed patterns. In this way, it is possible to investigate on the nucleosynthesis of those elements.


\begin{figure*}
\begin{center}
 \subfloat[]{\includegraphics[width=1\columnwidth]{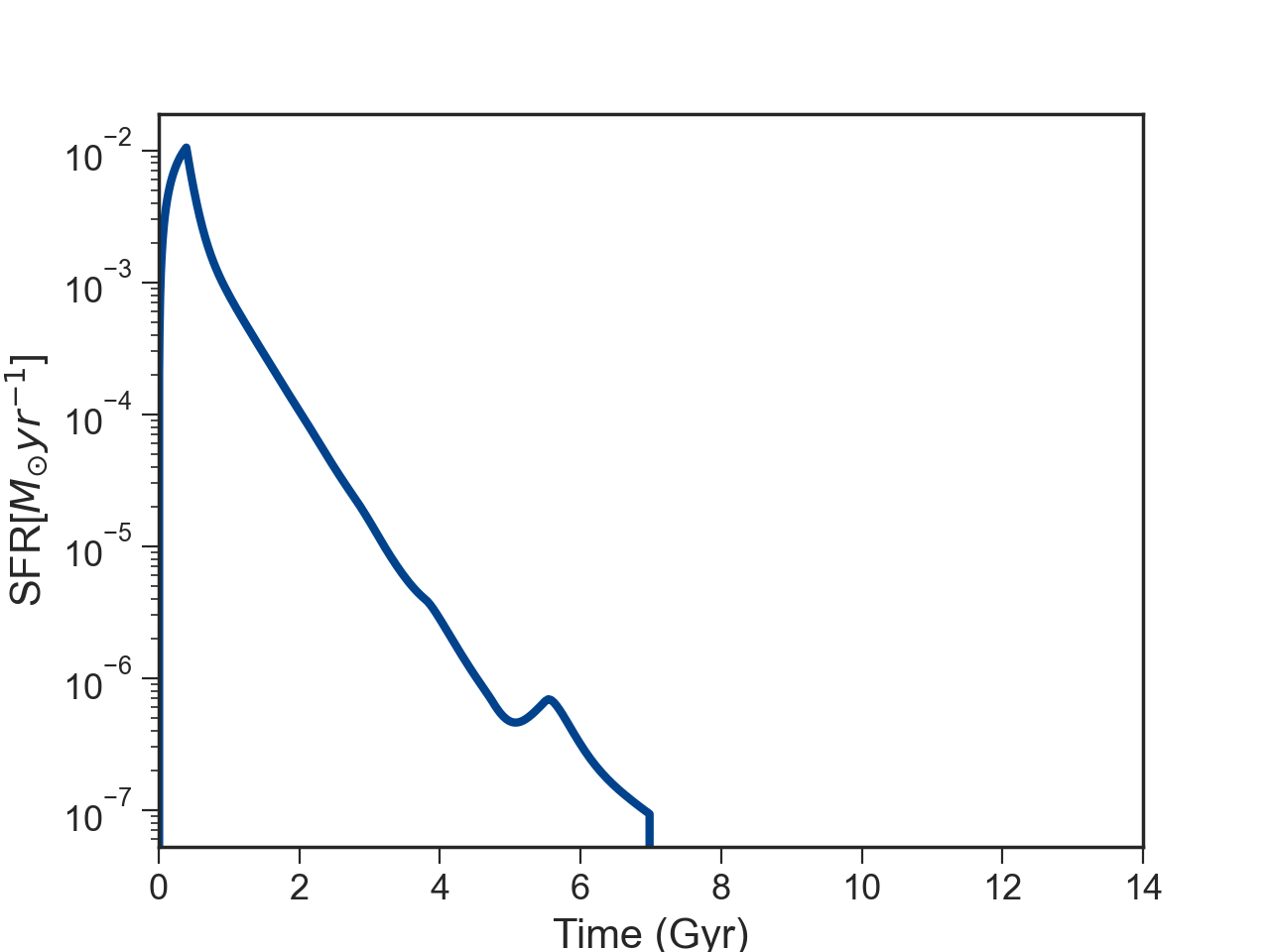}\label{fig:a}}
 \hfill
 \subfloat[]{\includegraphics[width=1\columnwidth]{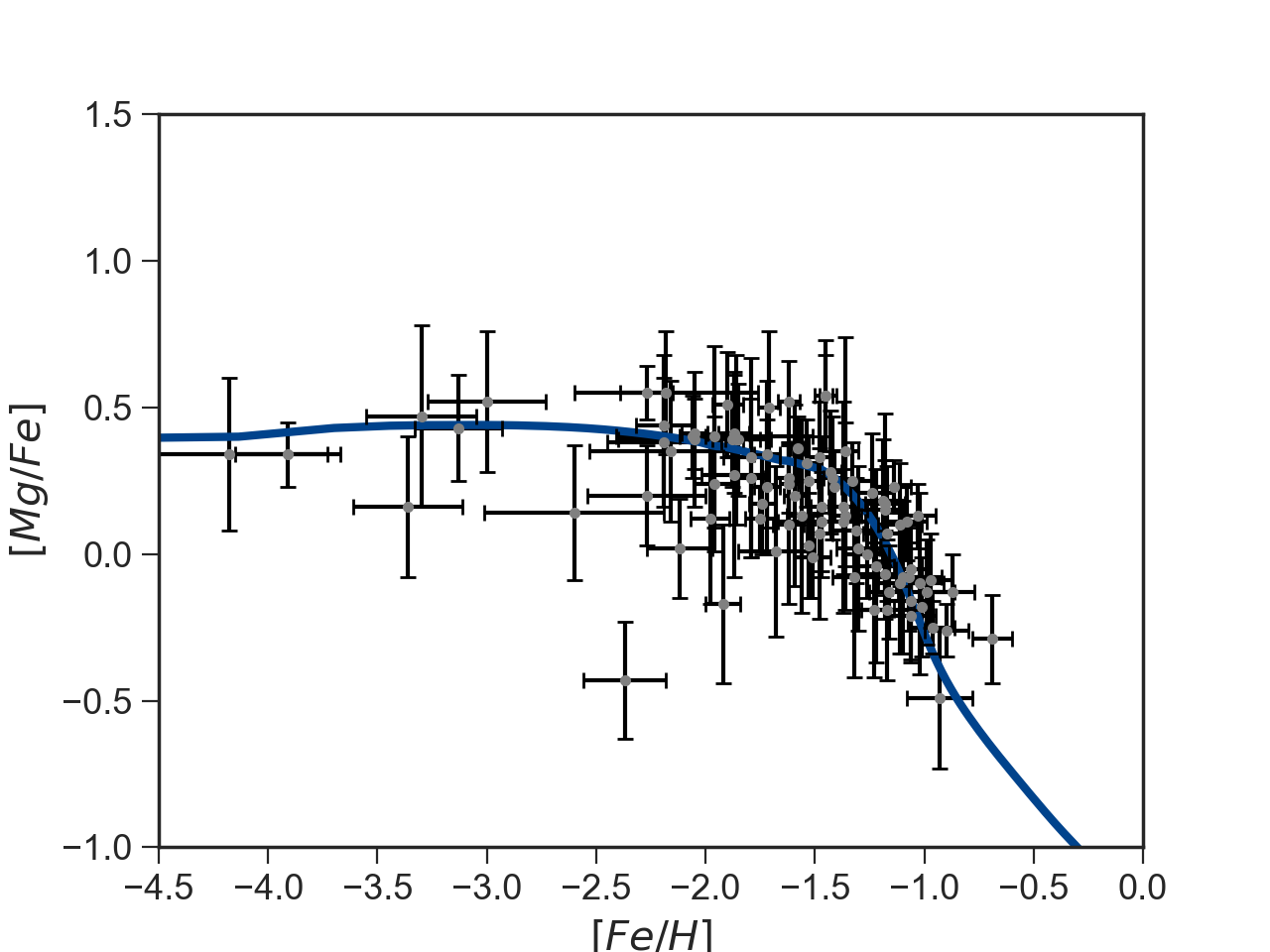}\label{fig:b}}
 \hfill
 \subfloat[]{\includegraphics[width=1\columnwidth]{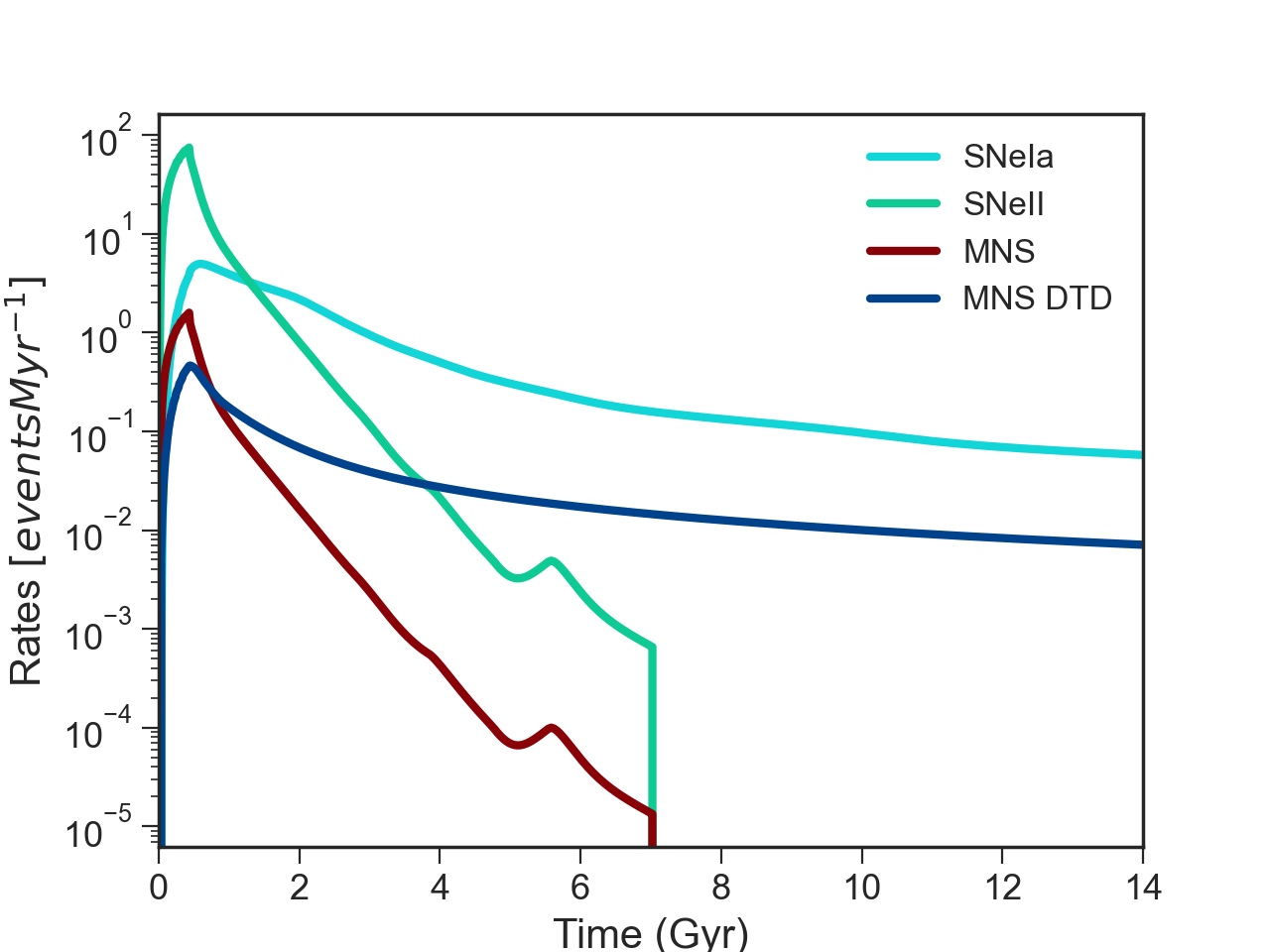}\label{fig:c}}
 \hfill
 \subfloat[]{\includegraphics[width=1\columnwidth]{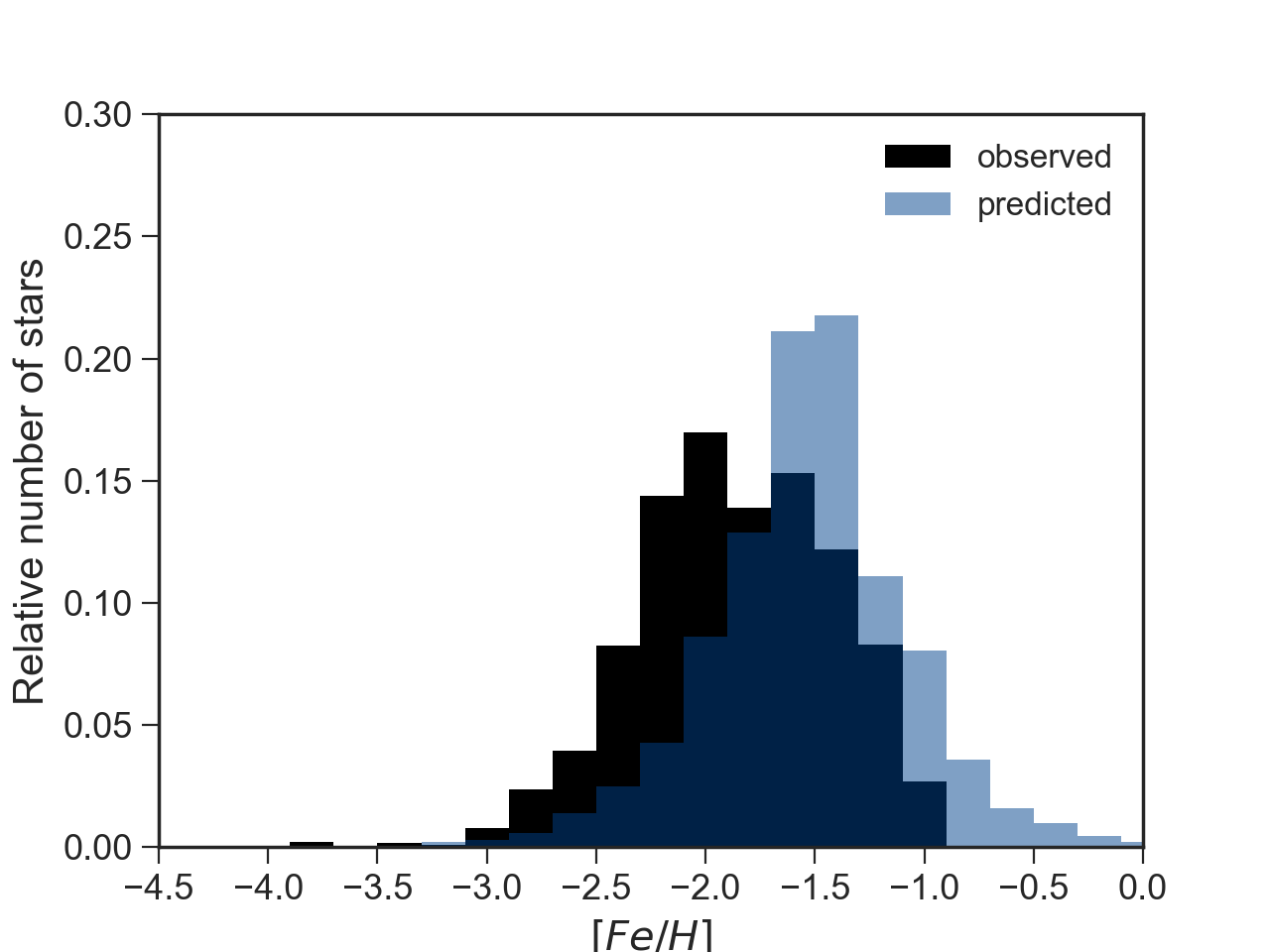}\label{fig:d}}%
 \caption{Results for Sculptor dSph. Panel (a): predicted SF history as a function of time; panel (b) predicted [Mg/Fe] vs [Fe/H] pattern together with observational data; panel (c) predicted rates of SNeIa (turquoise), SNeII (green), MNS with a constant delay time for merging (red) and MNS with a DTD (light blue); panel (d) our predicted MDF against the observed one.}%
 \label{fig: SFR+MDF_Scl}%
\end{center}
\end{figure*}

\begin{figure*}
\begin{center}
 \subfloat[]{\includegraphics[width=1\columnwidth]{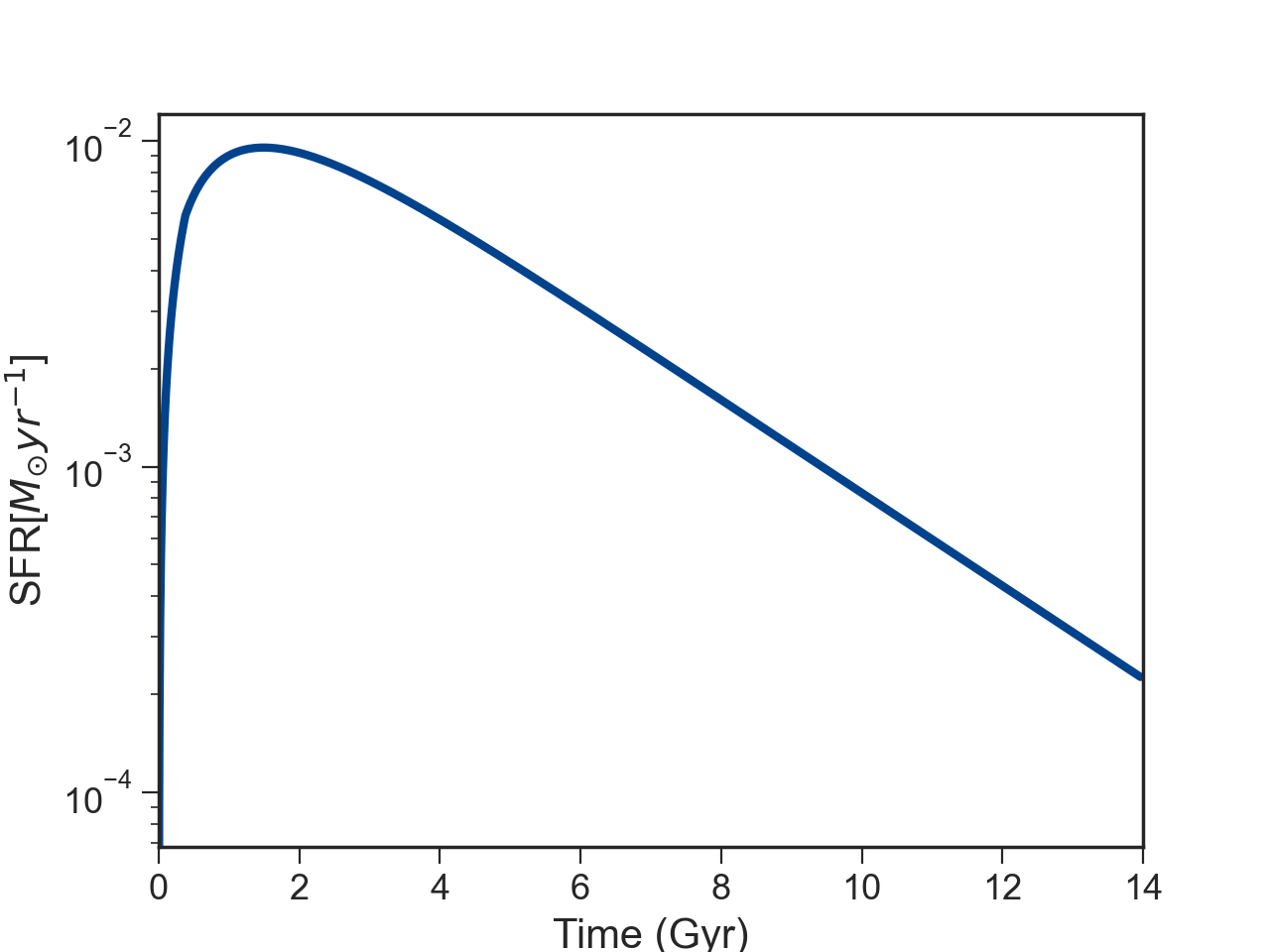}\label{fig:a}}
 \hfill
 \subfloat[]{\includegraphics[width=1\columnwidth]{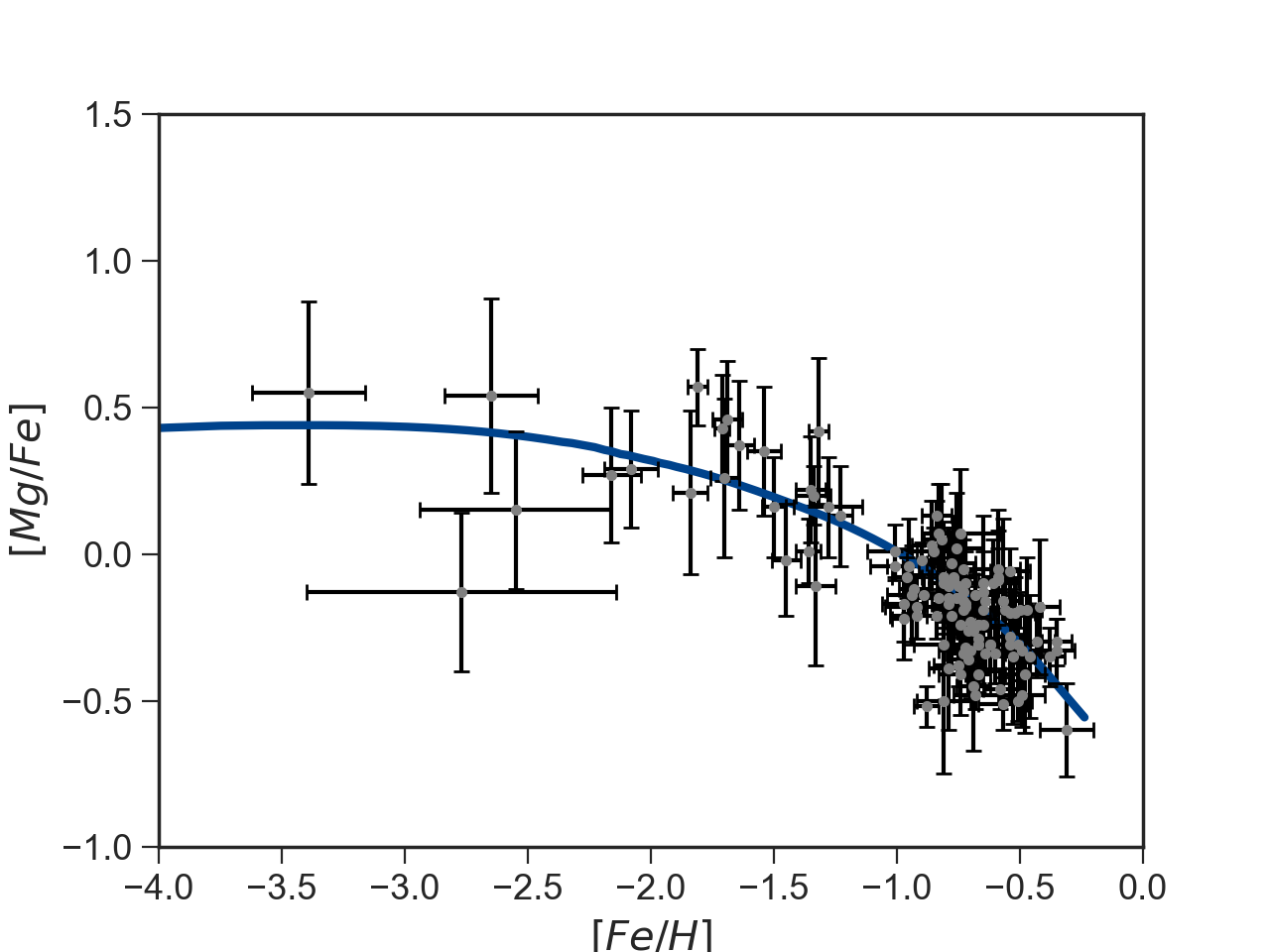}\label{fig:b}}
 \hfill
 \subfloat[]{\includegraphics[width=1\columnwidth]{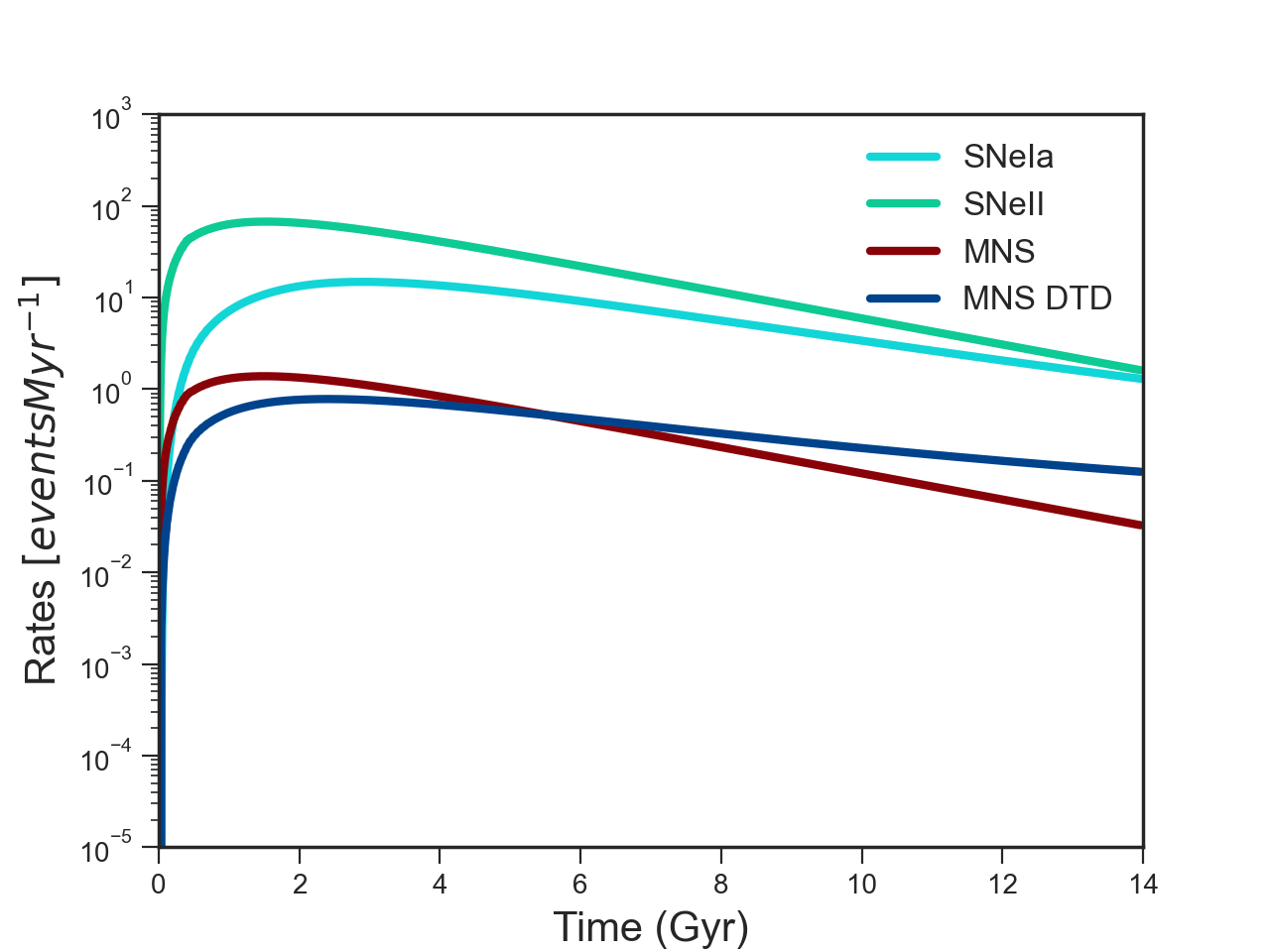}\label{fig:c}}
 \hfill
 \subfloat[]{\includegraphics[width=1\columnwidth]{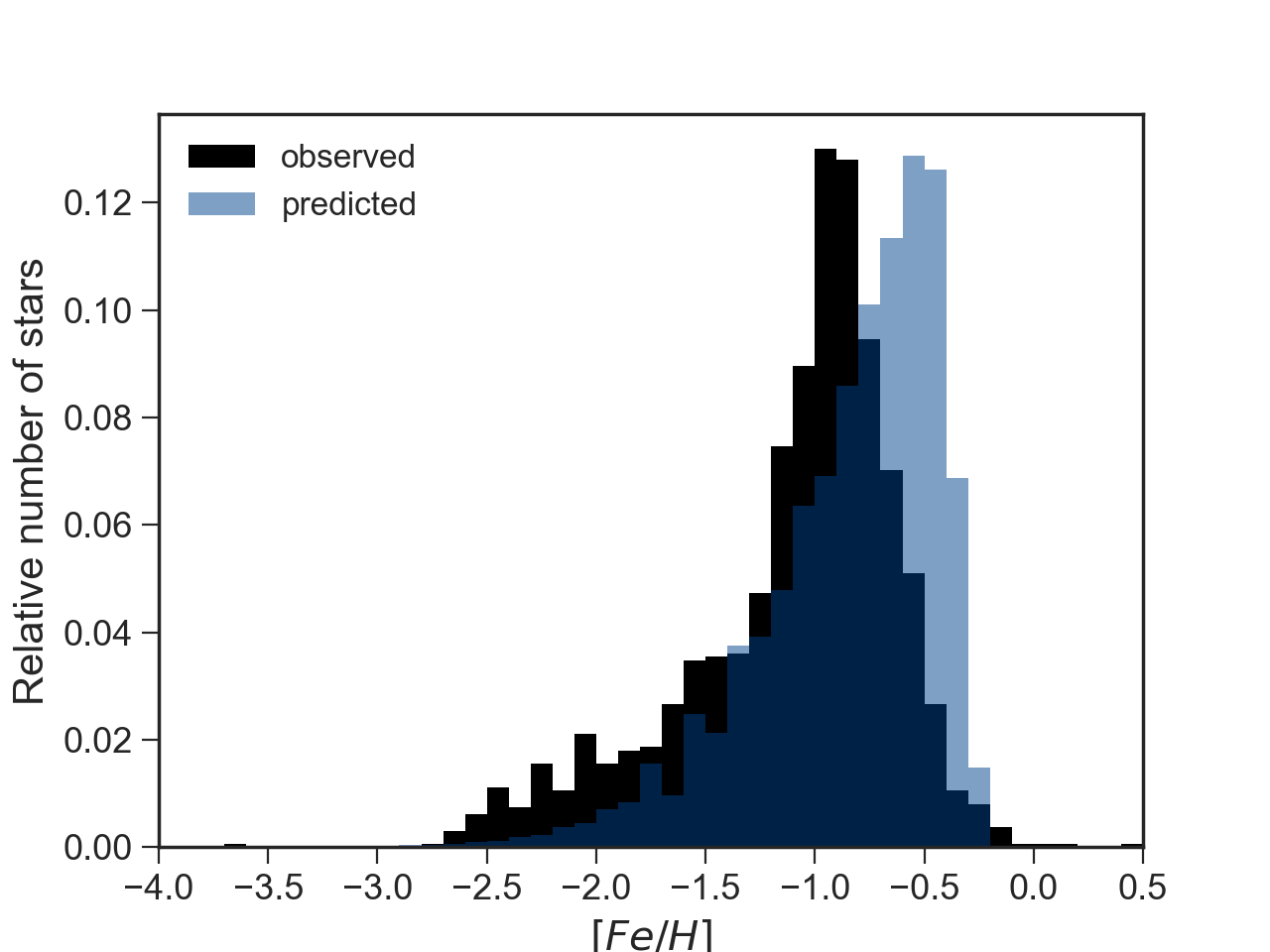}\label{fig:d}}%
 \caption{Same as Figure \ref{fig: SFR+MDF_Scl} but for Fornax.}%
 \label{fig: SFR+MDF_For}%
\end{center}
\end{figure*}

\subsection{Sculptor and Fornax}
\label{sec: sculptor/fornax}

For Sculptor dSph galaxy, we have adopted similar theoretical prescriptions to \cite{Lanfranchi&Matteucci04} (who first modelled the chemical evolution of Sculptor) and to \cite{Vincenzo2014}. We assumed a dark matter halo of mass $\mathrm{M_{DM}=3.4\times10^{8} M_{\odot}}$ (\citealp{Battaglia2008}) and a core radius $\mathrm{R_{DM}=1 kpc}$. The effective radius of the luminous component of the galaxy has been set to $\mathrm{R_L={260} pc}$ (\citealp{Walker2009}). The SF of Sculptor has been derived from the CMD fitting analysis by \cite{deBoer2012} and it consists of one episode of star formation which lasts 7 Gyr. Our predicted SFR as a function of time is shown in panel (a) of Figure \ref{fig: SFR+MDF_Scl}. It is characterized by an initial fast increase, due to the short time-scale of the infall, followed by a decline caused by the onset of the galactic wind. Our model predicts a final stellar mass of $\mathrm{M_{\star,f}}=2.6\times10^{6}\mathrm{M_\odot}$, similar to the observed one $\mathrm{M_{\star,f}}=1.2\times10^{6}\mathrm{M_\odot}$ derived by \cite{deBoer2012} by integrating the SFR up to the present time.

In panel (b) of Figure \ref{fig: SFR+MDF_Scl} we show the [Mg/Fe] vs [Fe/H] evolution together with the prediction of our model. The pattern is characterized by a flat plateau at low metallicities\footnote{The flat plateau is due to the assumption that stars more massive than 20 M$_\odot$ explode as hypernoavae. If all stars explode as CC-SNe, a [Mg/Fe] trend increasing with decreasing [Fe/H] is obtained instead (see \citealp{Romano2010}, their Figure 12). We note that a flat trend fits the data much better.} followed by a decrease for [Fe/H] $\geq$ -1.5 dex due to the fact that, for [Fe/H] $\geq$ -1.5 dex, SNeIa start contributing in a substantial way to the Fe enrichment (\citealp{Matteucci2001}). In fact, we remind that while $\alpha$-elements are mainly produced in Type II SNe on short time-scales, the majority of Fe and Fe-peak elements are produced by SNeIa on longer time-scales (see \citealp{2021palla} for a detailed discussion). Also, as the galactic wind is activated, the SF starts to decline until it stops at 7 Gyr. Consequently, the production of $\alpha$-elements by SNeII will decrease too. On the other hand, Fe-peak elements are continuously ejected into the ISM even when there is no SF activity, because of the long lifetimes of the progenitors of SNeIa. Because of these two facts, the [Mg/Fe] vs [Fe/H] trend will be strongly influenced by the efficiency of the SF ($\nu$) and by the wind parameter ($\omega$): the higher $\nu$ is, the longer the [Mg/Fe] plateau will be, while the higher $\omega$ is, more pronounced the [Mg/Fe] decrease will be . As seen, our model with $\nu=0.2 \mathrm{Gyr^{-1}}$ and with $\omega=9$ is able to perfectly fit the observed [Mg/Fe] vs [Fe/H] abundances.

In panel (c) of Figure \ref{fig: SFR+MDF_Scl} we report the rates of different phenomena predicted by our simulations. It is possible to see how SNeII follow the SF history of the simulated galaxy, while SNeIa continue to explode even after the quenching of the SF. Rates of MNS are also reported in the panel, showing both constant delay time and DTD. In the case of a constant total delay time, the rate of MNS follows the evolution of the SFR of Sculptor, so that no MNS event is predicted at the present time. On the other hand, when we assume a DTD, the dependence of the MNS rate on the SFR is not so important (see \citealp{Simonetti2019}, \citealp{2019cot} for an extensive discussion about the delay times of MNS in the Galaxy). In this case, the evolution of the MNS rate will be similar to that of SNeIa and its present time value will differ from zero, being equal to $\mathrm{R_{MNS}}\simeq7$  events Gyr$^{-1}$.

In panel (d) of Figure \ref{fig: SFR+MDF_Scl}, the observed MDF together with the prediction from our model is reported. There is a quite good agreement between model and the data, even if our results appear to be shifted towards higher metallicities. In order to predict a MDF peaked at lower metallicities one could lower the star formation efficiency. However, we point out that this would also lead to a higher MDF peak, as well as a shorter plateau in the [Mg/Fe] abundances ratio. Therefore, in order not to lose the really good agreement for the [Mg/Fe] evolution, we do not change our choice of the parameters.\\

Concerning the chemical evolution of Fornax, we assumed a dark matter halo of mass $\mathrm{M_{DM}=5 \times 10^9 \mathrm{M_\odot}}$ and a core radius $\mathrm{R_{DM}=15.5 kpc}$. The effective radius of the luminous component has been set to $\mathrm{R_L=1.55 kpc}$. For the SF history, we take into consideration that of \cite{fornaxdeBoer2012}, which is derived from the CMD fitting analysis, according to which Fornax formed stars at all ages, from as old as 14 Gyr to as young as 0.25 Gyr. In particular, they conclude that, even if stars are formed continuously during the evolution of the galaxy, most of the star formation takes place at intermediate ages (see also \citealp{coleman2008}). We model a continuous SF, characterized by one long episode lasting 14 Gyr, with a constant efficiency equal to $\mathrm{\nu=0.1 Gyr^{-1}}$. Our predicted star formation as a function of time is reported in panel (a) of Figure \ref{fig: SFR+MDF_For}. It is seen that in our model a high number of stars formed in the first Gyr, and then the gas gets depleted due to the star formation itself and to the action of galactic winds causing a gas loss until the present time. Our model predicts a final stellar mass of $\mathrm{M_{\star,f}}=2.9\times20^{7}\mathrm{M_\odot}$, similar to the one estimated by \cite{fornaxdeBoer2012} equal to $\mathrm{M_{\star,f}}=4.3\times20^{7}\mathrm{M_\odot}$.

Panels (b) and (d) of Figure \ref{fig: SFR+MDF_For} show that the results of our model are in agreement with both the observed [Mg/Fe] vs [Fe/H] and the MDF, respectively. A better agreement could have been obtained for the MDF by lowering the star formation efficiency in order to shift our MDF peak towards lower metallicities. However, as we already pointed out for Sculptor, that would also bring to a higher MDF peak and to a shorter plateau for the [Mg/Fe] vs [Fe/H]. 

Finally, in panel (c) of the same Figure we report the evolution of the rates of different phenomena. The present time value of the rate of MNS will be different from zero both in the case in which we adopt a constant total delay time for merging and in the case in which we adopt a DTD, because of the long and continuous episode of SF. The rate of MNS in the two cases will be $\mathrm{R}_\mathrm{MNS}\simeq33$  events Gyr$^{-1}$ and $\mathrm{R}_\mathrm{MNS}^{\mathrm{DTD}}\simeq125$  events Gyr$^{-1}$, respectively.

\begin{figure*}
\begin{center}
 \subfloat[]{\includegraphics[width=1\columnwidth]{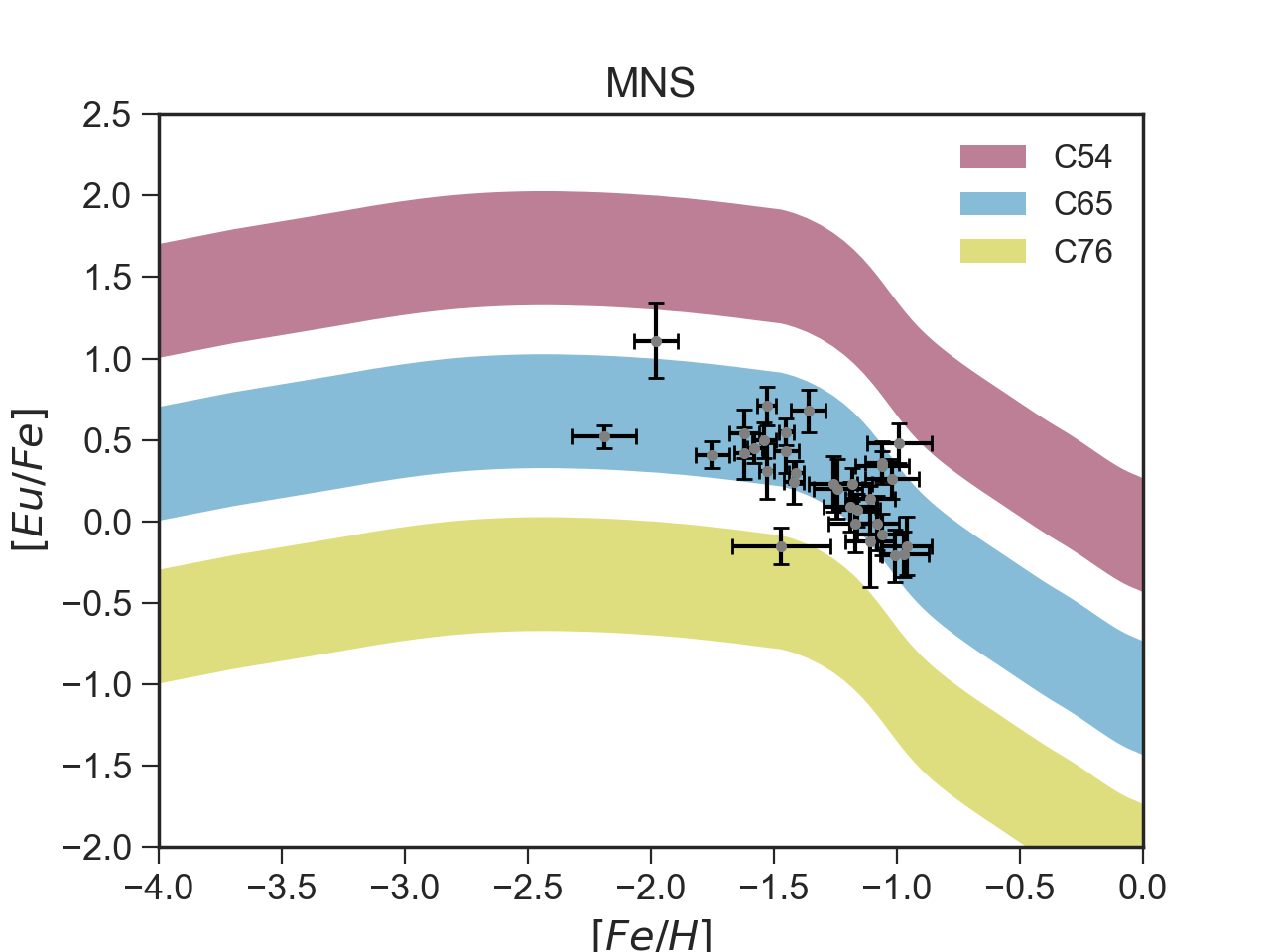}\label{fig:a}}
 \subfloat[]{\includegraphics[width=1\columnwidth]{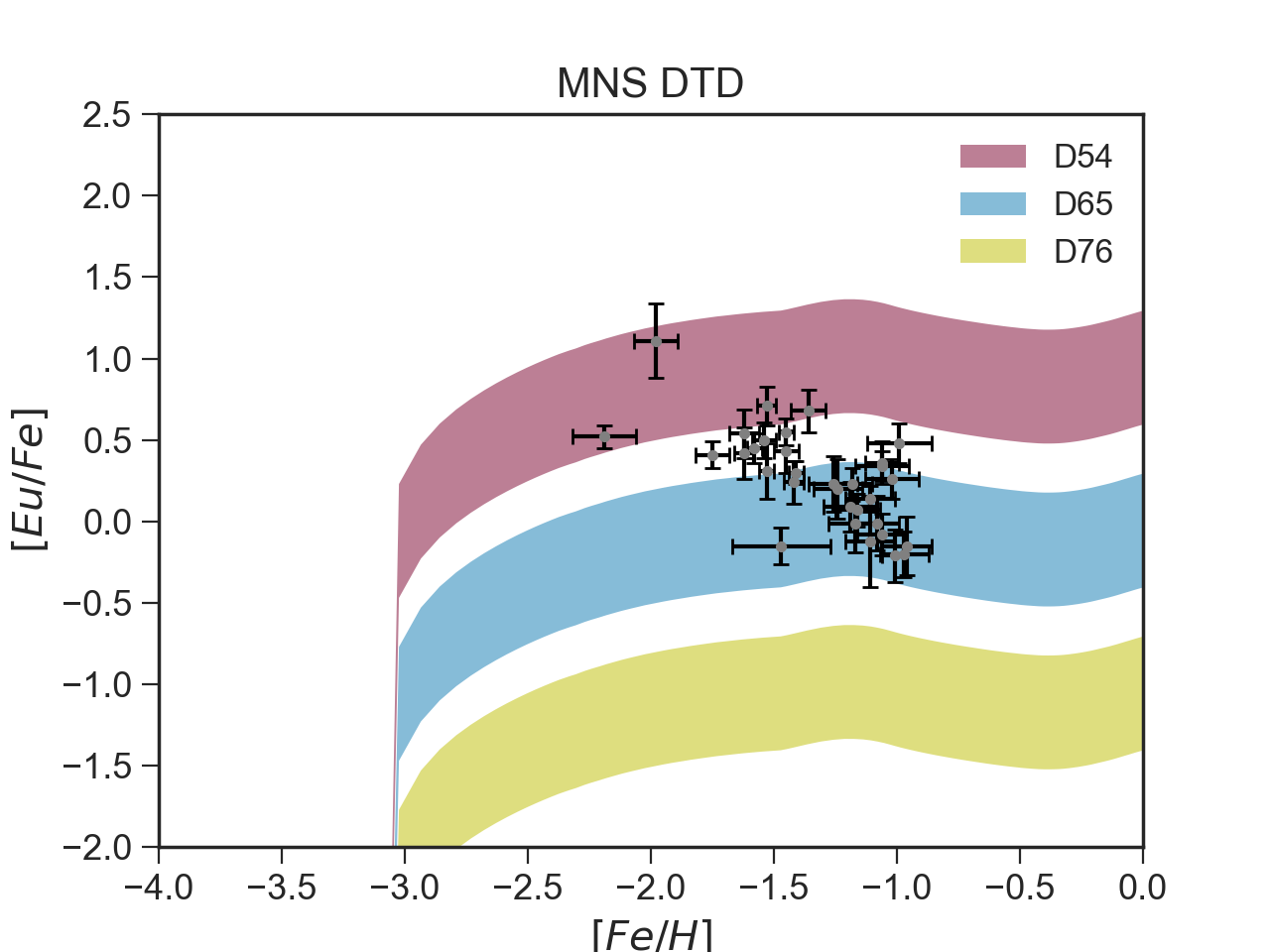}\label{fig:b}}
 \hfill
 \subfloat[]{\includegraphics[width=1\columnwidth]{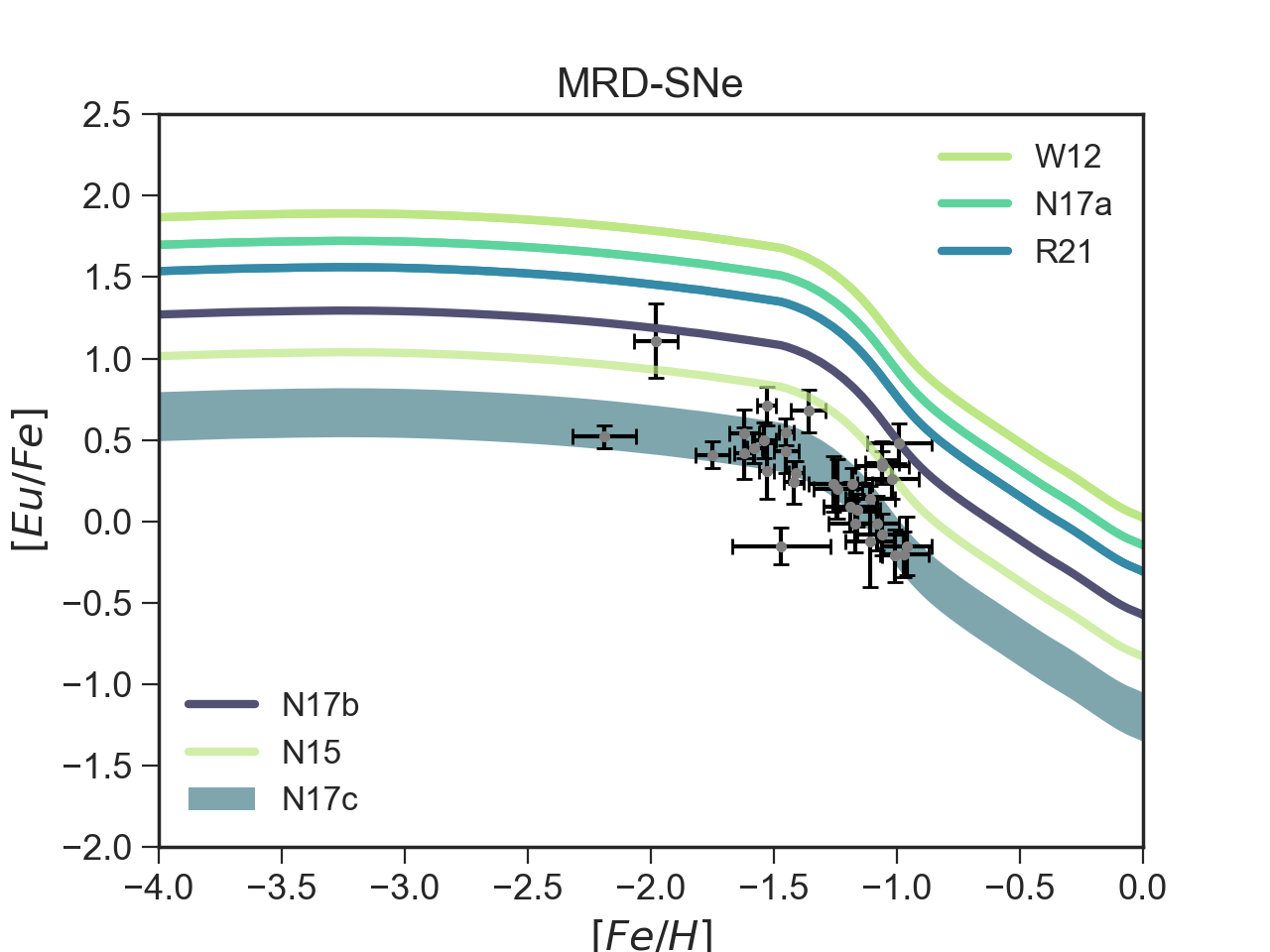}\label{fig:c}}
 \subfloat[]{\includegraphics[width=1\columnwidth]{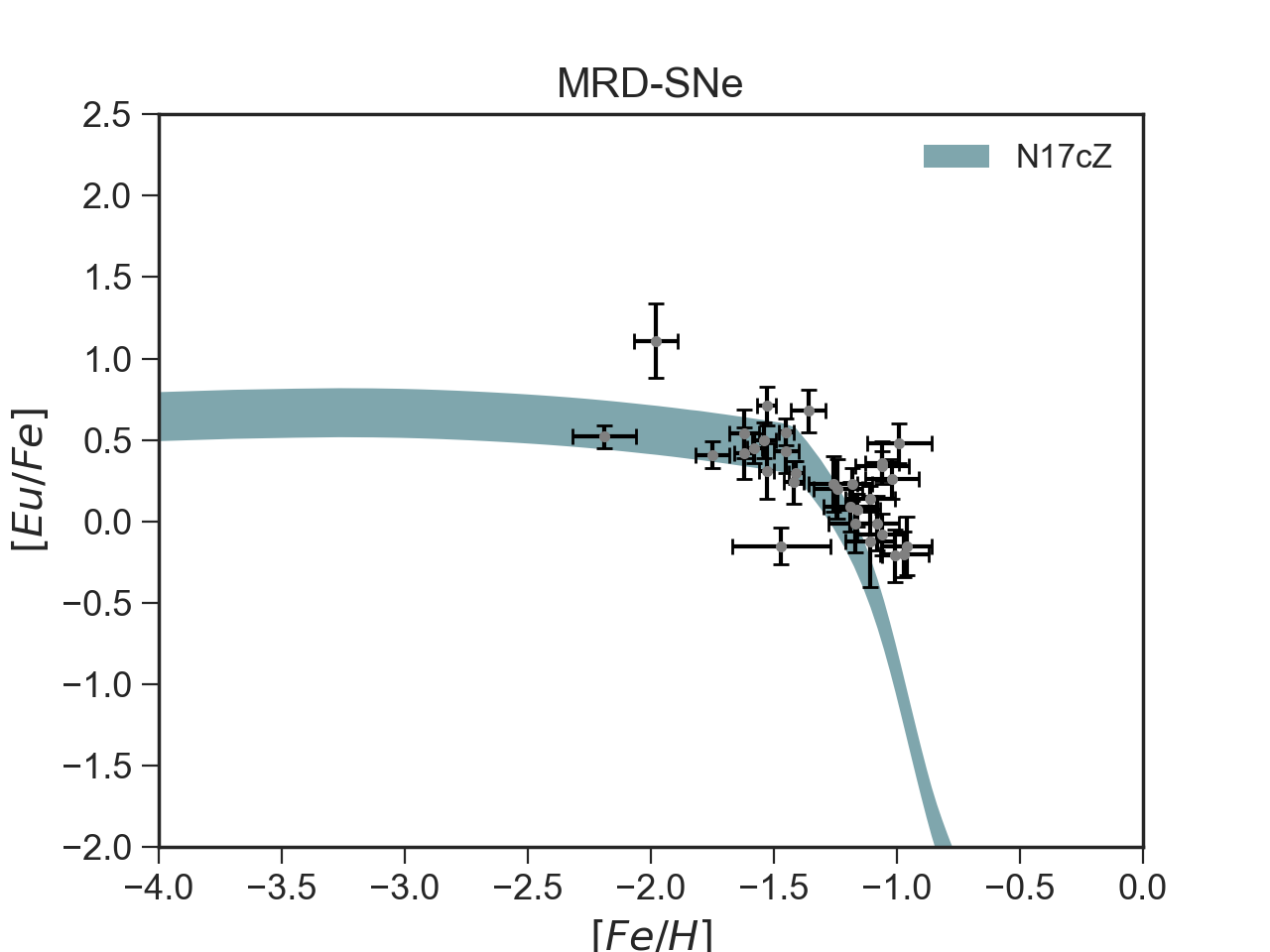}\label{fig:d}}
 \hfill
 \subfloat[]{\includegraphics[width=1\columnwidth]{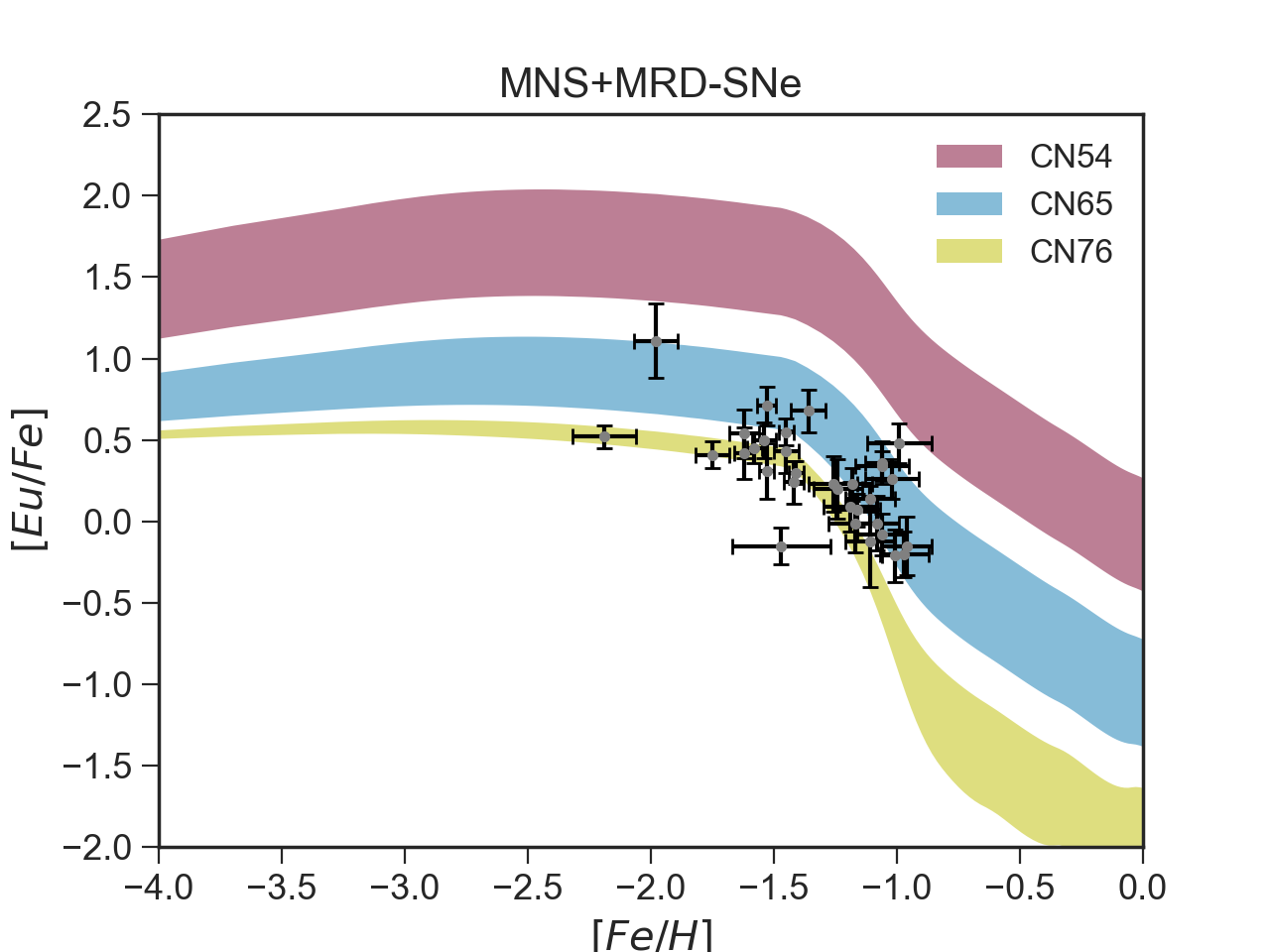}\label{fig:e}}
 \subfloat[]{\includegraphics[width=1\columnwidth]{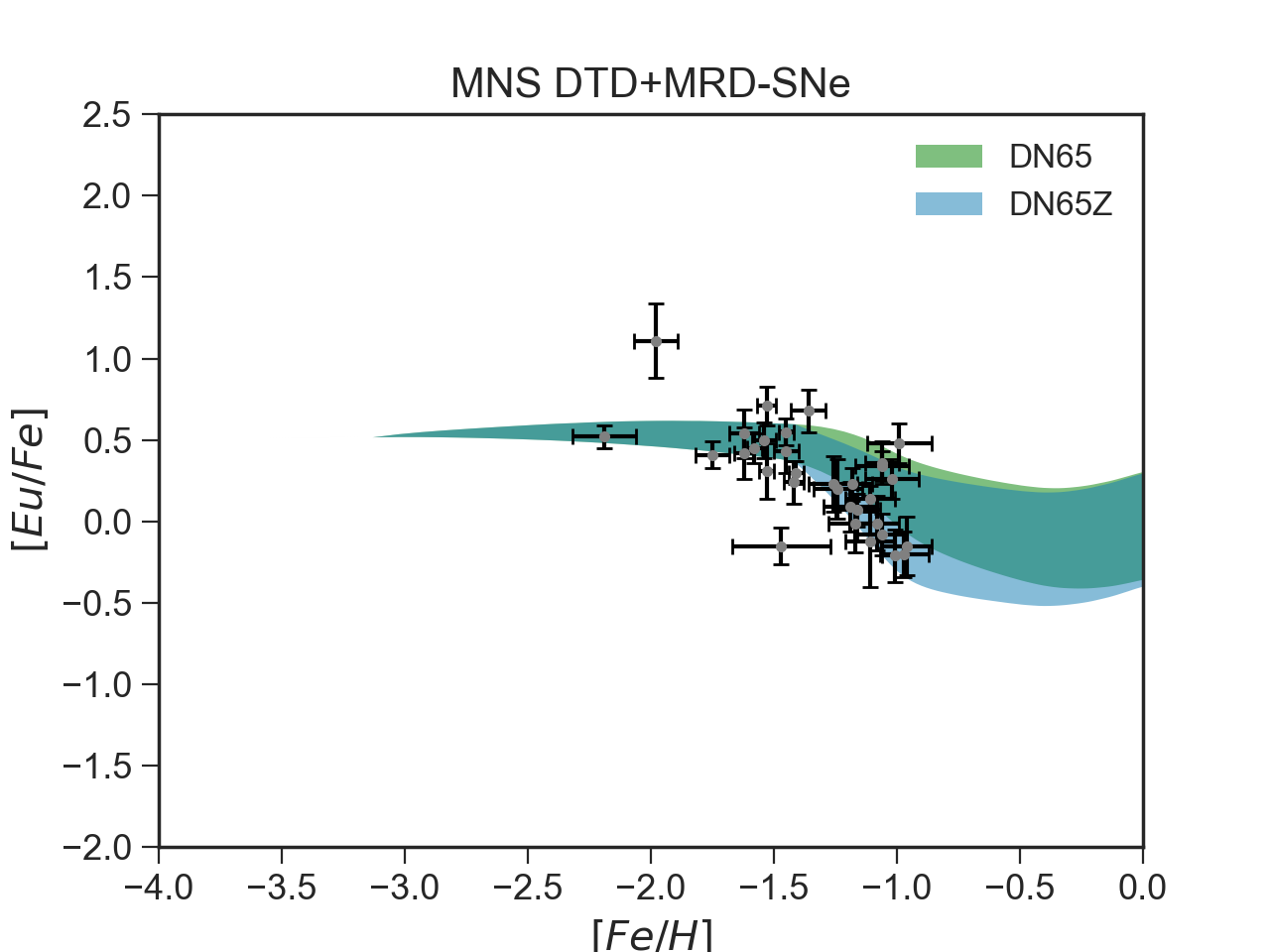}\label{fig:f}}
 \caption{Results of models that differ only in the adopted nucleosynthesis prescriptions for the [Eu/Fe] vs [Fe/H] pattern for Sculptor dSph. Panel (a): results of models in which only MNS produce Eu with a constant delay time for merging; panel (b): results of models in which only MNS produce Eu with a DTD; panel (c): results of models in which only MRD-SNe produce Eu; panel (d): results of models in which only MRD-SNe produce Eu for Z$\leq10^{-3}$; panel (e): results of models in  which Eu is produced by both MNS with a constant delay time for merging and MRD-SNe; panel (f): results of models in which Eu is produced both by MNS with a DTD and MRD-SNe acting both at low Z and and for all the range of metallicities. Details of models are in Table \ref{tab: models_Scl}.}%
 \label{fig: Scl_New}%
\end{center}
\end{figure*}

\begin{figure*}
\begin{center}
 \subfloat[]{\includegraphics[width=1\columnwidth]{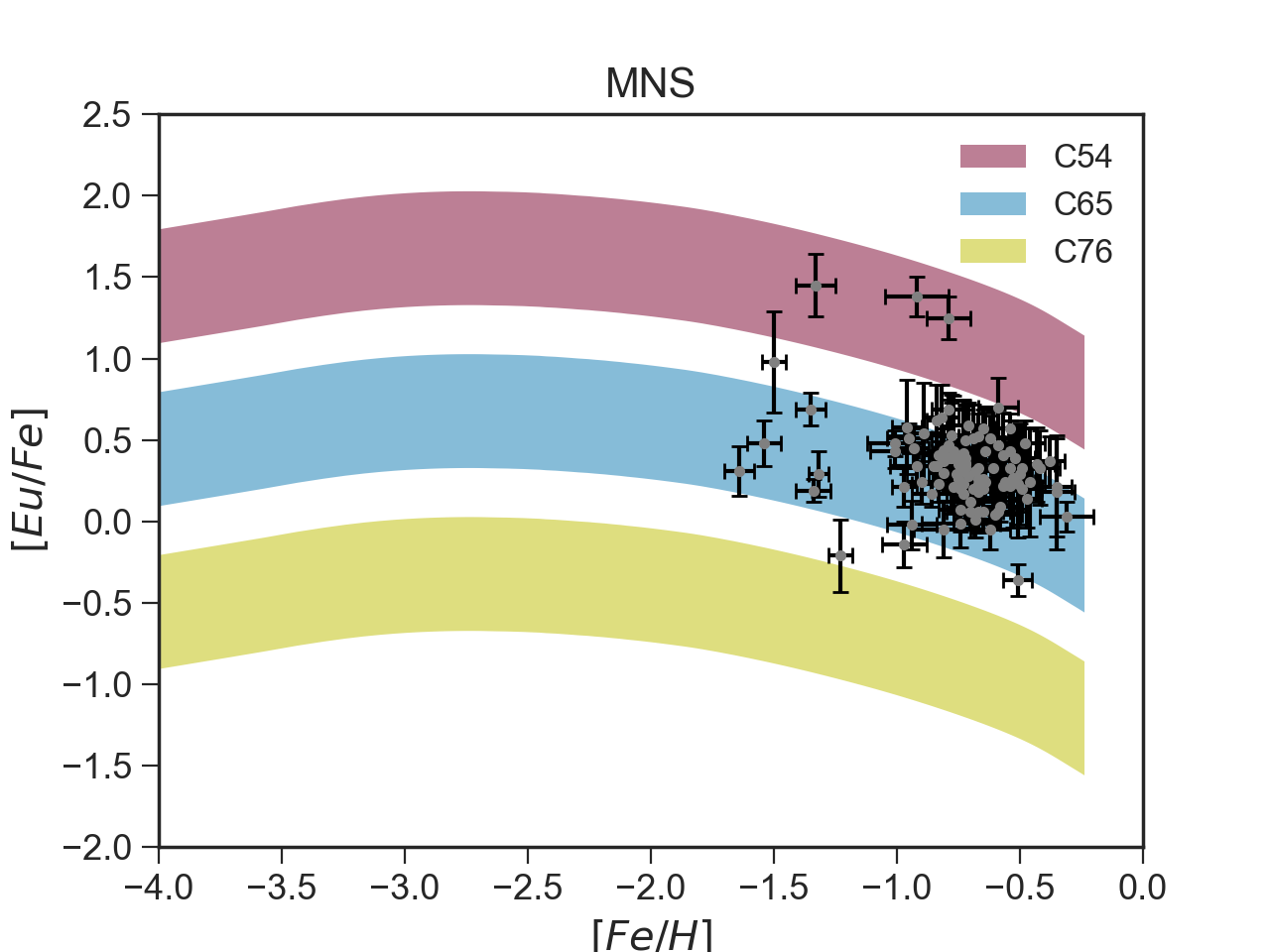}\label{fig:a}}
 \subfloat[]{\includegraphics[width=1\columnwidth]{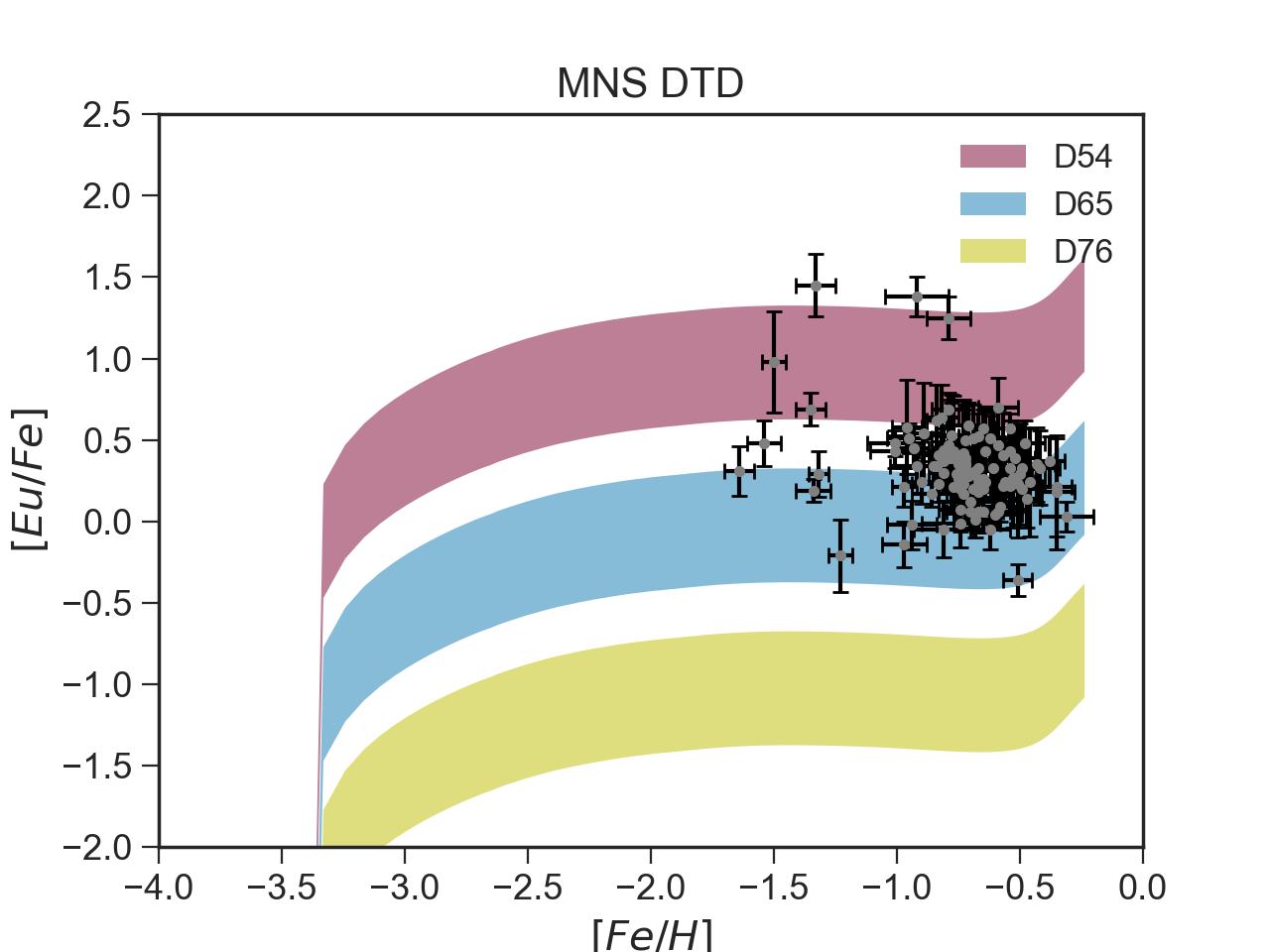}\label{fig:b}}
 \hfill
 \subfloat[]{\includegraphics[width=1\columnwidth]{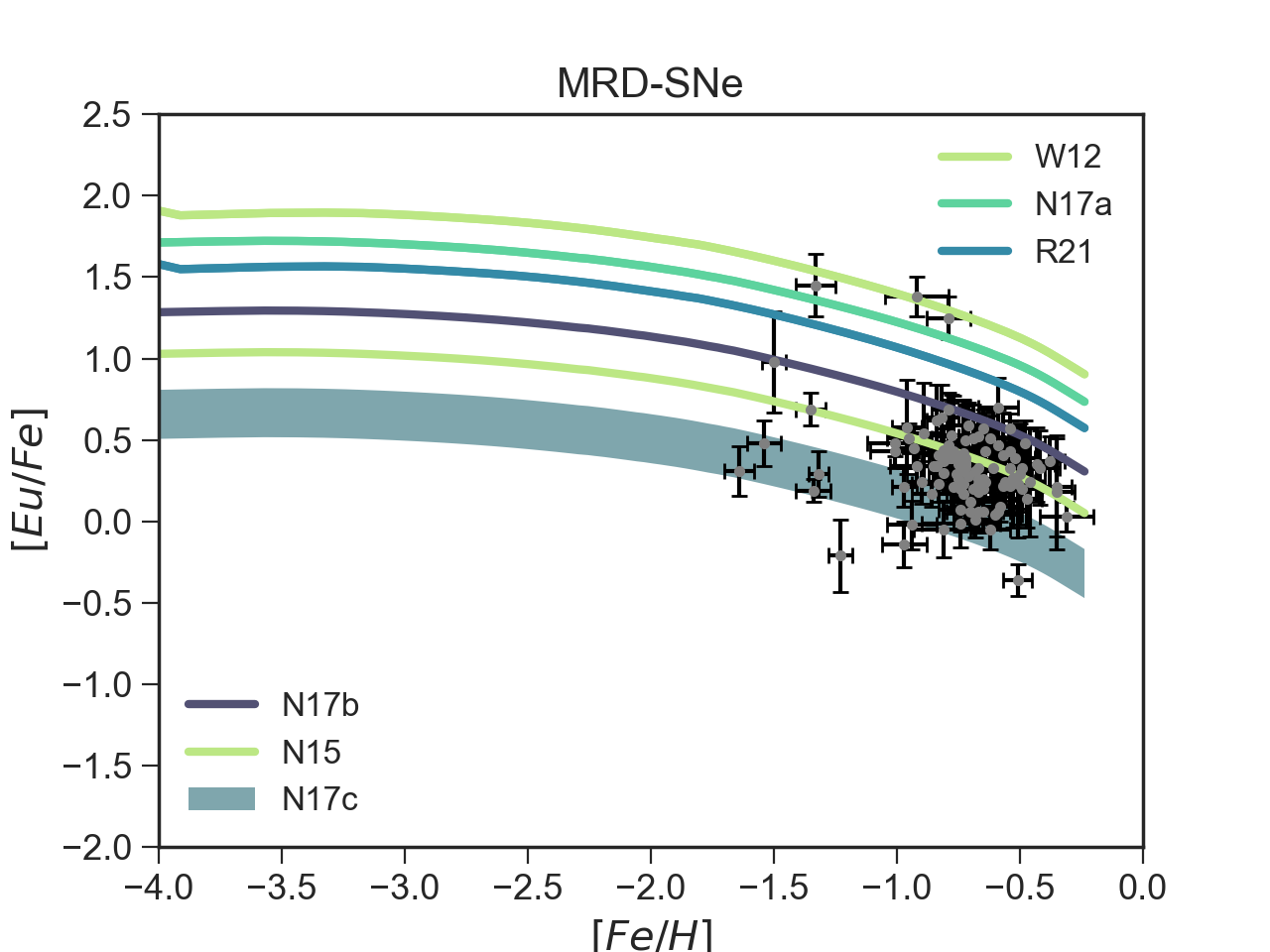}\label{fig:c}}
 \subfloat[]{\includegraphics[width=1\columnwidth]{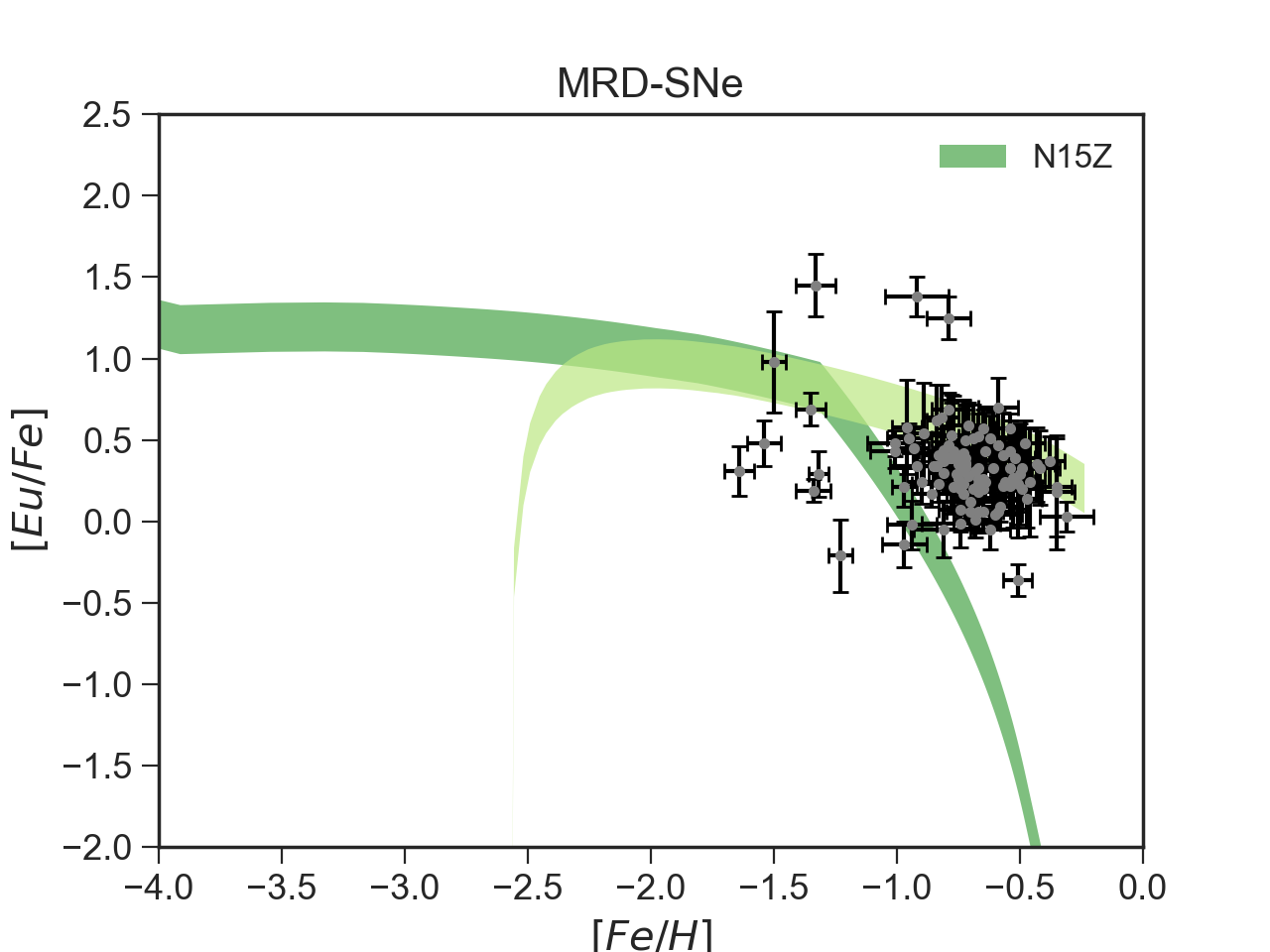}\label{fig:d}}
 \hfill
 \subfloat[]{\includegraphics[width=1\columnwidth]{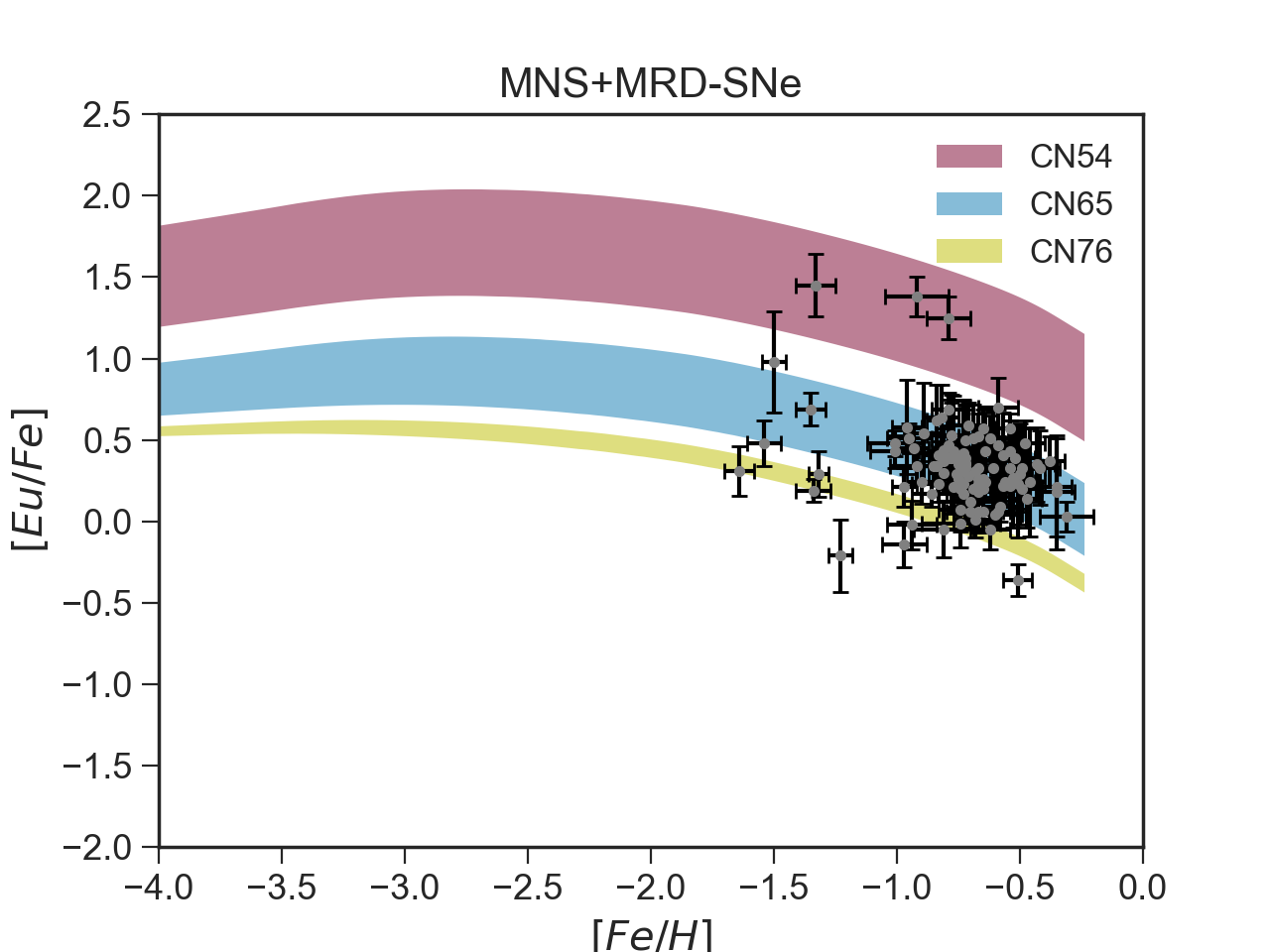}\label{fig:e}}
 \subfloat[]{\includegraphics[width=1\columnwidth]{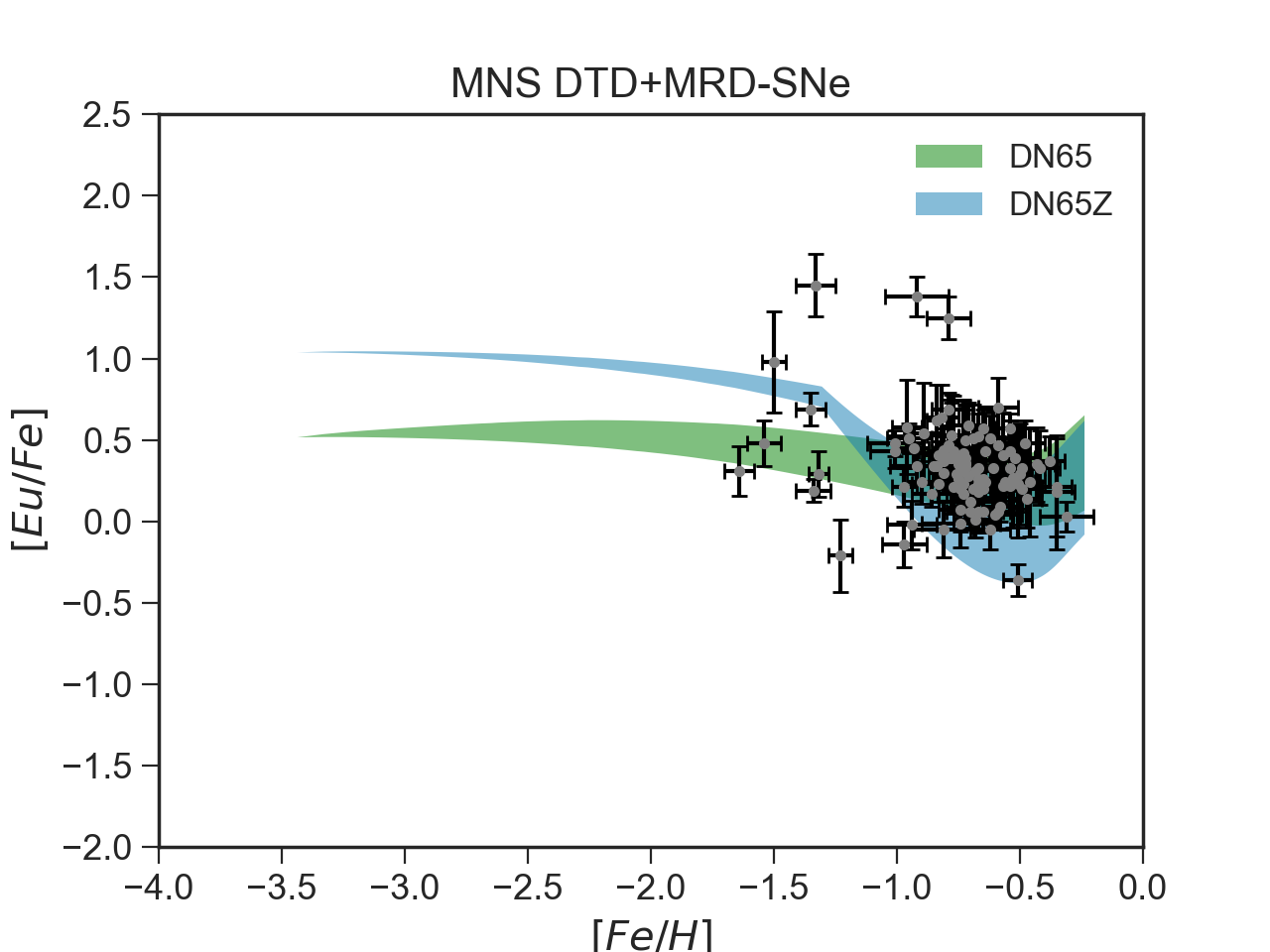}\label{fig:f}}
 \caption{Same of Figure \ref{fig: Scl_New} but for Fornax. Note that in panel (d) the light green curve refers to the case in which the MRD-SNe are producing r-process material only for metallicities higher than $10^{-3}$. See text for details.}%
 \label{fig: For_New}%
\end{center}
\end{figure*}

\subsubsection{Results for Europium in Sculptor and Fornax}
\label{results europium}

In Figures \ref{fig: Scl_New} and \ref{fig: For_New} we report the observed [Eu/Fe] vs [Fe/H] pattern together with predictions of our models for Sculptor and Fornax dSphs, respectively. We remind that details about different nucleosynthesis prescriptions implemented in the models are reported in Table \ref{tab: models_Scl} of section \ref{sec: Eu/Ba yields}.

Observationally, the evolution of [Eu/Fe] vs [Fe/H] shows the typical trend of Eu in the Galaxy, similar to that of an $\alpha$-element. Especially in the case of Sculptor, we can easily distinguish the plateau at low to intermediate metallicities (from $\sim-2.25$ to $\sim-1.25$ dex) and the decrease at higher [Fe/H]. In the case of Fornax it is more difficult to distinguish such a trend. The data appear to be more concentrated in the high metallicity range of the [Eu/Fe]-[Fe/H] diagram, so that the results of our models in this range must be considered just a prediction. We decided not to adopt more data from other authors, in order not to loose the homogeneity of our sample. We do however note that additional high-resolution data containing Ba and Eu are limited. Moreover, from a theoretical point of view, a plateau at low metallicities in Fornax is expected to be present because of the time-delay model (\citealp{2012matteucci}) which applies to any galaxy. As already discussed, in the early phases of galaxy evolution we expect a plateau in the [$\alpha$/Fe] vs [Fe/H] due to the sole contribution of CC-SNe, independently of the SFH. The [Eu/Fe] vs [Fe/H] usually shows a pattern similar to those of the $\alpha$-elements, so that a plateau at low metallicities also in the [Eu/Fe] of Fornax is expected. 

In panels (a) of both Figures \ref{fig: Scl_New} and \ref{fig: For_New}, we report results of models C54, C65 and C76, in which we consider Eu production only by MNS with a constant delay time for merging. The model which best reproduces the expected trend is model C65. In this case, the yield of Eu from MNS is in the range $(3.0\times10^{-6}-1.5\times10^{-5}) \mathrm{M_\odot}$, with a lower limit which is in agreement with the one predicted by \cite{Matteucci2014} for the chemical evolution of the MW. On the other hand, models C54 and C76 seem to overestimate and underestimate the expected trend, respectively. 

In panels (b) of the same Figures are reported results of models D54, D65 and D76 for which Eu is produced by only MNS with a DTD. Those models differ from the previous ones just by the adoption of the DTD. Because of the longer delay times assumed, there is an increasing trend rather then a plateau at low metallicities, as expected. Also, as discussed in the previous section, the adoption of a DTD causes NS continuing to merge until present time, so that the production of Eu from MNS will not stop, even when there is no SF activity (see panel (c) of Figure \ref{fig: SFR+MDF_Scl}). This results in producing a plateau or even an increasing trend at high metallicities for Sculptor and Fornax, respectively, rather than a decrease. Because of that, models D54, D65 and D76 are not able to reproduce the observed pattern, as seen from the Figure, in agreement with the findings of previous studies for the MW (\citealp{Simonetti2019}, \citealp{2019cot}). Moreover, models D54 and D76 overestimate and underestimate the general trend for all the range of metallicities, respectively. 

In panels (c) of Figures \ref{fig: Scl_New} and \ref{fig: For_New} we report the results from models for which we assume Eu production only by MRD-SNe. As already discussed, we try different yields of Eu proposed in literature. Among those, model N17c with yields from \cite{Nishimura2017} appears to be the best one for Sculptor, while model N15 with yields from \cite{Nishimura2015} better reproduce the [Eu/Fe] of Fornax. We remind that in both cases we assume that between 1\% and 2\% of all stars with mass in the $(10-80) \mathrm{M_\odot}$ range would explode as MRD-SNe. Furthermore, we stress that theoretical calculations of the r-process involve large uncertainties in the modelling (see e.g., \citealp{2019cowan}; \citealp{2019horowitx} for recent reviews).

In panels (d) we show the effect of activating the MRD-SNe channel only at metallicities lower than $10^{-3}$, without changing the Eu yield with respect to model N17c and N15 for Sculptor and Fornax, respectively. Model N17cZ reproduces the plateau at low metallicities in the Scl dSph as well as the decrease at higher [Fe/H]. The decrease is actually faster than that produced by model N17c, but the data are also well reproduced. On the other hand, activating MRD-SNe only at low metallicities in Fornax results in loosing the agreement with observations, as expected. Actually, because of the concentration of data at high [Fe/H], it seems that only a highly implausible scenario in which MRD-SNe are acting at high metallicities can reproduce the expected trend, as represented by the light green curve of Figure \ref{fig: For_New}. Therefore, if MRD-SNe are the only producers of Eu in the Fornax dSph, they must be active at all metallicities.

In panels (e) of Figures \ref{fig: Scl_New} and \ref{fig: For_New} we show results of models CN54, CN65 and CN76, in which we assume Eu produced by both MNS with a constant total delay time for merging together with MRD-SNe. Yield of Eu from MRD-SNe are those of \cite{Nishimura2017}, and the three models differ because of the different yields of Eu from MNS. Obviously, when more than one channel contribute to the Eu production, the Eu yields from each channel should be lower than in the case of only one active source, in order to maintain the fit. 

In panels (f), we show results of models DN65 and DN65Z in which both MNS with a DTD and MRD-SNe can produce Eu. For both models the yield of Eu from MNS is in the $(3.0\times10^{-6}-1.5\times10^{-5}) \mathrm{M_\odot}$ range, while that of MRD-SN is equal the one of \cite{Nishimura2017}. The two models differ only for the range of metallicities in which MRD-SNe are active: in model DN65 they act for the whole range, while in model DN65Z they act only at low metallicities. Both models seems to be able to reproduce the main trend. In particular, the lack of Eu from MNS at low metallicities, due to longer delay times for merging, is compensated by the production of Eu from MRD-SN which, in both models, are active at low metallicities. In the same way, when in model DN65Z MRD-SNe stop to produce Eu from metallicities higher than $10^{-3}$, MNS can compensate. For model DN65 we get Eu from both MNS and MRD-SNe also at high metallicities, resulting in a slightly higher trend with respect to model DN65Z which, in the case of Fornax, is more in agreement with the data. 

\begin{figure*}
\begin{center}
 \subfloat[]{\includegraphics[width=1\columnwidth]{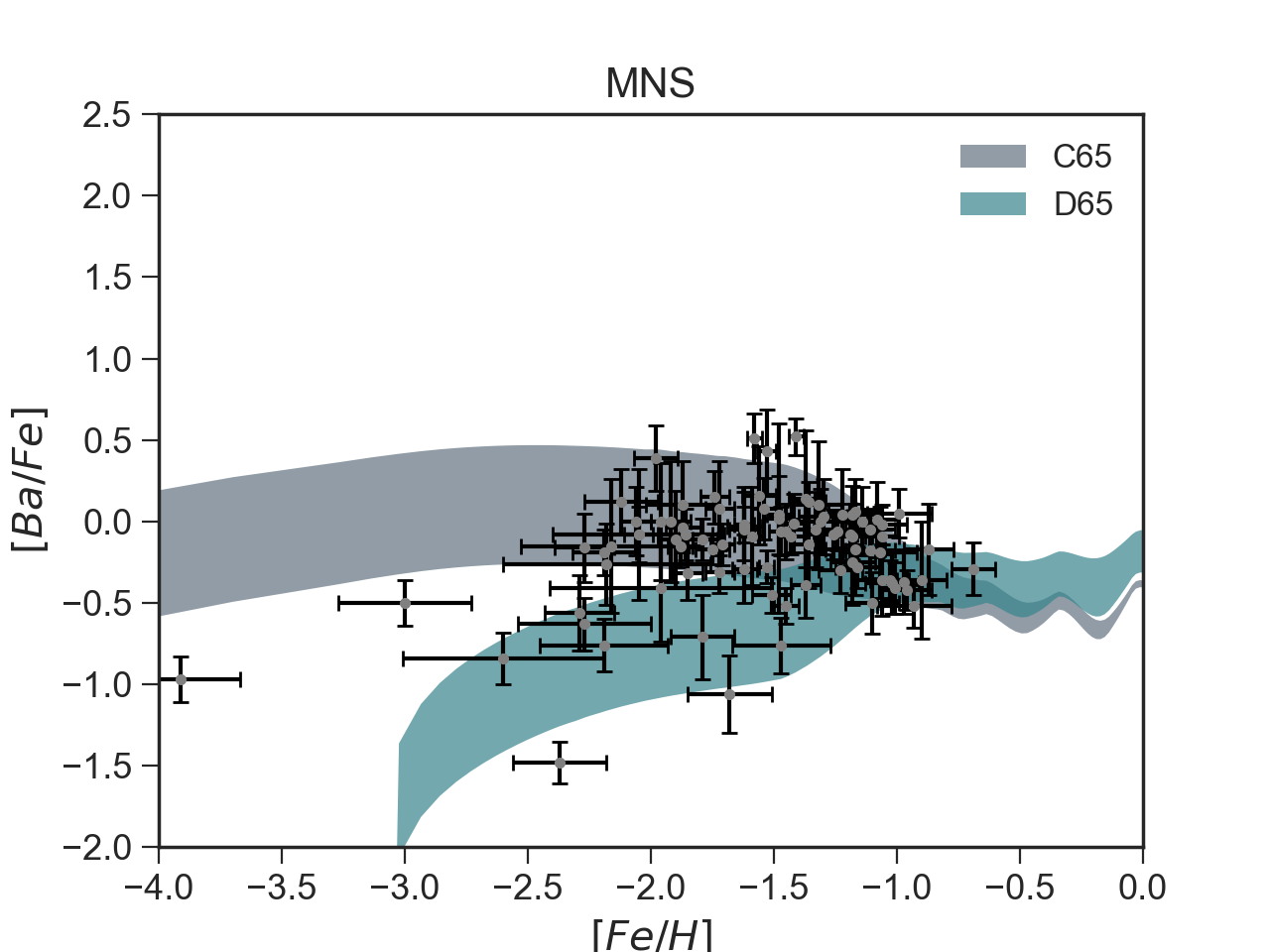}\label{fig:a}}
 \subfloat[]{\includegraphics[width=1\columnwidth]{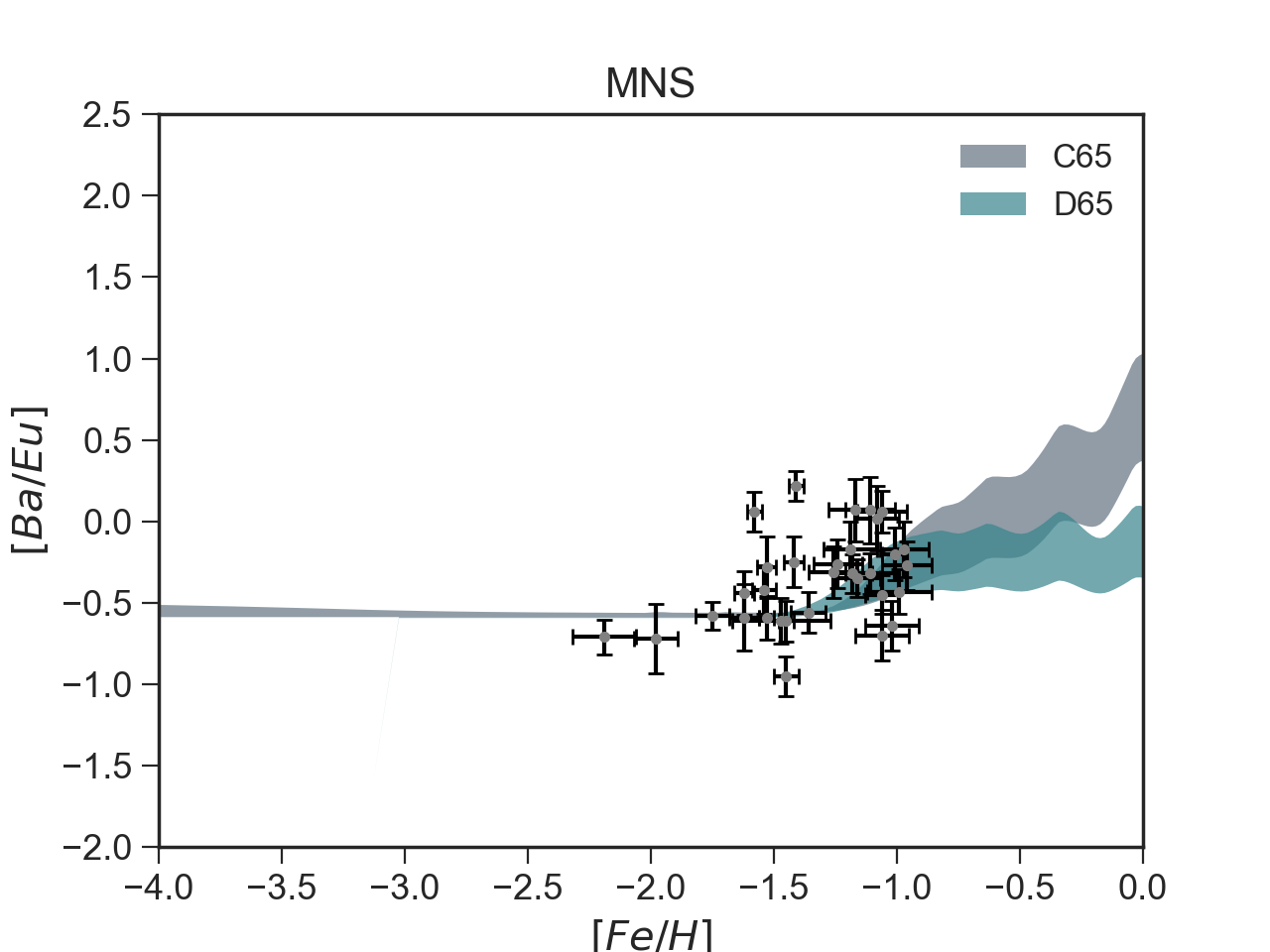}\label{fig:b}}
 \hfill
 \subfloat[]{\includegraphics[width=1\columnwidth]{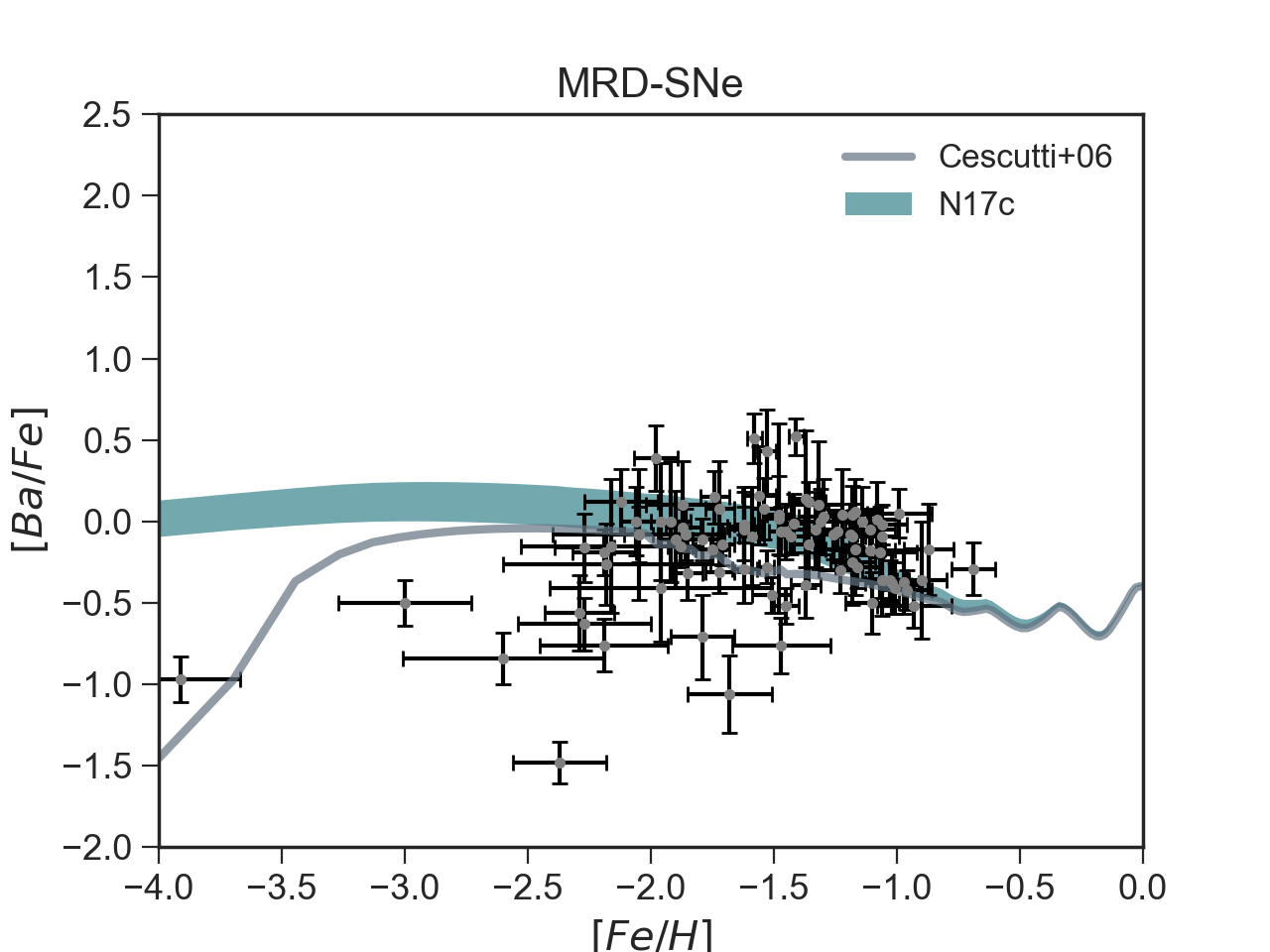}\label{fig:c}}
 \subfloat[]{\includegraphics[width=1\columnwidth]{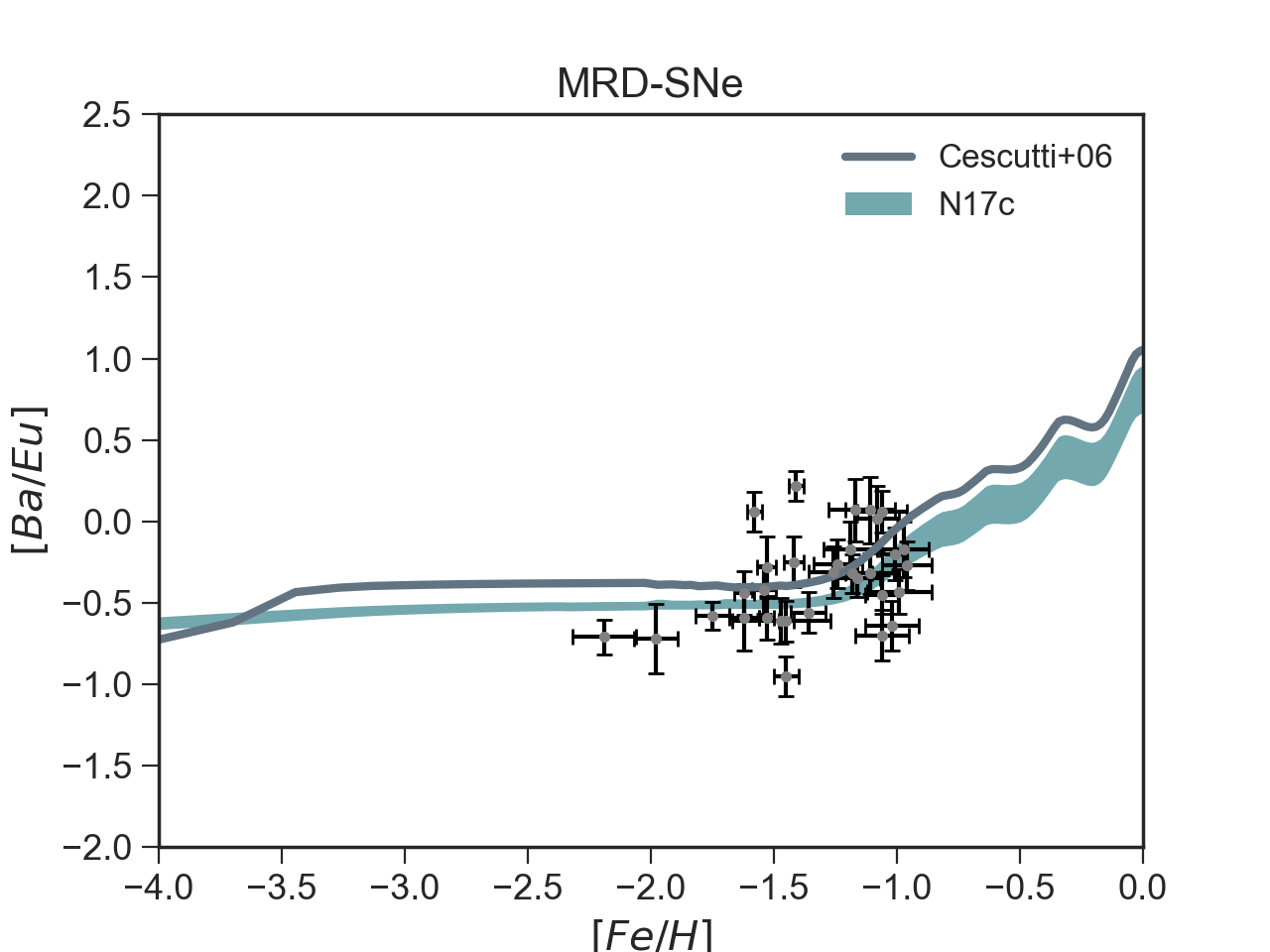}\label{fig:d}}
 \hfill
 \subfloat[]{\includegraphics[width=1\columnwidth]{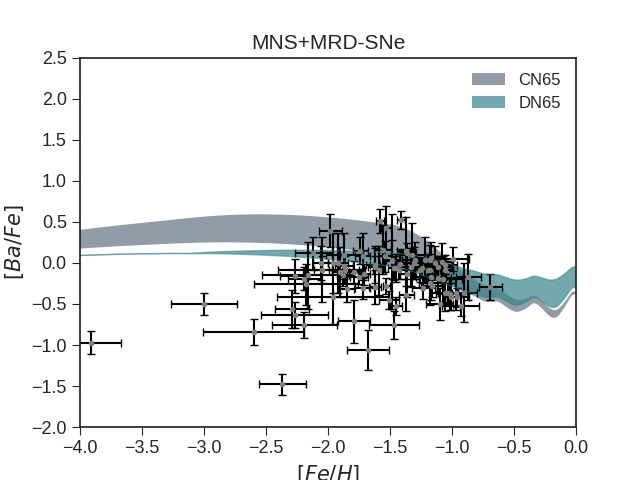}\label{fig:e}}
 \subfloat[]{\includegraphics[width=1\columnwidth]{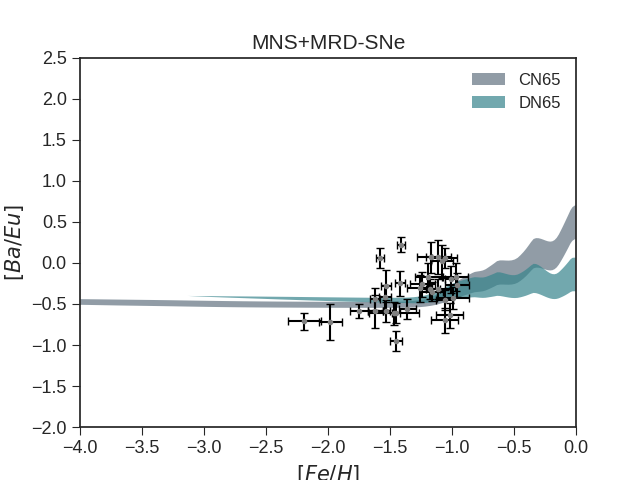}\label{fig:f}}
 \caption{Results of models for the evolution of [Ba/Fe] and [Ba/Eu] vs [Fe/H] for Sculptor dSph. Panels (a) and (b): results of models for which r-process Ba and Eu are produced only by MNS with and without a DTD; panels (c) and (d): results of models in which r-process Ba and Eu are produced only by massive stars; panels (e) and (f): results of models in which r-process Ba and Eu are produced both by MRD-SNe and by MNS (with and without a DTD). For all models s-process Ba production comes from LIMS. Details of models are reported in Table \ref{tab: models_Scl}}%
 \label{fig: BaFe_Scl}%
\end{center}
\end{figure*}

\begin{figure*}
\begin{center}
 \subfloat[]{\includegraphics[width=1\columnwidth]{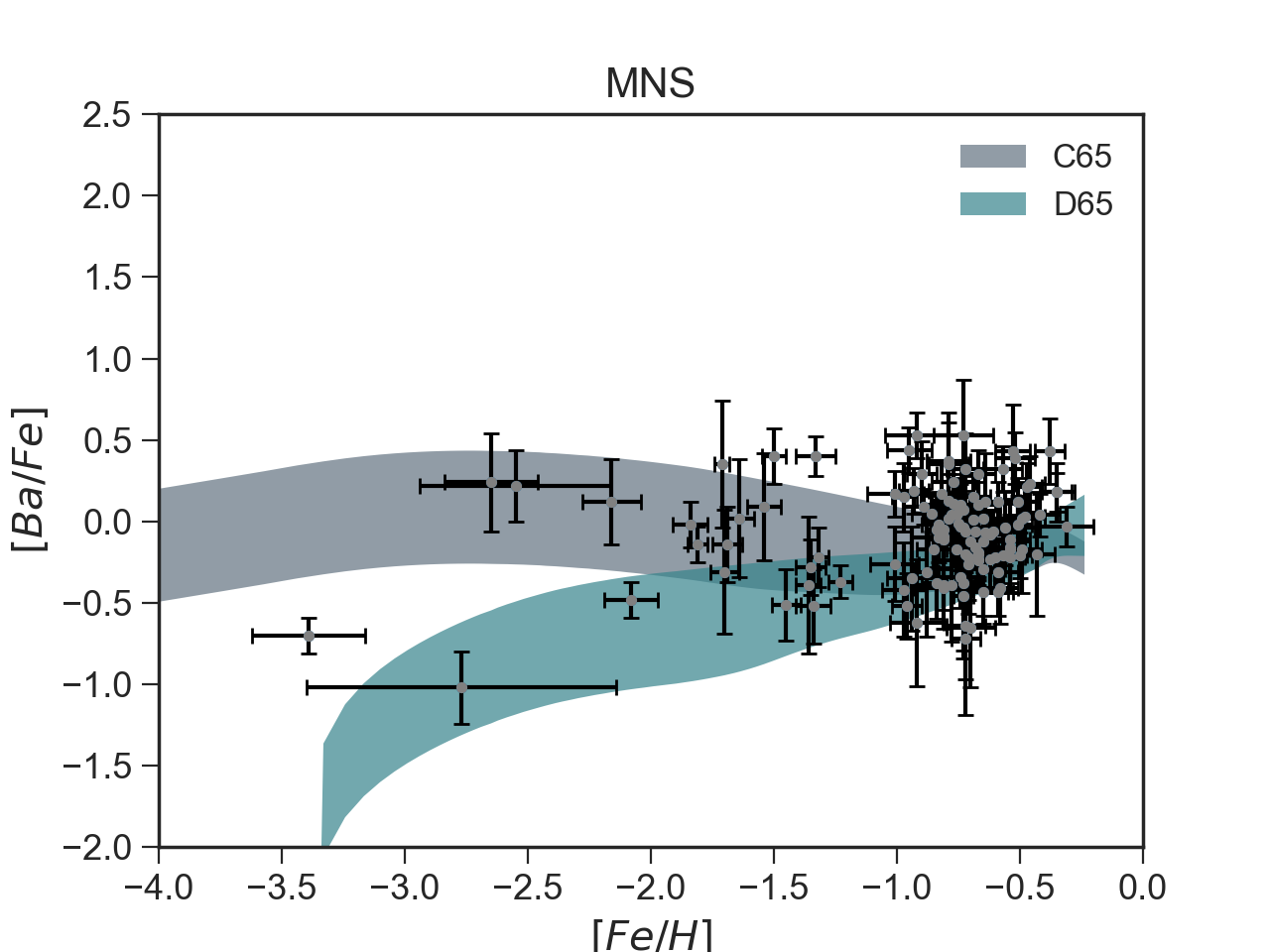}\label{fig:a}}
 \subfloat[]{\includegraphics[width=1\columnwidth]{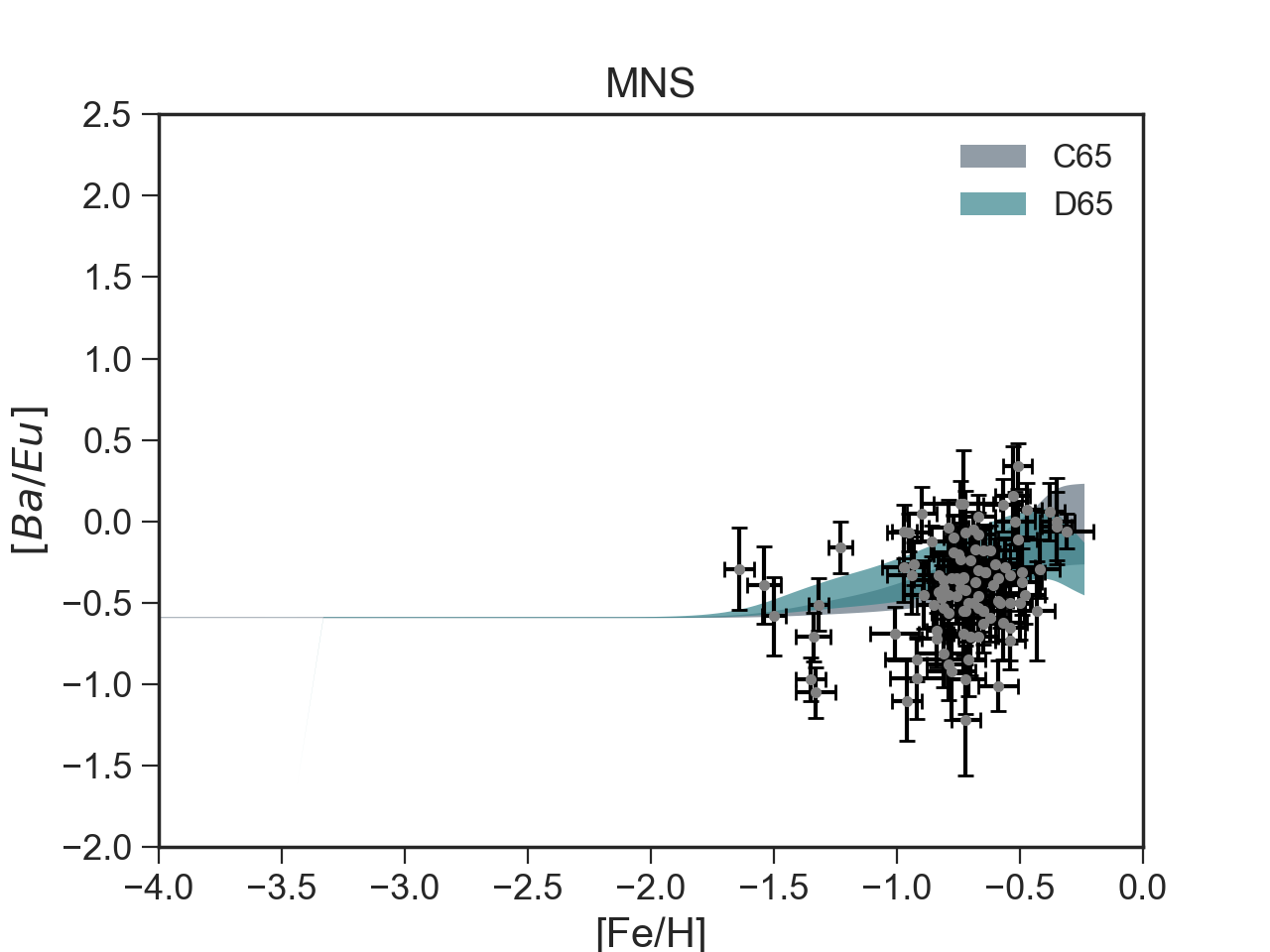}\label{fig:b}}
 \hfill
 \subfloat[]{\includegraphics[width=1\columnwidth]{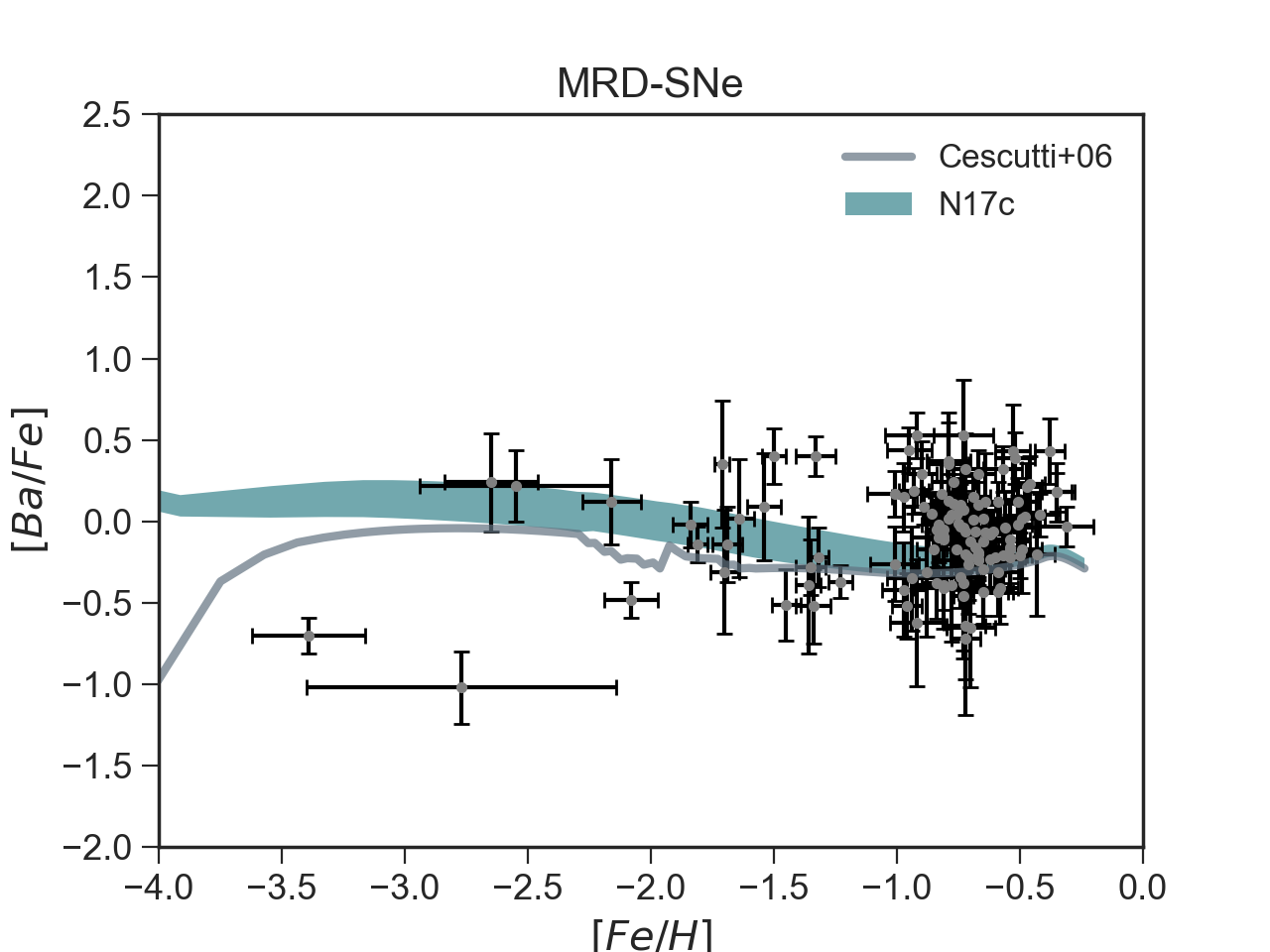}\label{fig:c}}
 \subfloat[]{\includegraphics[width=1\columnwidth]{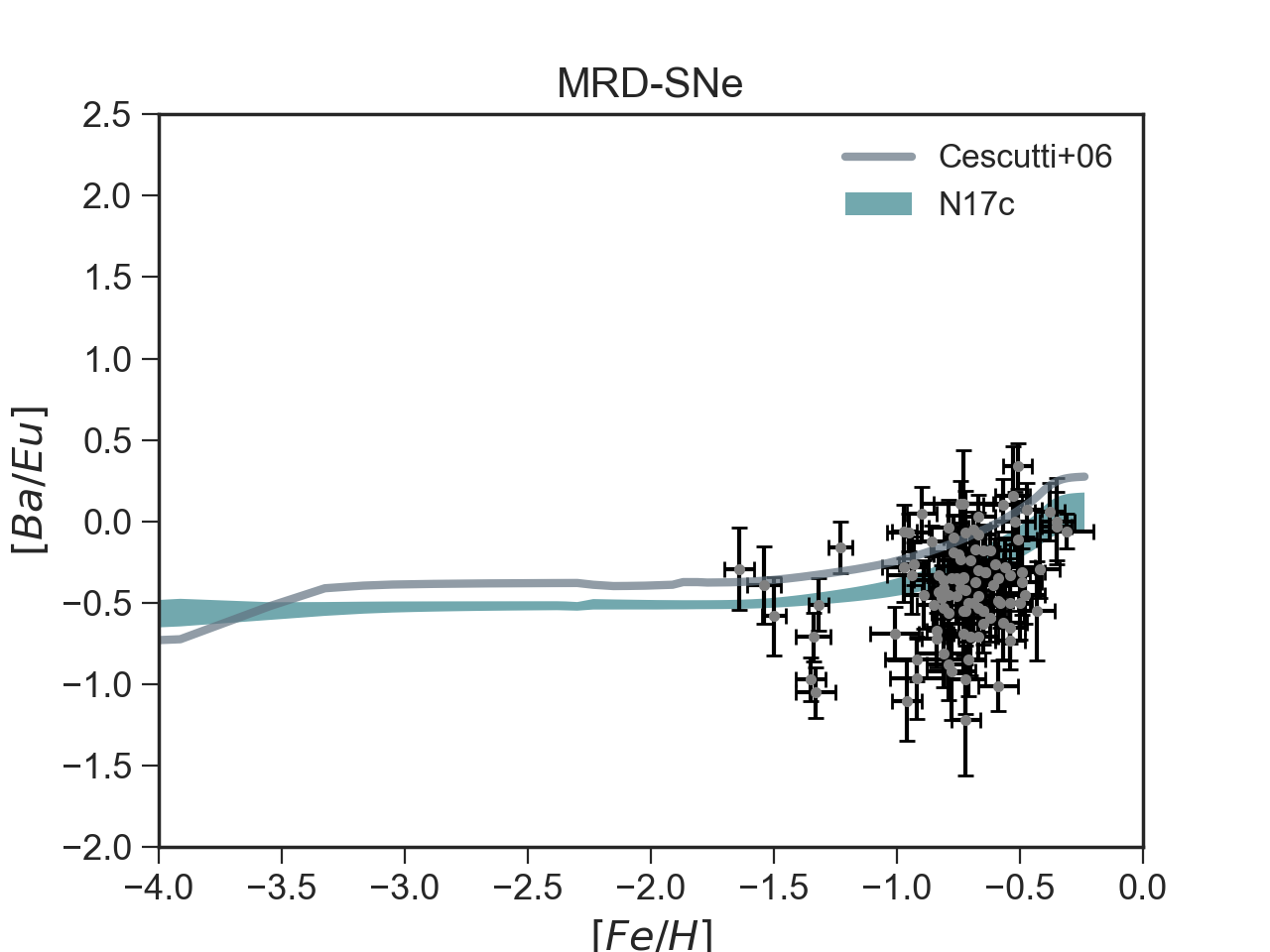}\label{fig:d}}
 \hfill
 \subfloat[]{\includegraphics[width=1\columnwidth]{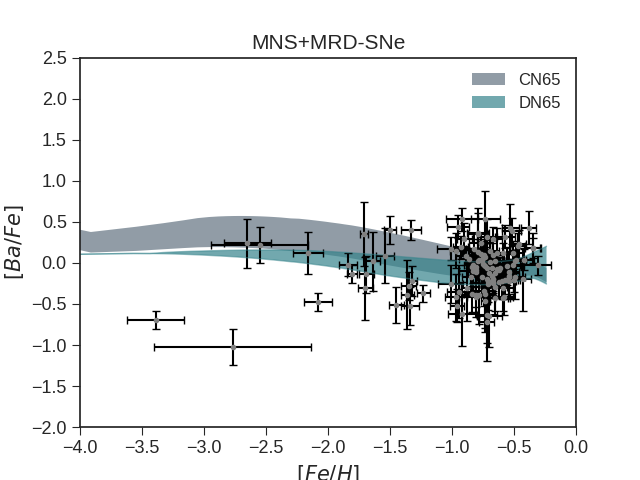}\label{fig:c}}
 \subfloat[]{\includegraphics[width=1\columnwidth]{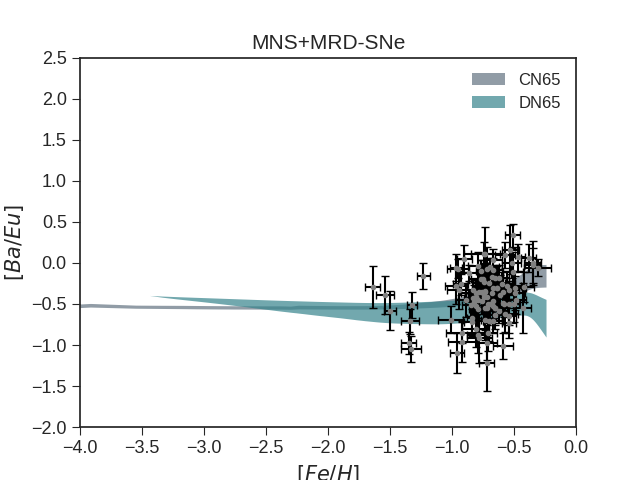}\label{fig:d}
 }
 \caption{Same of Figure \ref{fig: BaFe_Scl} but for Fornax.}%
 \label{fig: BaFe_For}%
\end{center}
\end{figure*}

\subsubsection{Results for Barium in Sculptor and Fornax}
\label{results barium}

In Figures \ref{fig: BaFe_Scl} and \ref{fig: BaFe_For}, we report predictions for the [Ba/Fe] and [Ba/Eu] vs [Fe/H] together with the observational data. 

The observed [Ba/Fe] vs [Fe/H], is characterized by a low abundance of Ba at low metallicities ([Fe/H]$\leq$-2.25) and by almost solar values from intermediate to high metallicities, suggesting different mechanisms for the production of the s- and r- process fractions of Ba. In fact, at low metallicities Ba is mostly created by r-process, but as more LIMS go through the AGB phase, the s-process becomes more important and the [Ba/Fe] ratio increases with increasing [Fe/H] until a plateau is reached (\citealp{Skuladottir2020}). For the [Ba/Eu] vs [Fe/H], the data are characterized by a plateau at lower metallicities, followed by an increase of the [Ba/Eu] at higher [Fe/H]. The plateau is indicative of the fact that the Ba and Eu elements are growing at the same rate at low metallicities as a function of Fe. This does not necessarily means that the two elements must be produced by the same events, but they must be produced at least with the same time delay (\citealp{Reichert2020}). On the other hand, the increasing trend of the [Ba/Eu] at higher metallicities sets in when the production of s-process Ba from LIMS starts to be non negligible. We note that for all of our simulations, we fixed the yields of Ba from the s-process and varied only the contribution from the r-process.

In panels (a) and (b) of Figures \ref{fig: BaFe_Scl} and \ref{fig: BaFe_For}, we show results of models C65 and D65 in which we adopt MNS as the only producers of r-process Ba with and without a DTD, respectively. In both cases yields of r-process Ba are in the $(3.20\times10^{-5}-1.58\times10^{-4}) \mathrm{M_{\odot}}$ range while those of Eu are in the $(3.0\times10^{-6}-1.5\times10^{-5}) \mathrm{M_\odot}$ range. For the [Ba/Fe] (panels (a)), in the case of a constant delay time for merging models are able to fit the data from intermediate to high metallicities, but fails at lower ones. On the other hand, if we adopt a DTD for MNS the agreement at low metallicities is improved, but the data are underestimated at intermediate ones (-2.6 $\le$ [Fe/H] $\le$ -1.5), suggesting that a second source should be active. For the [Ba/Eu], it is possible to see that both our models are able to reproduce the plateau in the data, thanks to the same delay assumed for the production of Eu and r-process Ba, as well as the increase when the production of s-process Ba from LIMS sets in.

In panels (c) and (d) of the same Figures, we report results of model N17c in which we assume r-process Ba and Eu produced only by MRD-SNe with yields from \cite{Nishimura2017}. Also models in which we adopt yields of \cite{Cescutti2006} for the r-process production by massive stars are shown. For the [Ba/Fe], both models fit the data at high metallicities, but fail at lower ones overproducing the data. In particular, model N17c produces almost a plateau rather than an increasing trend at low [Fe/H], because of the production of r-process Ba from stars with initial masses in a too wide range (10-80 M$_\odot$). In fact, if a more narrow range is assumed, as in the case of models with yields of \cite{Cescutti2006} (12-30 M$_\odot$), we can predict a more intense increase, which sets in too early with respect to the data however. In the case of the [Ba/Eu], models N17c can reproduce the expected trend for all the range of metallicities for Sculptor, but underestimate the data in the case of Fornax. On the other hand, if the yields of \cite{Cescutti2006} are adopted, the models overestimate the expected abundance trends in both galaxies. However, the general trend is reproduced in all cases, since Eu and r-process Ba are produced by the same event and therefore with the same delay. 

In panels (e) and (f), we show results of models CN65 and DN65 in which r-process Ba and Eu are produced by both MRD-SNe and MNS. In model CN65 we assume a short and constant delay time for MNS, while in model DN65 a DTD is adopted. As expected, both models are not able to reproduce the low data of [Ba/Fe] at low metallicities. For both models, in fact, the production of r-process Ba sets in too early and a too high trend is produced at low [Fe/H]. The models are able to reproduce only the [Ba/Eu], producing the expected plateau at low metallicities and the increase at later [Fe/H] thanks to the production of s-process Ba by LIMS. The plateau is reproduced not only because of the same delay assumed for the production of the two elements in the case of model CN65, but also because of the similar r-process Ba/Eu yields between MRD-SNe and MNS.

\begin{figure*}
\begin{center}
 \subfloat[]{\includegraphics[width=1\columnwidth]{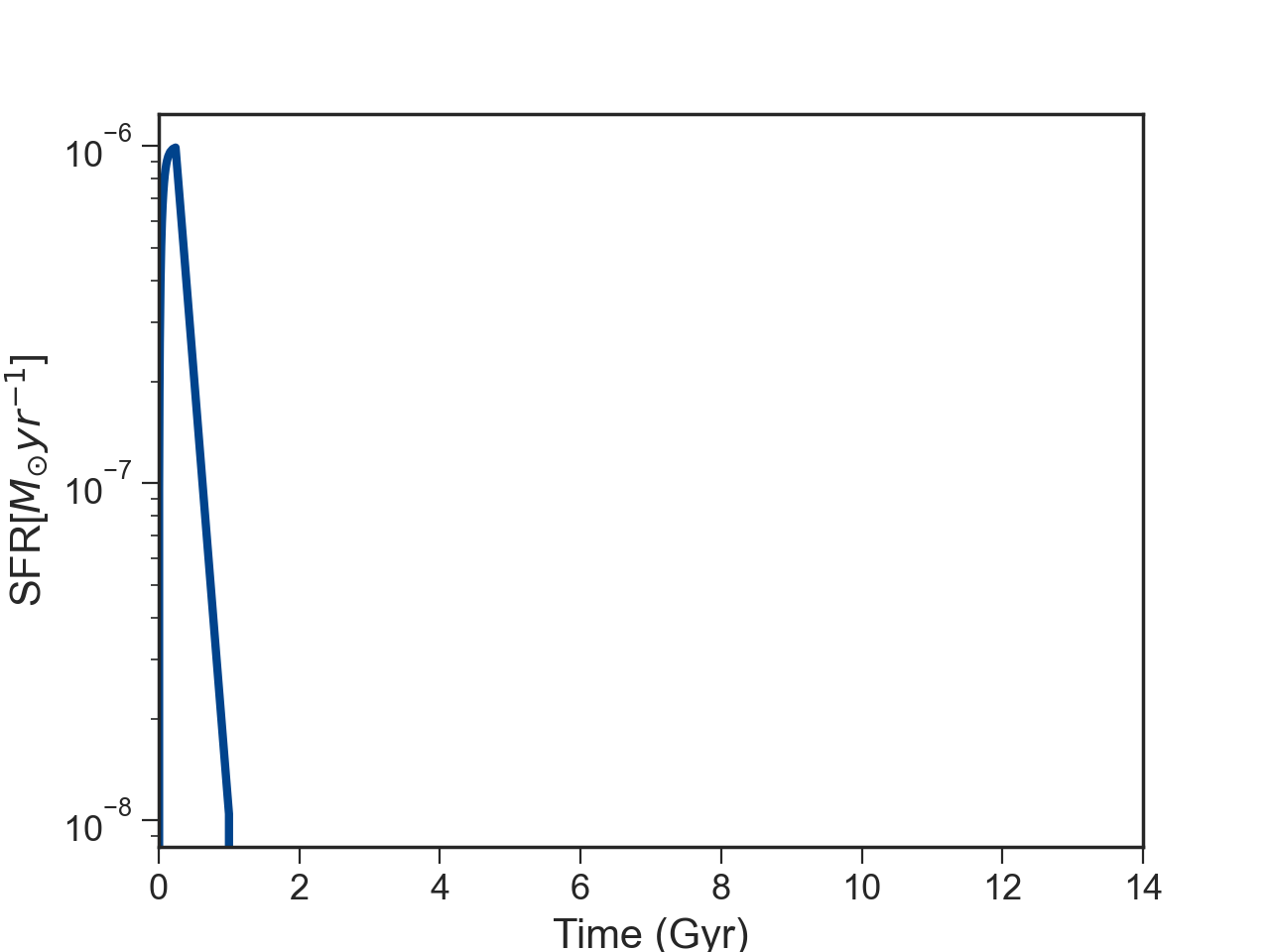}\label{fig:a}}
 \hfill
 \subfloat[]{\includegraphics[width=1\columnwidth]{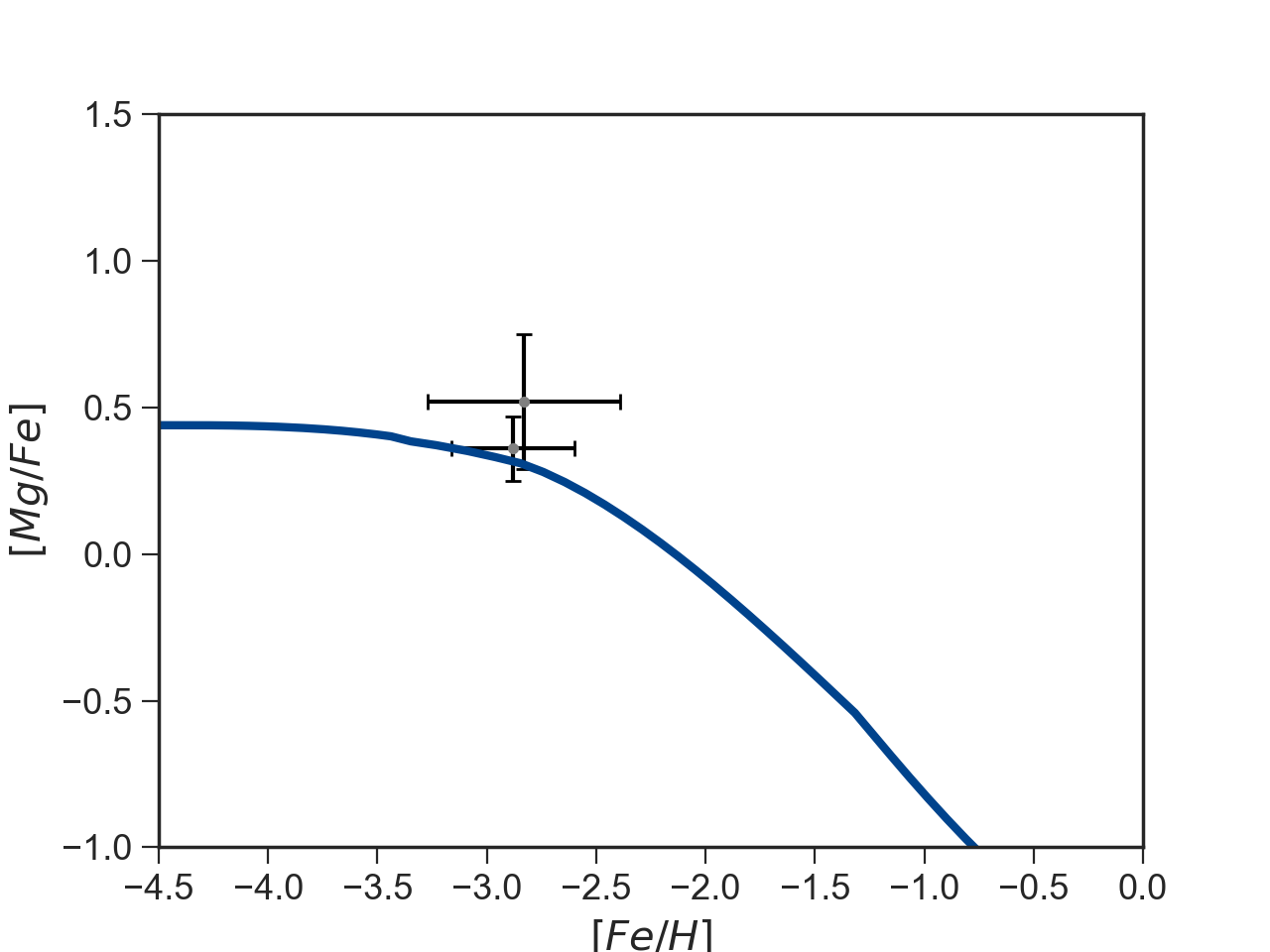}\label{fig:b}}
 \hfill
 \subfloat[]{\includegraphics[width=1\columnwidth]{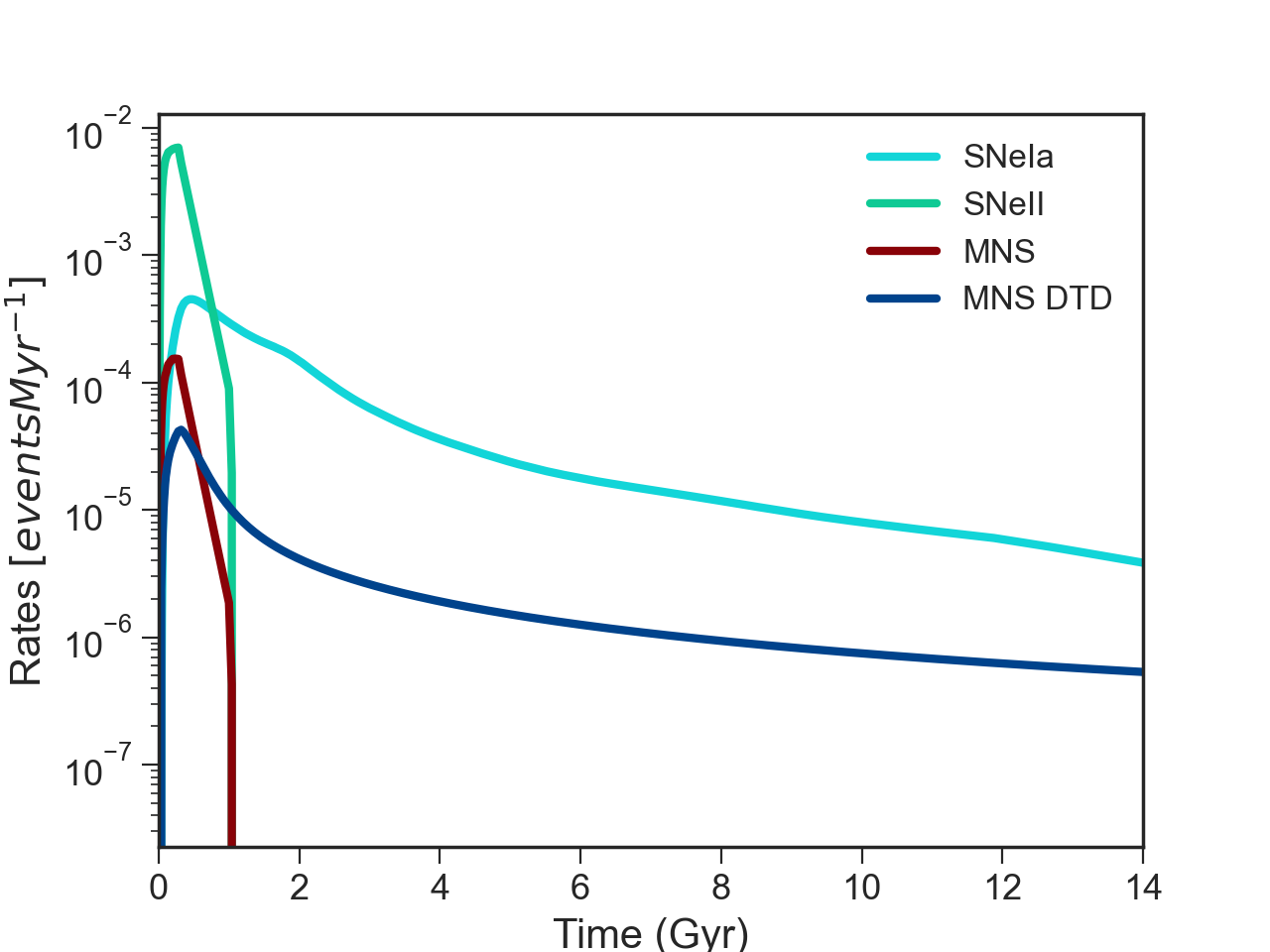}\label{fig:c}}
 \hfill
 \subfloat[]{\includegraphics[width=1\columnwidth]{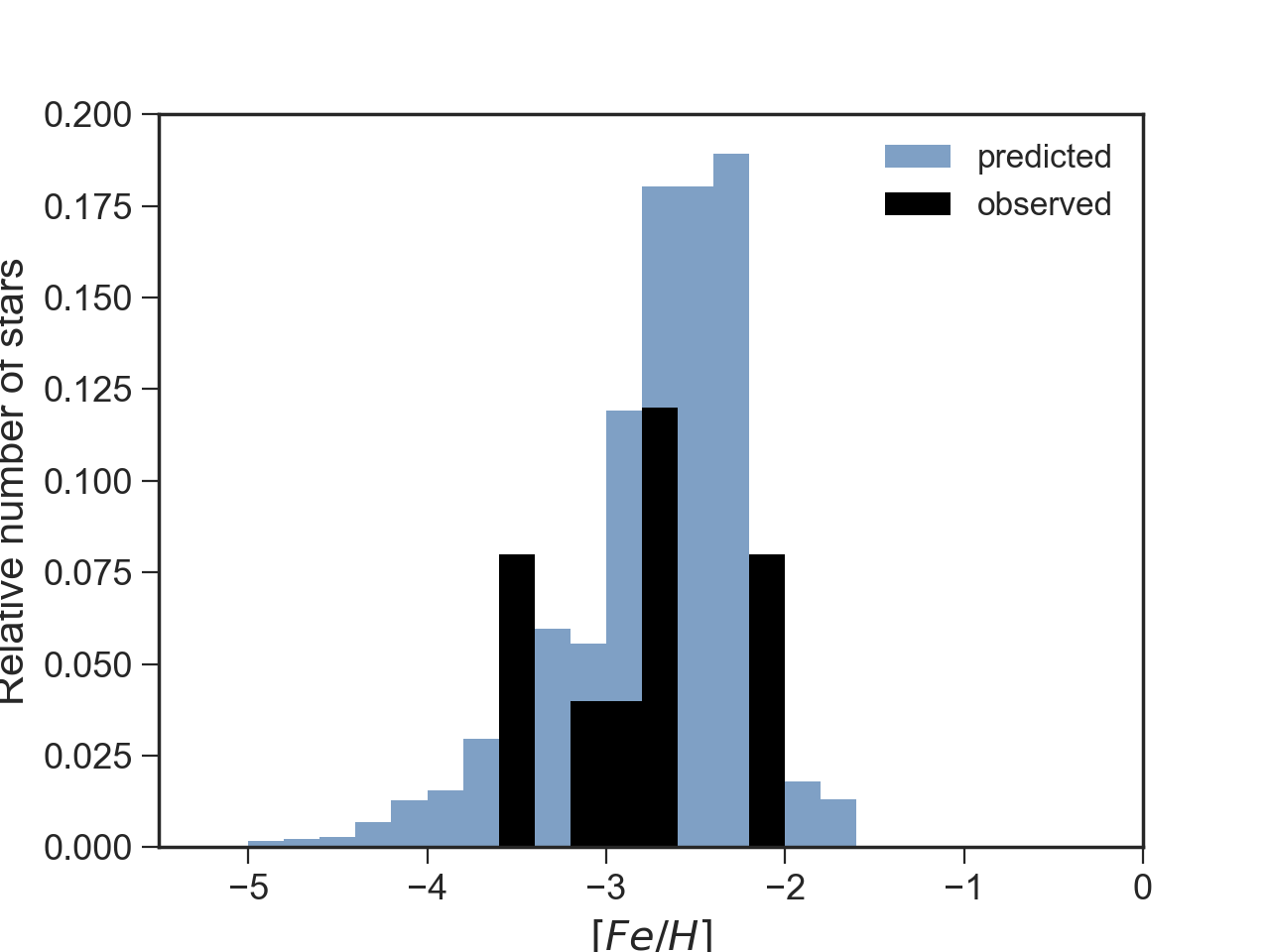}\label{fig:d}}%
 \caption{Same of Figures \ref{fig: SFR+MDF_Scl} and \ref{fig: SFR+MDF_For} but for Reticulum II.}%
 \label{fig: SFR+MDF_Ret}%
\end{center}
\end{figure*}

\subsection{Reticulum II}
\label{reticulum}

For the chemical evolution of Reticulum II UFD, we assume a dark matter halo of mass $\mathrm{M_{DM}}=3.0\times10^6 \mathrm{M_\odot}$ and a core radius $\mathrm{R_{DM}}=170 \mathrm{pc}$. The effective radius of the luminous component of the galaxy has been set at $\mathrm{R_L}=50 \mathrm{pc}$. We predict a present time stellar mass of $\mathrm{M_{\star,f}}=0.6\times10^{3}\mathrm{M_\odot}$, similar to the one observed by \cite{2015bechtol} equal to $\mathrm{M_{\star,f}}=2.6\times10^{3}\mathrm{M_\odot}$.

In panel (a) of Figure \ref{fig: SFR+MDF_Ret}, we show our assumed SFR as a function of time. It consists of one short episode of SF which lasts 1 Gyr. 

In panel (b) of the same Figure, we report the [Mg/Fe] vs [Fe/H] together with the prediction of our model. Because of the poor dataset, it is not possible to derive strong conclusions on the observed trend. Therefore, we model the [Mg/Fe] in order to reproduce the typical evolution of an $\alpha$-element, characterized by a plateau at low metallicities and by a decrease which set in when SNeIa start contributing in a substantial way to the Fe enrichment. 

In panel (c) of Figure \ref{fig: SFR+MDF_Ret}, we report the rates of different phenomena. Also in this case, it is seen that the rate of MNS follows the evolution of the star formation only in the case of a constant delay time for merging, leading to a predicted present time rate of MNS equal to zero. In the case of DTD, instead, the present time rate of MNS will be equal to $\mathrm{R_{MNS}}\simeq5\times10^{-4}$  events Gyr$^{-1}$.

Finally, in panel (d) we report the observed MDF together with the prediction from our model. Because of the low number of stars observed in Reticulum II, it is very difficult to asses the quality of the fit. The observational sample is likely incomplete. Therefore, our theoretical MDF has to be regarded as a prediction, to be confirmed (or disproved) by future observations, rather than a fit to the existing data.


\begin{figure*}
\begin{center}
 \subfloat[]{\includegraphics[width=1\columnwidth]{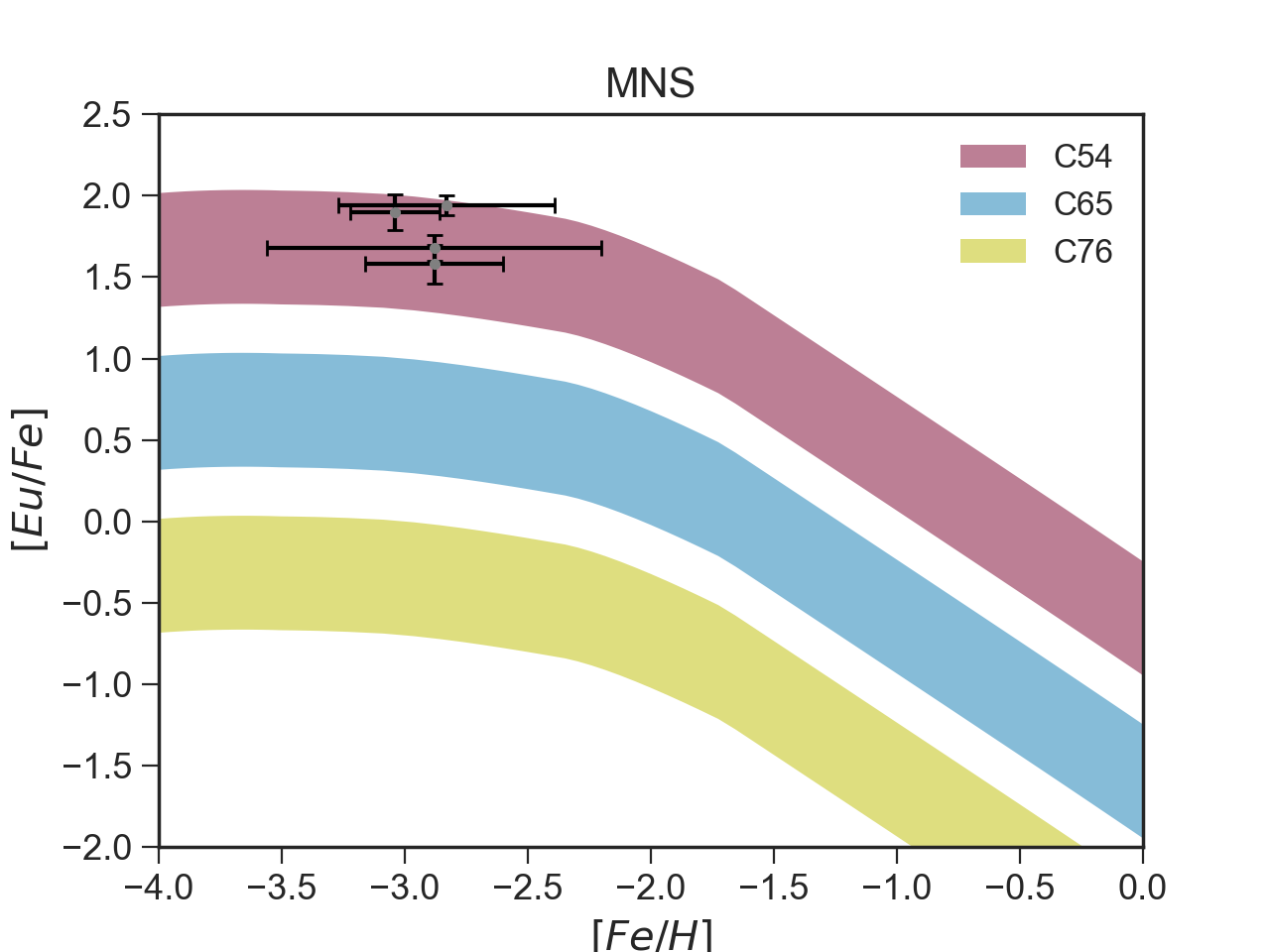}\label{fig:a}}
 \subfloat[]{\includegraphics[width=1\columnwidth]{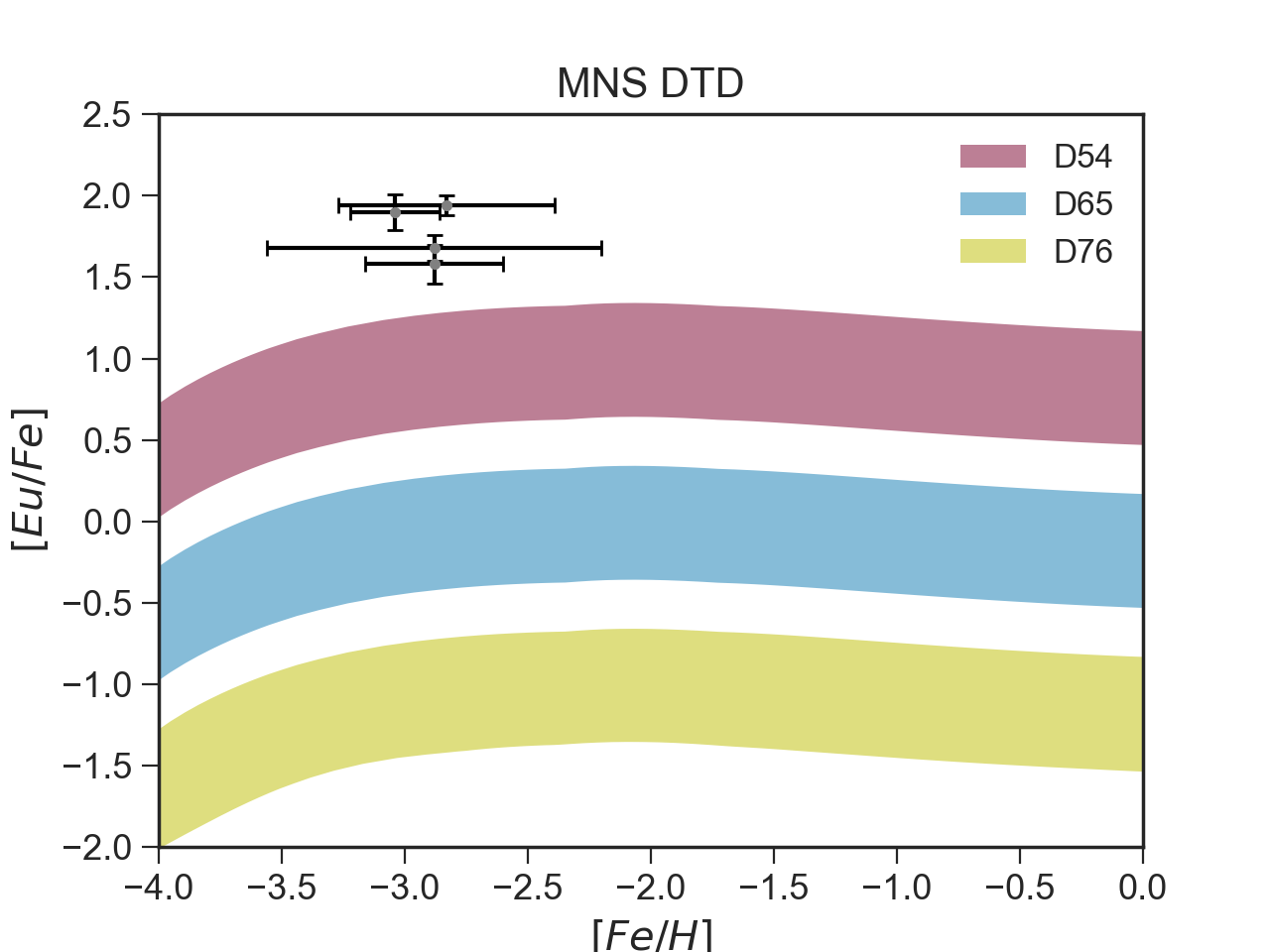}\label{fig:b}}
 \hfill
 \subfloat[]{\includegraphics[width=1\columnwidth]{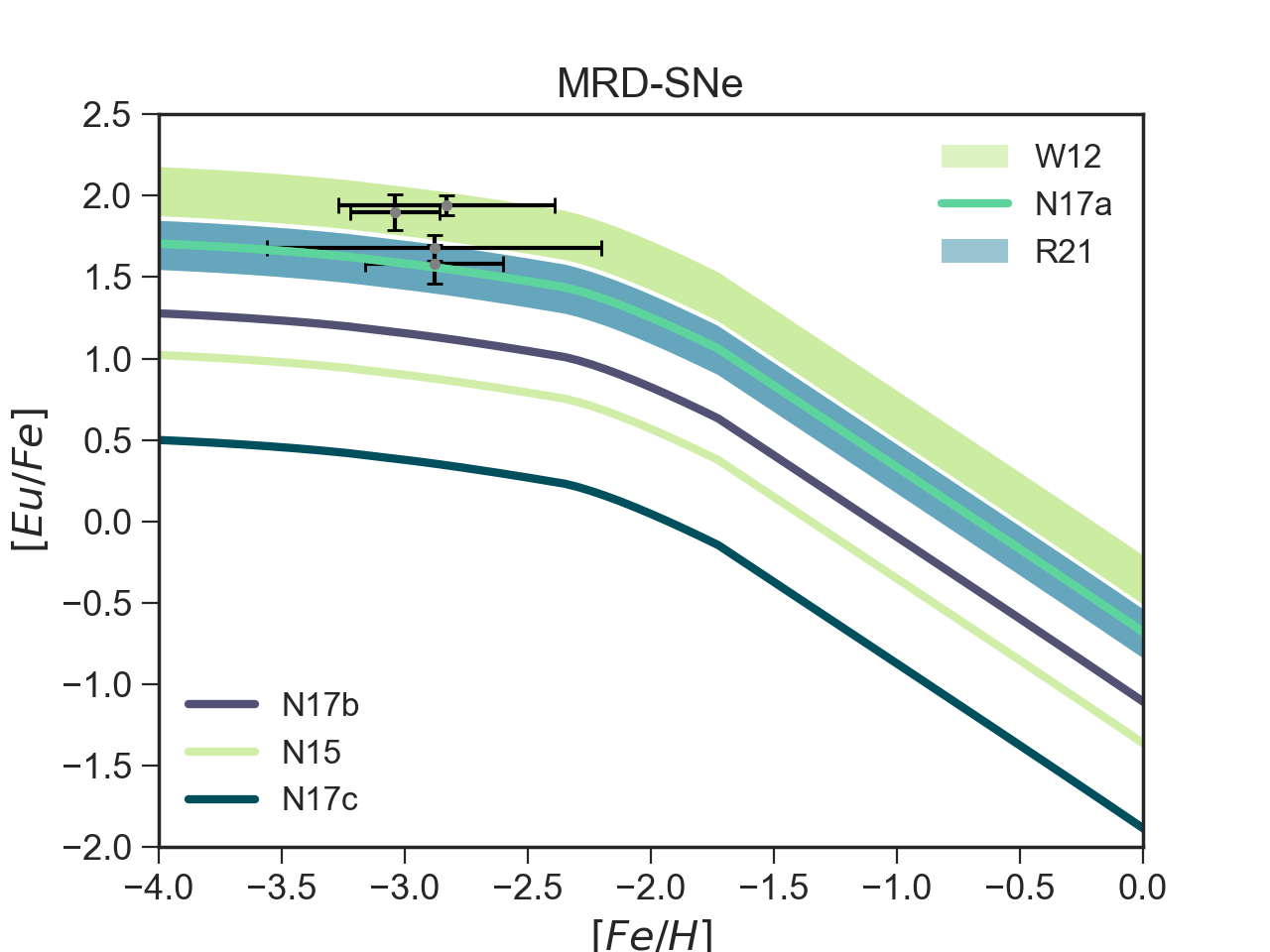}\label{fig:c}}
 \subfloat[]{\includegraphics[width=1\columnwidth]{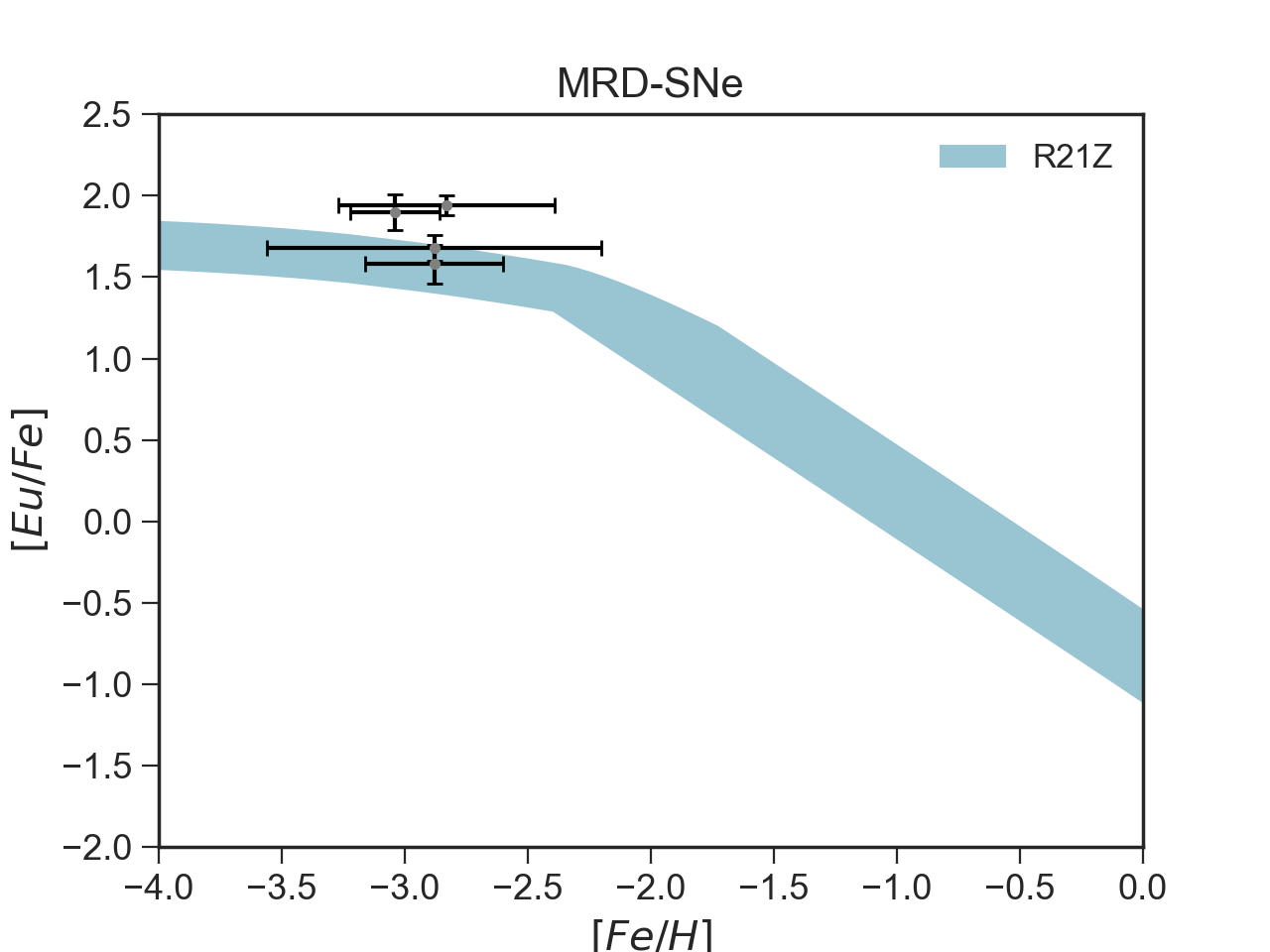}\label{fig:d}}
 \hfill
 \subfloat[]{\includegraphics[width=1\columnwidth]{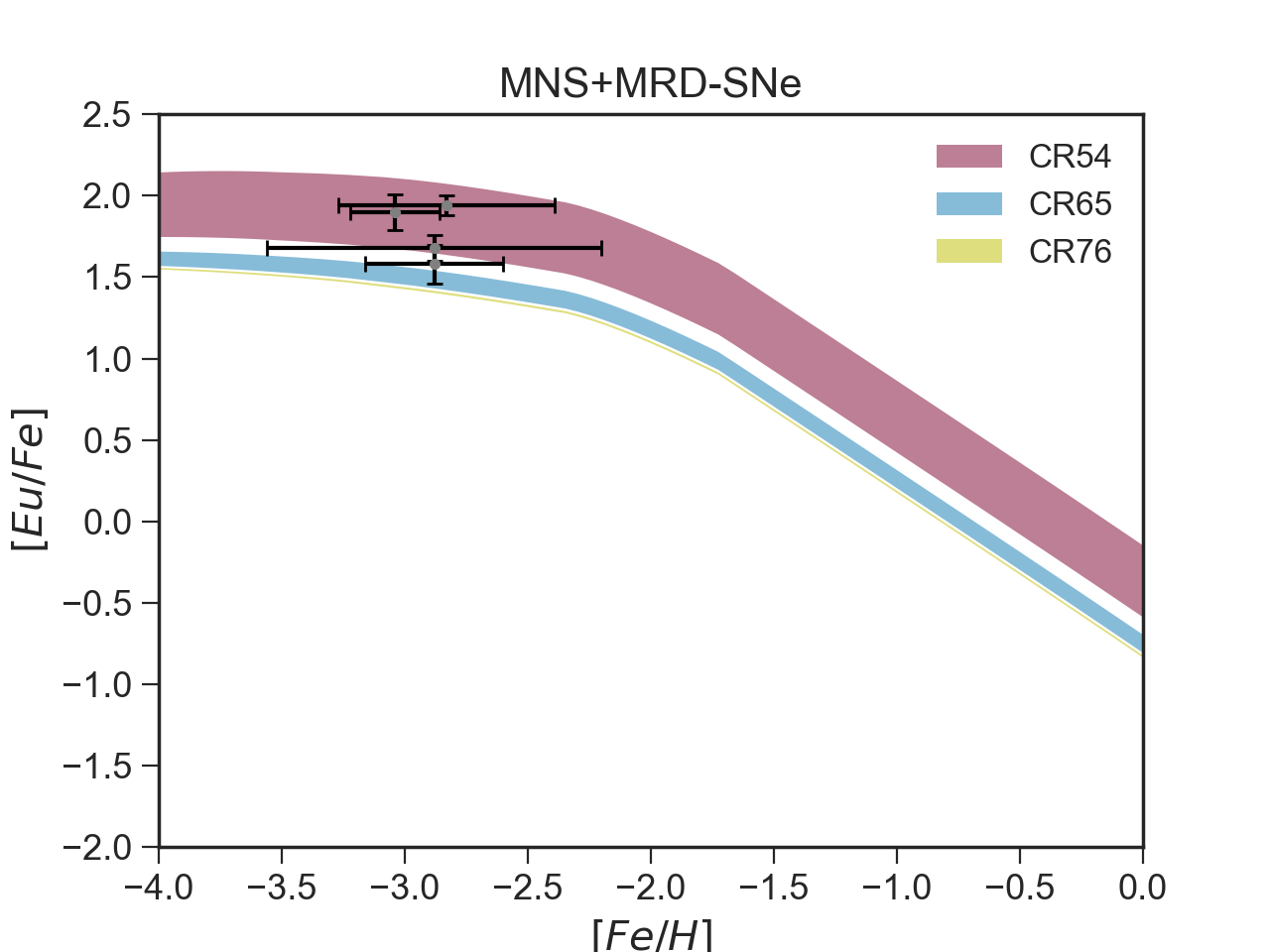}\label{fig:e}}
 \subfloat[]{\includegraphics[width=1\columnwidth]{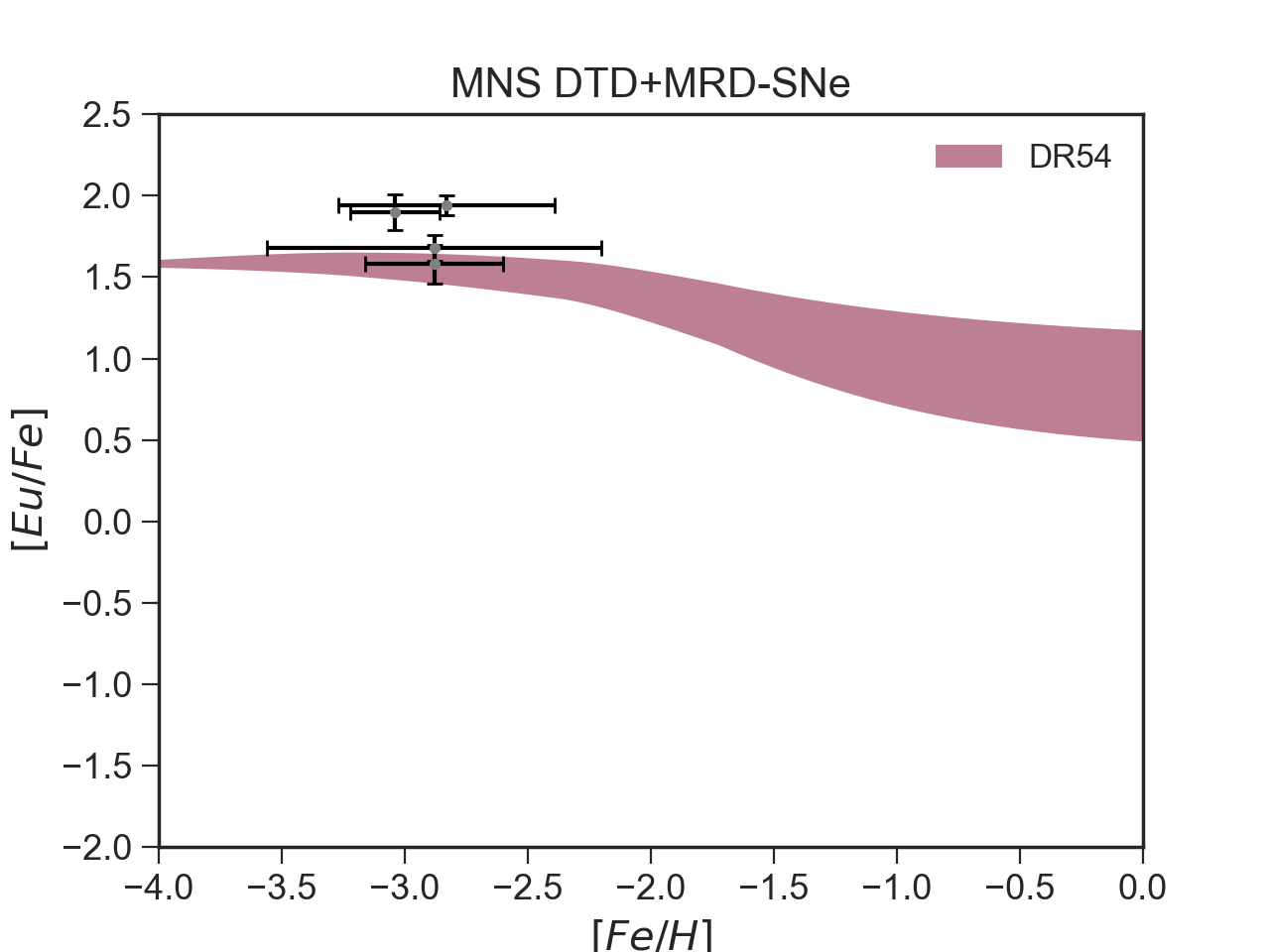}\label{fig:f}}
 \caption{Same of Figures \ref{fig: Scl_New} and \ref{fig: For_New}, but for Reticulum II.}%
 \label{fig: Ret_New}%
\end{center}
\end{figure*}

\subsubsection{Results for Europium in Reticulum II}
\label{results europium reticulum}

In Figure \ref{fig: Ret_New} we report our results together with the observational data for the [Eu/Fe] vs [Fe/H] in the Reticulum II UFD.

Concerning the observational data, Reticulum II stands out among all the other galaxies because of its peculiar Eu and Ba abundances. The data are concentrated at low metallicities and also show strong enhancements, which is about 2 orders of magnitude higher than what is observed in the other dwarf galaxies.

In panel (a), we report results of models in which MNS are the only Eu producers and their delay time for merging is assumed to be short and constant. In this case model C54, in which the yield of Eu from MNS is in the $(3.0\times10^{-5}-1.5\times10^{-4}) \mathrm{M_\odot}$ range, well reproduce the high [Eu/Fe] abundance ratio. On the other hand, when we hypothesize a DTD for MNS (panel (b) of the same Figure), we are no more able to reproduce the observational constraints because of the longer delay assumed for merging. 

In the case in which we assume that Eu is produced only by MRD-SNe, models R21 and W12 are able to reproduce the observed [Eu/Fe] both in the case in which MRD-SNe are active at all metallicities (panel (c)) and in the case in which they are active only at the low end (panel (d)). Actually, because of the really short SF assumed for Reticulum II, there are small differences between these two cases. For these two models the yield of Eu from MRD-SNe has been set equal to that predicted by \cite{2021reichert} for model R21 and by \cite{Winteler2012} for model W12. In panel (c) we report also model N17c for which the yield of Eu is equal to that of \cite{Nishimura2017}, showing how high the yield of Eu is required to be to fit the data in Reticulum II with respect to the other galaxies.

In panel (e) of the same Figure, we report results of models in which we assumed Eu produced by both MNS (with no DTD) and MRD-SNe. The yield of Eu from MRD-SNe has been set equal to that of model R21. As expected model CR54, is the one which best reproduce the data. However, models CR65 and CR76 are only slightly below the observations, showing once again that if a second channel other than MNS is activated then the yield of Eu from MNS can be lower in order to fit data. 

In panel (f) of Figure \ref{fig: Ret_New}, we report results of models in which both MNS (with a DTD) and MRD-SNe are producing Eu. In particular, we assumed that MRD-SNe are acting for all metallicities, but we note that we would have obtained basically the same results even if MRD-SNe would have been activated only at low metallicities. The yield of Eu from MNS is equal to $(3.0\times10^{-5}-1.5\times10^{-4}) \mathrm{M_\odot}$ and that from MRD-SNe is that of model R21, equal to $5.19\times10^{-6} \mathrm{M_\odot}$ for each event. Because of the contribution from MRD-SNe we are now able to fit the data at low metallicities despite the longer delay assumed for MNS. An even better agreement would have been obtained if we adopted even higher yield of Eu from MRD-SNe (for example those of model W12). However one should note that, because of the lack of data at high metallicities, it is impossible to distinguish which is the best model between CR54 and DR54, since they differ only in the absence/presence of a DTD for MNS. 

\begin{figure*}
\begin{center}
 \subfloat[]{\includegraphics[width=1\columnwidth]{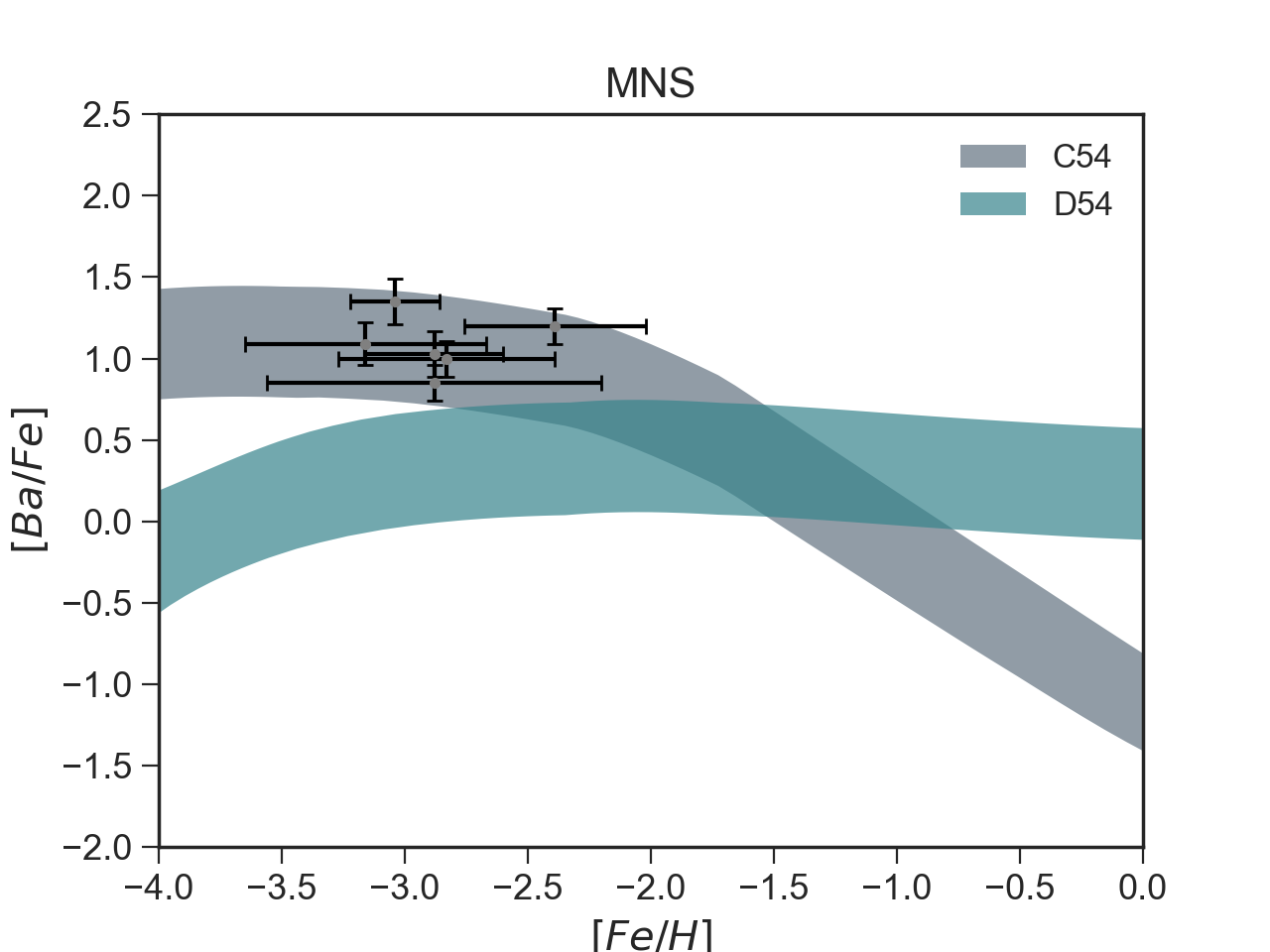}\label{fig:a}}
 \subfloat[]{\includegraphics[width=1\columnwidth]{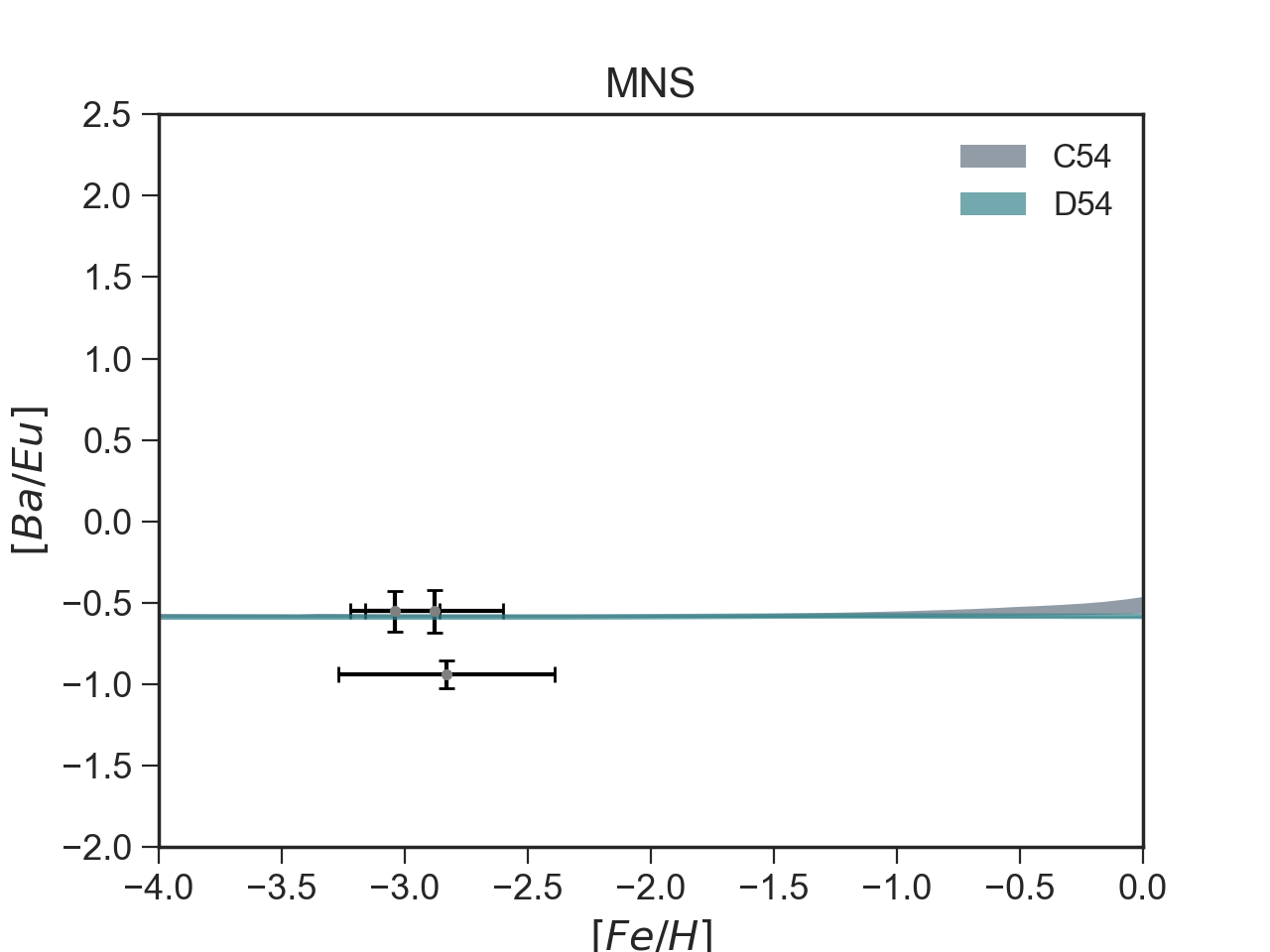}\label{fig:b}}
 \hfill
 \subfloat[]{\includegraphics[width=1\columnwidth]{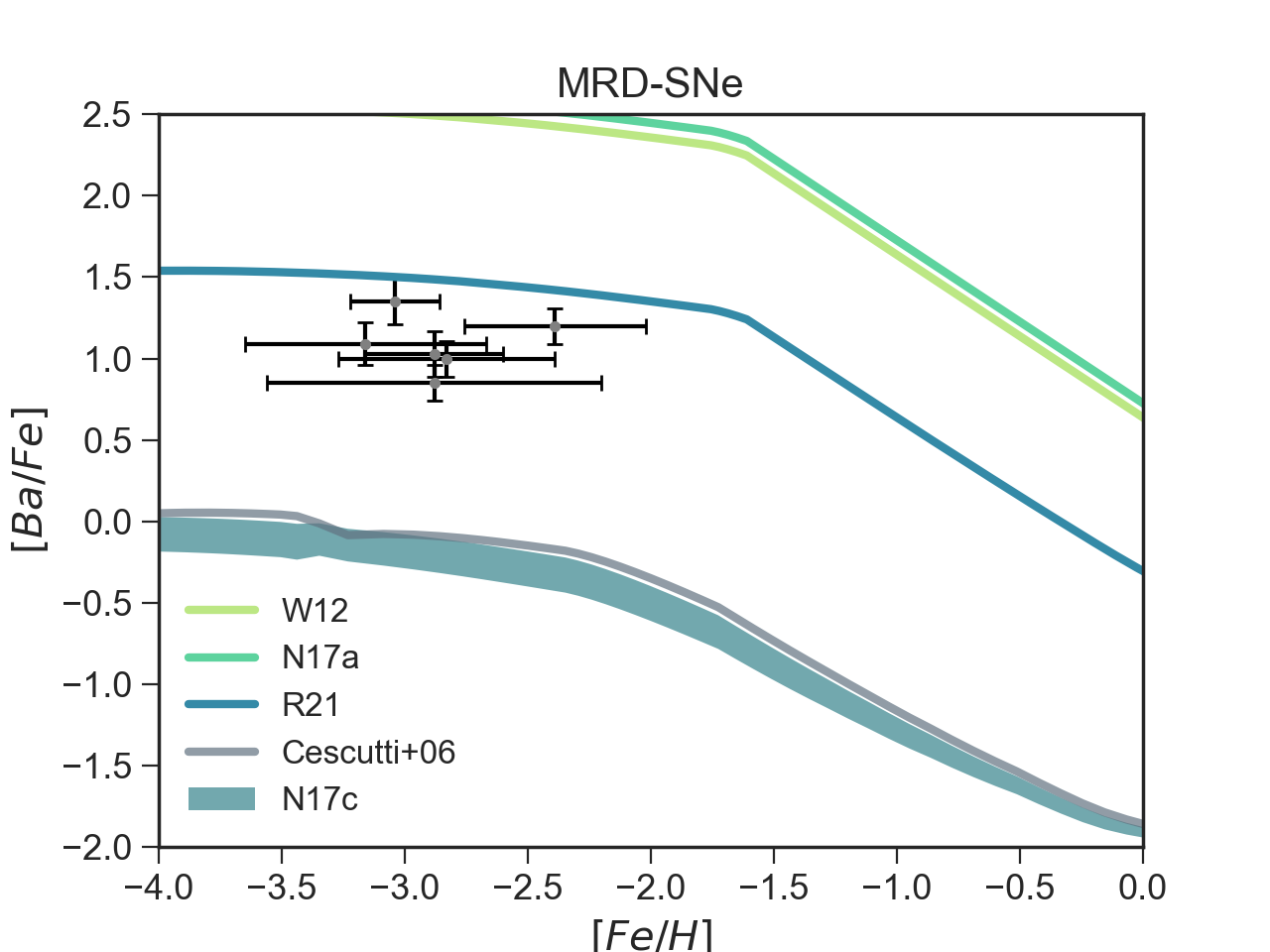}\label{fig:a}}
 \subfloat[]{\includegraphics[width=1\columnwidth]{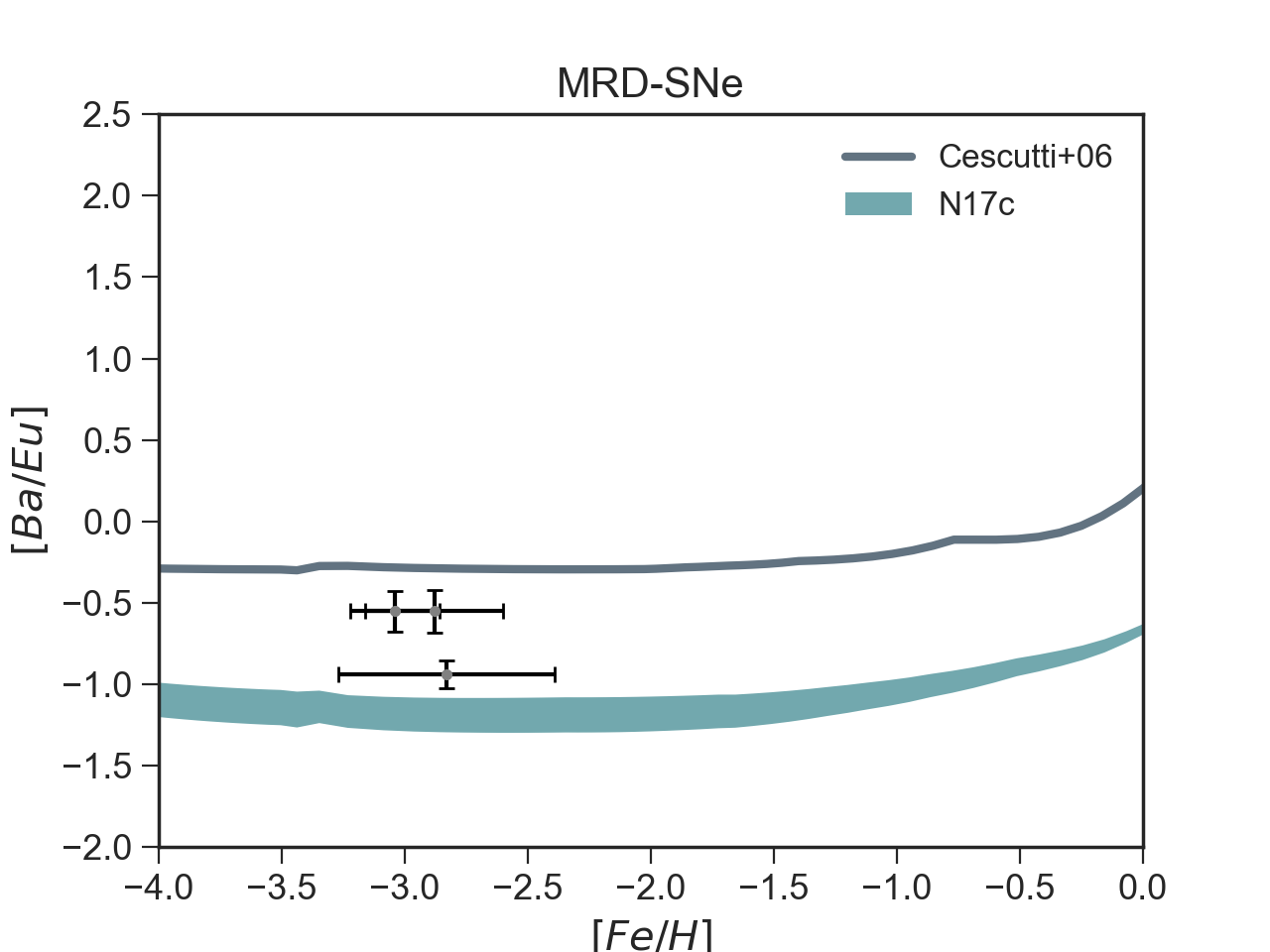}\label{fig:d}}
 \hfill
 \subfloat[]{\includegraphics[width=1\columnwidth]{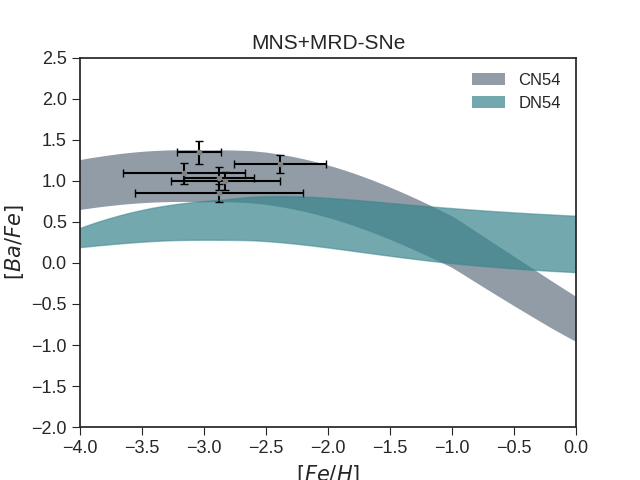}\label{figc:e}}
 \subfloat[]{\includegraphics[width=1\columnwidth]{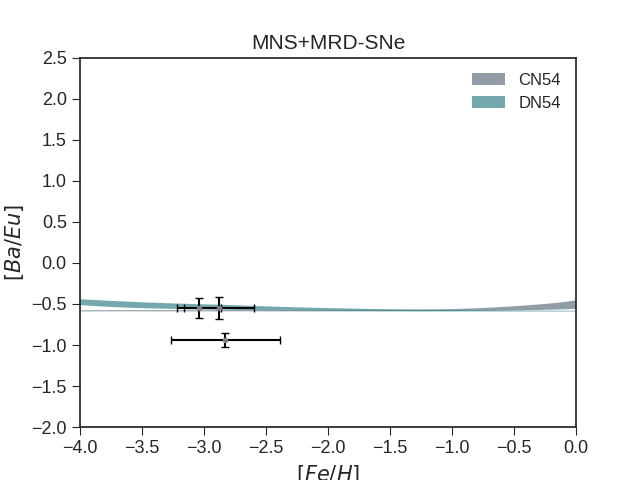}\label{fig:f}}
 \caption{Same of Figures \ref{fig: BaFe_Scl} and \ref{fig: BaFe_For} but for Reticulum II.}%
 \label{fig: BaFe_Ret}%
\end{center}
\end{figure*}

\subsubsection{Results for Barium in Reticulum II}
\label{results barium reticulum}

In Figure \ref{fig: BaFe_Ret} we report predictions for the [Ba/Fe] and [Ba/Eu] vs [Fe/H] together with the observational data. We note that for all models the production of the s-process fraction of Ba comes from LIMS and the adopted yields of \cite{Busso2001}, as for the other dwarfs.

In panels (a) and (b), we show results of models C54 and D54 in which we adopt MNS as the only producers of the r-process Ba and Eu with and without a DTD, respectively. For both models, we chose higher yields of r-process Ba from MNS with respect to those adopted for the other galaxies, fixing them in the $(3.20\times10^{-4}-1.58\times10^{-3})\mathrm{M_\odot}$ range. Yields of Eu from MNS are in the $(3.0\times10^{-5}-1.5\times10^{-4}) \mathrm{M_\odot}$ range. As seen from the [Ba/Fe], model C54 is able to fit the data at low metallicities. Then it predicts a constantly decreasing trend, as expected. On the other hand, model D54 is not able to fit the data, because of the delay in the production of r-process Ba. For the [Ba/Eu] vs [Fe/H], since we are assuming that both Eu and r-process Ba are produced by the same event (and therefore on the same timescale), the two models are both producing the expected plateau at low metallicities and are able to fit the observed data. Then, model C54 predicts an increasing pattern towards high metallicities, because of the s-process Ba production from LIMS. On the other hand, model D54 in which we have a DTD for MNS, predicts a constant plateau for all the range of metallicities. This is due to the fact that, because of the high yields of r-process Ba and of the delay in its production by MNS, the contribution to Ba from LIMS at high metallicity appears to be negligible. This happens only when we adopt a DTD, because in this case the contribution to the Ba production from MNS is stronger at high metallicities (see also panel (a) of the same Figure) with respect to the case in which we adopt a constant delay time for merging. This is the case, even though we adopt the same r-process yields for Ba and Eu in both models. We note that we cannot comment on the nature of Ba and Eu in Reticulum II at higher metallicities owing to the lacking observational data.

In panels (c) and (d) of the same Figure, we report results of model N17c in which we assume r-process Ba and Eu produced only by MRD-SNe with yields from \cite{Nishimura2017}. Also the case in which we adopt yields of \cite{Cescutti2006} for the production of both elements is shown. It clearly appears that, both models are not able to fit the high abundances of the [Ba/Fe], underproducing Ba by more than one order of magnitude. We tested also models W12, N12a and R21 for the [Ba/Fe], since they matched the [Eu/Fe] described in the previous sections. However, all of them overproduce the observed Ba abundances. In the case of the [Ba/Eu] vs [Fe/H], the two models produce a similar trend, but none of them is able to fit the observed data, underproducing or overproducing the expected abundances, respectively.

In panels (e) and (f), we show results of models DN54 and CN54 in which r-process elements are produced by both MRD-SNe and MNS (with and without a DTD, respectively). Yields for MRD-SNe are those of \cite{Nishimura2017} and yields for MNS are in the $(3.20\times10^{-4}-1.58\times10^{-3})\mathrm{M_\odot}$ range. For the [Ba/Fe], model CN54 in which two fast sources are producing r-process elements are able to fit the high observed abundances, while model DN54 in which also a delayed source is active underproduces the data. For [Ba/Eu], model CN54 can produce the expected plateau at low metallicities and the increasing pattern at higher [Fe/H], because the production of Eu and r-process Ba happens on the same timescales. On the other hand, model DN54 produces almost a constant plateau both at low metallicities, because of the similar r-process Ba/Eu yields between MRD-SNe and MNS, and at higher ones because of the same reasons explained for model D54 (panel (b) of the same Figure).

\subsubsection{A single r-process event}

The generally accepted explanation for the high r-process abundances observed in Reticulum II is that a single nucleosynthetic event produced a large quantity of r-process material ($\sim10^{-4.5}$ of Eu $\mathrm{M_\odot}$ according to \citealp{2016Ji}). As we showed in the previous sections, the amount of r-process material produced in our model by MNS should be in the $(10^{-5}-10^{-4}) \mathrm{M_\odot}$ range for Eu, in agreement with \cite{2016Ji} estimation, and in the $(10^{-4}-10^{-3}) \mathrm{M_\odot}$ range for Ba. 
However, these yields are 1-2 order of magnitude higher than those estimated for the other galaxies. The reason why we need high r-process yields in our model is that we are actually working with a fraction of one enrichment event. During the first Gyr of SF, in fact, we have a total of $2.39\times10^{-2}$ events of MNS when a DTD is adopted and of $5.65\times10^{-2}$ events in the case of a constant delay.

\begin{figure*}
\begin{center}
 \subfloat{\includegraphics[width=1\columnwidth]{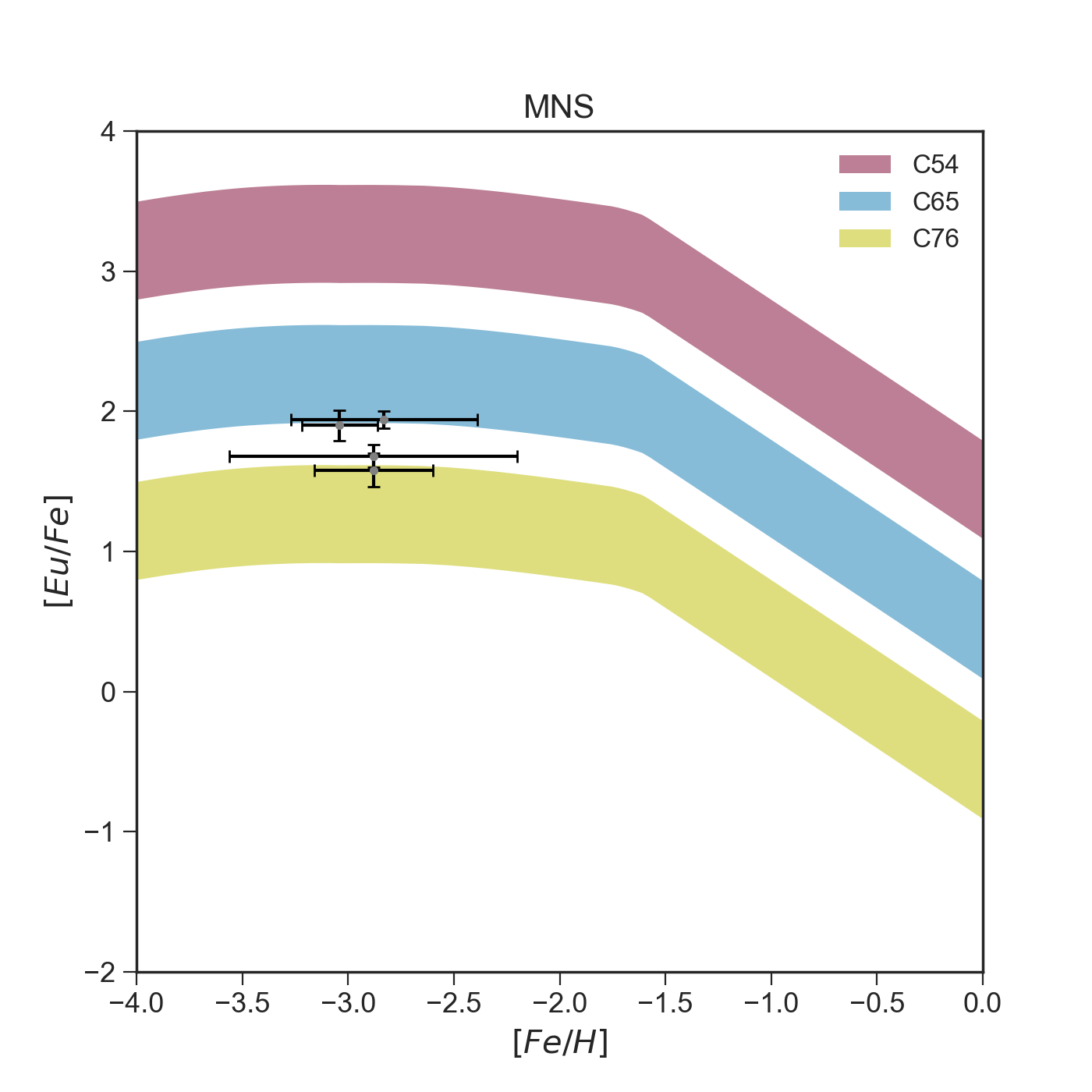}}
 \hfill
 \subfloat{\includegraphics[width=1\columnwidth]{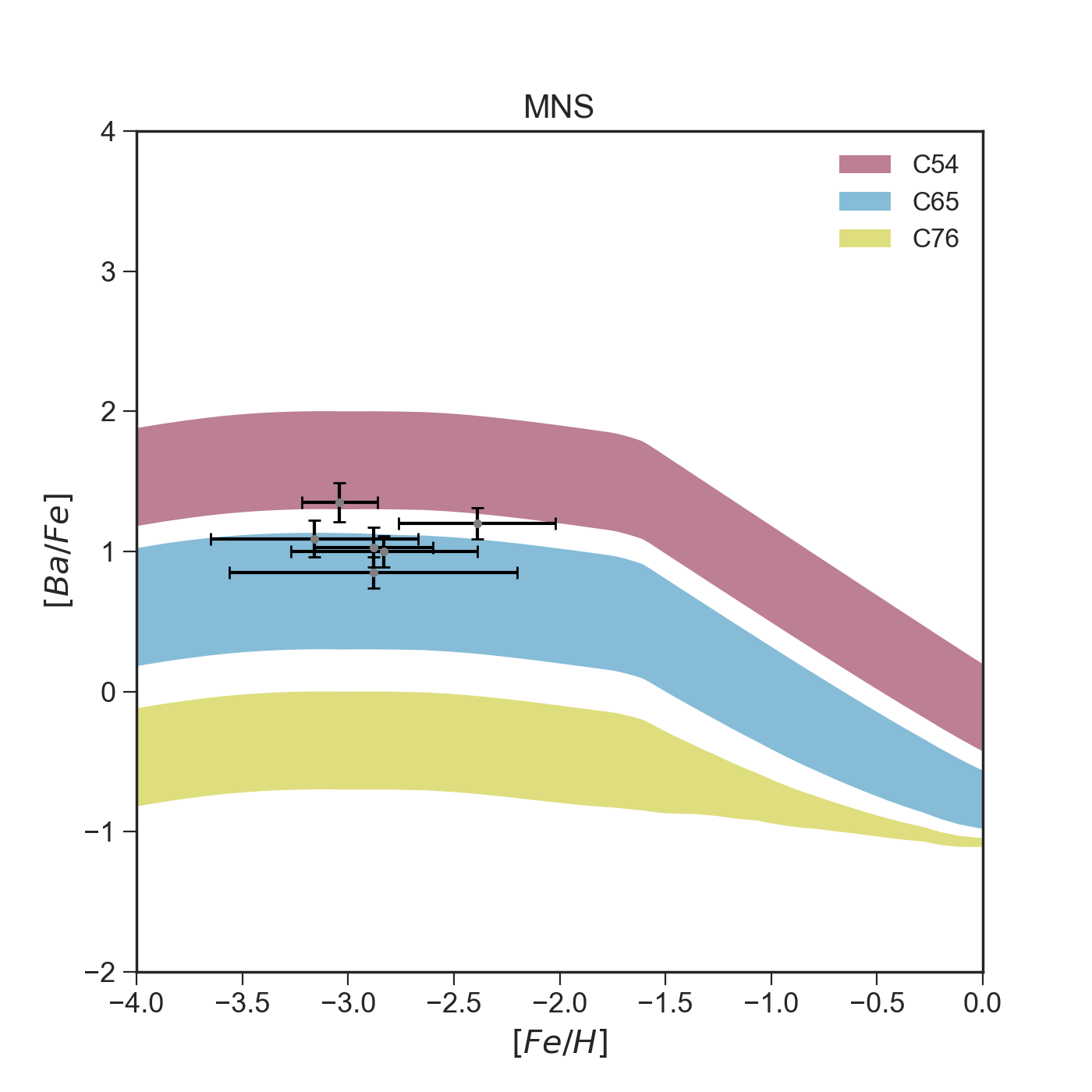}}%
 \caption{Results of models C54, C65 and C76 for Reticulum II for [Eu/Fe] and [Ba/Fe] vs [Fe/H] in the case of 1 event of MNS in the first Gyr.}%
 \label{fig: point1}%
\end{center}
\end{figure*}

Therefore, we performed a test in which we increased the value of the $\alpha_{\mathrm{MNS}}$ parameter to 1 in order to artificially obtain a total of 1 event of MNS in the first Gyr. A probability of 100\% of having a MNS event is a strong condition, but it is justified by the low stellar mass content of Reticulum II. We then computed the [Eu/Fe] and the [Ba/Fe] vs [Fe/H] abundances for the three sets of yields reported in Table \ref{tab: Yields of Sr, Eu, Ba}. The results are shown in Figure \ref{fig: point1}. As one can see, the observational data can now be reproduced by models which assume more reasonable r-process yields, similar to those of the other dwarfs. The yields can be in the $(1.50-3.00)\times10^{-6}\mathrm{M_{\odot}}$ range for Eu and around $1.50\times10^{-5}\mathrm{M_\odot}$ for Ba.

\section{Conclusions}
\label{conclusions}

We modelled the chemical evolution of seven dSph and two UFD galaxies in order to study the evolution of their Eu and Ba abundances. In the main text of the present work, we focused on the results obtained for Sculptor and Fornax, which can be taken as representative of those obtained for the others dSphs. Reticulum II UFD was shown for its peculiar elemental abundances. The results for the other galaxies are provided as Supplementary Material and available online only. We adopted new nucleosynthesis prescriptions for the production of Eu and the r-process Ba produced in MNS, scaled to the yields of Sr measured in the spectra of the kilonova AT2017gfo (\citealp{Watson2019}). We also tested different nucleosynthesis prescriptions for MRD-SNe r-process elements. Here, we summarize our main results and conclusions:\\

-For both Sculptor and Fornax we can conclude that:

\begin{itemize}

\item Models in which r-process elements are produced only by a unique quick source, such as MNS with a constant and short delay for merging or MRD-SNe, are able to reproduce the [Eu/Fe] vs [Fe/H]. However, those models fail in reproducing the low-metallicity data for [Ba/Fe];

\item On the contrary, models in which r-process elements are produced only with longer delays, namely by MNS with a DTD, have difficulties in reproducing the [Eu/Fe] vs [Fe/H], but succeed in reproducing the low-metallicity data for [Ba/Fe];

\item If both a quick source and a delayed one are adopted for the production of r-process elements, the [Eu/Fe] vs [Fe/H] is successfully reproduced. In particular, the quick source can be represented by MRD-SNe and the delayed one by MNS with a DTD. However, those models still fail in reproducing the low-metallicity data for [Ba/Fe].

\end{itemize}

It is reasonable to presume that a possible scenario is one in which NS merge with a DTD and produce Eu together with MRD-SNe. In this case, MRD-SNe can produce Eu at all metallicities or only at low ones, without making any significant difference in the final results. This allows us to reproduce the [Eu/Fe] vs [Fe/H] abundances, in agreement with what has been proposed by several authors (e.g.: \citealp{Simonetti2019, 2019cot, 2020skuladottirevidence, Molero2021}). In particular, the amount of Eu produced by each MNS event would be in the $(3.0\times10^{-6}-1.5\times10^{-5}) \mathrm{M_\odot}$ range, while that produced by MRD-SNe would be in the range of the theoretical calculations of \cite{Nishimura2017} and equal to $4.69\times10^{-7} \mathrm{M_\odot}$. Here we assume that only 1\%-2\% of all stars with initial mass in the (10-80)M$_\odot$ range would explode as MRD-SNe (according also to \citealp{WoosleyHeger2006}). However, within this scenario the low metallicity data of [Ba/Fe] cannot be reproduced. The only way to reproduce them is if only MNS (with DTD) are producing the r-process fraction of Ba, with yields in the $(3.20\times10^{-5}-1.58\times10^{-4}) \mathrm{M_\odot}$ range. If also MRD-SNe participate to this process, the agreement with the data is lost. Nevertheless, excluding MRD-SNe from the production of Ba cannot be physical motivated. Moreover, models in which r-process Ba is produced only by MNS with a DTD, still underestimate the [Ba/Fe] at intermediate metallicities, suggesting that a source for the production of the "weak" s-process fraction must be included. In particular, this second source for the production of s-process elements could be rotating massive stars, which have already been included in several studies to successfully explain the evolution of neutron capture elements. In particular, \cite{2013cescutti}, \cite{2014cescuttichiappini}, \cite{Cescutti2015} and more recently \cite{2021rizzuti}, showed that including the s-process from rotating massive stars in chemical evolution models is fundamental in order to explain the heavy element enrichment, in particular of Sr and Ba.\\

-For Reticulum II we conclude that:

\begin{itemize}

\item A quick source for the r-process production of both Eu and r-process Ba is needed in order to reproduce both the [Eu/Fe] and [Ba/Fe] vs [Fe/H] trend. This quick source can be represented either by NS with a constant and short delay time for merging or by MRD-SNe. However, the yields must be 1-2 order of magnitude higher than those estimated for the other galaxies.
\item If only one quick event of MNS is assumed to happen, a more realistic r-process yield can be adopted in order to reproduce both the [Eu/Fe] and [Ba/Fe] vs [Fe/H]. However, in this case the probability of having a MNS event must be 100\%.

\end{itemize}

Therefore, our conclusions for Reticulum II are different from those for the other galaxies, because of the peculiar r-/s-process elements pattern which characterizes this galaxy. 
Actually, a way to reproduce the high abundances observed is to adopt higher yields of Eu and r-process Ba. Moreover, for this galaxy we are inclined to discard models which adopt a DTD for MNS because of their inability to fit the [Ba/Fe] at low metallicities. Therefore, a scenario which well reproduces the Eu and Ba evolution in Reticulum II is the one in which a quick source pollutes the ISM really fast and with large amount of r-process elements. This source can be represented either by MRD-SNe or by NS which merge in a very short time, contrary to what happens in other galaxies. In particular, the quantity of r-process material produced should be in the $(10^{-5}-10^{-4}) \mathrm{M_\odot}$ range for Eu (in agreement with previous estimate of $\sim10^{-4.5} \mathrm{M_\odot}$ by \citealp{2016Ji}). 

However, the assumption that the same nucleosynthesis events produce different total amounts of r-process material in different environments needs further discussion. As also analysed by \cite{2019simon}, the only way the same mechanism which enriched Reticulum II could account for lower r-process abundances in other dwarfs is if the gas masses of those systems were much larger than in Reticulum II or if the retention fraction of r-process ejecta were much lower. However, analytical calculations (e.g.: \citealp{2017safarzadeh}; \citealp{2018beniamini}; \citealp{2019Safarzadeh}; \citealp{2020Tarumi}) excluded these possibilities. At the moment, the most common accepted theory is that a single nucleosynthetic event polluted the galaxy at early times with copious amount of r-process material. We therefore computed a test in which the rate of MNS was forced to be equal to 1 in the first Gyr of SF. This allowed us to adopt realistic r-process yields, similar to those obtained for other dwarfs/UFDs. In order to obtain such a rate of MNS we had to set $\alpha_{\mathrm{MNS}}=1$, namely we had to assume a probability of 100\% of having a MNS event. This is a strong assumption, which can be justified by the low stellar mass content of Reticulum II. However, in our opinion, the peculiar trend in Reticulum II needs to be further investigated. In particular, a more realistic explanation for the high abundances observed could be given by a poor mixing of metals into the galaxy gas (see \citealp{2020emerick}; \citealp{2020Tarumi}) and/or by a low Fe content due to the small number of SN. If this is the case, it would be difficult to prove it in the framework of a homogeneous model so that stochastic chemical evolution simulations, which take inhomogeneous mixing into account, would be required.

\section*{Acknowledgements}

We thank the anonymous referee for the useful comments and suggestions. M.M. thanks Deutsche Forschungsgemeinschaft through SFB 1245, which partially supported this work. D.R. acknowledges the financial support of INAF through the Main Stream grant CRA 1.05.01.86.28 assigned to the project “SSH: the Smallest Scale of Hierarchy. A.A. thanks ERC Starting Grant EUROPIUM-677912. G.A.L. thanks FAPESP grant 2017/25779-2.

\section*{Data Availability}

The data underlying this article will be shared upon request.



\bibliographystyle{mnras}
\bibliography{example} 





\appendix

\section{The other galaxies}

Results for Bootes I, Carina, Sagittarius, Sextan and Ursa Minor are presented as Supplementary Material in the online version of the journal. We show the same plots discussed in the main text of this work and do not further discuss those results, as they are analogous of those obtained for Sculptor, Fornax and Reticulum II.


\bsp	
\label{lastpage}
\end{document}